\documentclass[12pt,letterpaper,oneside]{article}
\usepackage{amsmath}
\usepackage{amssymb}
\usepackage[left=2cm,right=2cm,top=2.5cm,bottom=2.5cm]{geometry}
\usepackage{graphicx}
\usepackage{xcolor}	
\usepackage{feynmp}
\usepackage{slashed} 
\usepackage[colorlinks=false,linkcolor=blue,citecolor=blue]{hyperref}
\usepackage{subfigure}
\usepackage{booktabs}

\definecolor{lightgray}{gray}{0.9}

\usepackage[framemethod=default]{mdframed}
\newmdenv[skipabove=10pt,
skipbelow=10pt,
rightline=false,
leftline=false,
topline=false,
bottomline=false,
backgroundcolor=gray!20,
linecolor=gray,
innerleftmargin=10pt,
innerrightmargin=10pt,
innertopmargin=10pt,
innerbottommargin=10pt,
leftmargin=0cm,
rightmargin=0cm,
linewidth=4pt]{eBox}

\def\be{\begin{equation}}
\def\ee{\end{equation}}
\def\ba{\begin{eqnarray}}
\def\ea{\end{eqnarray}}
\def\bi{\begin{itemize}}
\def\ei{\end{itemize}}	
\def\l{\left}
\def\r{\right}

\def\d{\partial}

\def\mpl{M_{\rm p}}

\def\mbf{\mathbf}
\newcommand{\high}[1]{\underline{\emph{#1}}}

\begin{document}

\begin{fmffile}{graphs}

\title{{\bf An Introduction to Effective Field Theories}\vspace{.5cm}}
\author{ Riccardo Penco\thanks{Please contact me at \texttt{rpenco@cmu.edu} for comments or corrections.}}
\date{ \vspace{-.5cm} {\small \it Department of Physics, Carnegie Mellon University, Pittsburgh, PA} \\[5ex] {\small \today}}

\maketitle

\abstract{These notes are an overview of effective field theory (EFT) methods. I discuss toy model EFTs, chiral perturbation theory, Fermi liquid theory, and non-relativistic QED, and use these examples to introduce a variety of EFT concepts, including: matching a tree and loop level; summation of large logarithms; naturalness problems; spurion fields; coset construction; Wess-Zumino-Witten terms; chiral and gauge anomalies; field rephasings; and method of regions.  These lecture notes were prepared for the 2nd Joint ICTP-Trieste/ICTP-SAIFR school on Particle Physics, Sao Paulo, Brazil, June 22 - July 3, 2020.}

\newpage

\tableofcontents

\newpage

\section{Introduction}

A remarkable fact about nature is that interesting phenomena occur over a wide range of energy, length, and time scales. Over the last century alone, we have managed to stretch the scope of our investigations from microscopic scales as small as one billionth of the size of an atom ($10^4$ GeV) to macroscopic scales as large as the size of our observable universe ($10^{-42}$ GeV). This huge energy range is key to the success of physical sciences. If all physical phenomena had taken place at the same scale, we would have likely had to understand them all at once: progress in physics would have been impossible! Instead, in a first approximation we can take advantage of this separation of scales to set to infinity (zero) all scales that are much larger (smaller) than the typical energy $E$ of whatever physical process we happen to be interested in. In this limit, physics at scales much different from $E$ becomes irrelevant and can be neglected; if needed, its effect can be reintroduced in perturbation theory. To this day, this basic strategy is adopted by physicists in all subfields in a more or less implicit way. 

Effective field theories (EFTs) are a model building tool that explicitly implements the strategy outlined above and turns it into a precise, quantitative framework. Particle physicists like to think in terms of energy scales, but EFTs can be built by leveraging hierarchies between all sorts of  dimensionful quantities: lengths, times, velocities, momenta, angular momenta, etc... In fact, one could go as far as claiming that, when properly formulated, 
\vspace{.3cm}
\begin{quotation}
\noindent \hspace{3cm} {\it all physical theories are effective theories.}
\end{quotation}
\vspace{.3cm}

In what follows, I will introduce some of the main weapons in the EFT arsenal by showing how they work in a variety of different contexts. These notes are aimed at graduate students and junior postdocs with a high energy physics background. In fact, I tried to write the notes that I would have liked to read as a graduate student. Most of the content will therefore be familiar to the EFT practitioners,  although I hope that they too might still find something valuable here and there.

\subsection{Main ingredients of EFTs}

I will not follow a historical order of presentation, which would likely start with a discussion of Wilson's approach to the renormalization group (RG). Instead, I will jump right in and briefly summarize our present understanding of EFTs, since the main principles are very easy to state. An EFT is defined by an effective action, which in turn is completely specified by the following three ingredients: 
\begin{enumerate}
	\item {\it Degrees of freedom}. The first step when building an EFT is to figure out what are the degrees of freedom that are relevant to describe the physical system one is interested in. These are the variables that will appear in the effective action. The key word here is ``relevant'': you can always complicate the description of any phenomenon by adding additional structures, but you'll soon reach a point of diminishing return. Conversely, you can strive for the most economical description but, in the words of Albert Einstein, {\it ``everything should be made as simple as possible, but no simpler''}. Sometimes the degrees of freedom to be used are suggested by symmetry considerations, as is the case for Nambu-Goldstone modes; more often, though, the choice of degrees of freedom is an independent input, and it's a matter of art as much as science.  
	\item {\it Symmetries}. The second step in building an EFT consists in identifying the symmetries that constrain the form of the effective action, and therefore the dynamics of the system. Symmetries can come in many different flavors: they can be global, gauged, accidental, spontaneously broken, anomalous, approximate, contracted, etc... . We will encounter examples of all these possibilities in due time. For now, I will just mention that any term that is compatible with the symmetries of the system should in principle be included in the effective action. As a result, effective actions generically contain an infinite number of terms.
	\item {\it Expansion parameters}. The key to handling an action with an infinite number of terms lies in the fact that all EFTs feature one or more expansion parameters. These are small quantities controlling the impact that the physics we choose to neglect could potentially have on the degrees of freedom we choose to keep. For example, in particle physics these expansion parameters are often ratios of energy scales $E / \Lambda$, where $E$ is the characteristic energy scale of the process one is interested in, and $\Lambda$ is the typical energy scale of the UV physics one is neglecting. Other examples of expansion parameters include ratios of velocities ({\it e.g.} $v/c$ in a non-relativistic limit), angular momenta (as in a semi-classical expansion), or small dimensionless couplings. Observable quantities are calculated in perturbation theory as series in these small parameters. For this strategy to work, it is crucial to have an explicit power counting scheme, meaning that we should be able to assign a definite order in the expansion parameter to each term in the effective action. This ensures that only a finite number of terms contribute at any given order in perturbation theory, and that we can decide upfront which terms to keep in the action based on the desired level of accuracy.
\end{enumerate}

These three elements are fairly easy to state, but ensuring that they are properly implemented in an EFT can be a subtle matter. The way in which this is achieved in practice often depends on the problem at hand. A great deal of know-how has been accumulated over the last half a century, and the goal of these notes is to introduce you to the main tools of the~trade.

\subsection{Main advantages of EFTs}

Sometimes it is necessary to resort to an EFT description of a physical system because a more fundamental theory is lacking. Other times such a theory might be available, but extracting exact or even approximate predictions might be difficult, for example because the theory is strongly coupled. In these circumstances, EFTs can provide a more convenient, weakly coupled description of the same physics. Finally, even when EFTs are not forced upon us by necessity, they still remain a convenient framework. The main advantages of an EFT approach can be summarized as follows: 
\begin{itemize}
	\item EFTs drastically simplify calculations by focusing on the relevant degrees of freedom, and neglecting from the get-go those aspects that are not important for the problem at hand.
	\item When isolating the relevant degrees of freedom, new symmetries may become manifest that otherwise would have remained obscured. Often times, these symmetries can be used to draw general conclusions based on little or no calculations.  
	\item By working with an EFT, we essentially factorize calculations into two parts: one that involves the degrees of freedom that we are keeping (the actual EFT calculation), and another one that depends on the physics we are neglecting (the matching calculation). By adopting this ``modular'' approach, one can avoid repeating similar calculations over and over again.
	\item When dealing with problems that feature several different scales, there may be observables that depend on logarithms of their ratios. When these scales are widely separated, the logarithms become large and can worsen the accuracy of perturbative calculations. By focusing on one scale at a time, EFTs can use RG running to sum up these large logarithms. 
	\item EFTs contain a systematic parametrization of the neglected physics. When such physics is unknown or poorly understood, one can place model-independent constraints on it by comparing EFT predictions with experiments. 
	\item Observable quantities in an EFT are series in one or more expansion parameters. How far one goes in perturbation theory determines the precision of EFT results. 
	 \item Related to this, EFTs are kind enough to let you know when you should stop trusting their predictions. When you enter a regime where the expansion parameters are of $\mathcal{O}(1)$, perturbation theory stops working. This means that the degrees of freedom you have been using are no longer the relevant ones, and you should come up with a different EFT.
\end{itemize}
Some of these remarks might sound a bit obscure at this point. If so, I encourage you to continue reading, and come back to this section after we've discussed a few examples of EFTs. \\

\noindent {\it Conventions:} I will be following the conventions of~\cite{Srednicki:2007ab}, unless otherwise specified. This means in particular that I will be using a metric with ``mostly plus'' signature and I will be working in units such that $\hbar = c = 1$. I will also work in Gaussian units for electromagnetism. A pair of equal indices should always be understood as contracted when one is raised and the other one is lowered. I will \high{underline} the key concepts I introduce throughout the these lectures. I will only cite sources to provide additional background material or additional details, not to attribute credit. Apologies in advance to those who may feel that their contributions haven't been properly acknowledged. I will assume familiarity with quantum field theory, and refer the reader to~\cite{Srednicki:2007ab} for standard material.

\newpage

\section{A toy effective field theory} \label{sec: toy model}

In order to introduce some of the basic ideas and techniques surrounding effective field theories, we will start by considering the following toy model: 
\begin{align} \label{full theory}
	S = \int d^4 x \left\{ - \tfrac{1}{2} \d_\mu \phi \, \d^\mu \phi - \tfrac{1}{2} M^2 \phi^2 + \bar \Psi (i \gamma^\mu \d_\mu - m ) \Psi +  i y \phi \bar  \Psi \gamma^5 \Psi - \tfrac{1}{4!} \lambda \phi^4 \right\} .
\end{align}
This action is invariant under parity provided $\phi$ is a pseudoscalar, {\it i.e.} it transforms like $\phi \to - \phi$ under a parity transformation. A cubic self-interaction for $\phi$ would not be invariant and has therefore been omitted. We will consider a region of parameter space where $y, \lambda \ll 1$, so that we can work in perturbation theory. Moreover, we will assume that the scalar is much heavier than the fermion, {\it i.e.} $M \gg m$.

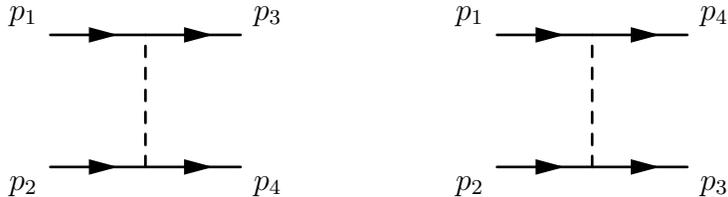
\begin{figure}
	\begin{center}
	\begin{fmfgraph*}(90,50) 
	\fmfleft{i1,i2} \fmfright{o1,o2} 
	\fmflabel{$p_1$}{i2}\fmflabel{$p_3$}{o2}
	\fmflabel{$p_2$}{i1}\fmflabel{$p_4$}{o1}
	\fmf{fermion}{i1,v1,o1} \fmf{fermion}{i2,v2,o2} 
	\fmffreeze
	\fmf{dashes}{v1,v2}
	\end{fmfgraph*}
	\hspace{2.5cm}
	\begin{fmfgraph*}(90,50) 
	\fmfleft{i1,i2} \fmfright{o1,o2} 
	\fmflabel{$p_1$}{i2}\fmflabel{$p_4$}{o2}
	\fmflabel{$p_2$}{i1}\fmflabel{$p_3$}{o1}
	\fmf{fermion}{i1,v1,o1} \fmf{fermion}{i2,v2,o2} 
	\fmffreeze
	\fmf{dashes}{v1,v2}
	\end{fmfgraph*}
	\vspace{.3cm}
	\caption{\small $2 \to 2$ fermion scattering at tree-level.}
		\label{fig fermion scattering full}
		\end{center}
\end{figure}

Imagine now that we are interested in describing a simple process such as the elastic scattering of two fermions. At lowest order in perturbation theory, {\it i.e.} at tree-level, this interaction is mediated by the exchange of a single scalar. The relevant Feynman diagrams are shown in Fig. \ref{fig fermion scattering full}, and the resulting scattering amplitude can be obtained using standard Feynman rules (see {\it e.g.}~\cite{Srednicki:2007ab}): 
\begin{align}
	i \mathcal{M} = (- y)^2 \left\{ \bar u_{\lambda_3, p_3} \gamma^5 u_{\lambda_1, p_1} \frac{-i}{(p_3-p_1)^2 +M^2} \,  \bar u_{\lambda_4, p_4} \gamma^5 u_{\lambda_2, p_2} \, - \, (3 \leftrightarrow 4) \right\} . 
\end{align}
If the typical energy $E$ of the fermions involved in this process is such that 
\begin{align} \label{energy range}
	m \ll E \ll M , 
\end{align}
the scalar propagator can be expanded in powers of momenta, and at lowest order our amplitude reduces to
\begin{align} \label{4pt amplitude EFT}
	i \mathcal{M} \simeq - i \, \frac{y^2}{M^2} \left[ \bar u_{\lambda_3, p_3} \gamma^5 u_{\lambda_1, p_1}  \bar u_{\lambda_4, p_4} \gamma^5 u_{\lambda_2, p_2} \, - \, (3 \leftrightarrow 4) \right] ,
\end{align}
up to corrections of $\mathcal{O}(E^2 /M^2)$. Interestingly, one could have also obtained this result starting from an \high{effective field theory} (EFT) described by the action
\begin{align} \label{4pt tree effective theory}
	S_{\rm eff} = \int d^4 x \left\{ \bar \Psi (i \gamma^\mu \d_\mu - m ) \Psi - \frac{y^2}{2 M^2} (\bar  \Psi \gamma^5 \Psi)^2 \right\} 
\end{align}
which makes no reference whatsoever to the heavy scalar field. This makes intuitive sense. Within the energy range in Eq. \eqref{energy range}, conservation of energy prevents heavy scalars from being created, and if there are no scalars in the initial state we should be able to describe what's going on solely in terms of fermions. This intuition is based on our successful track record in modeling all sorts of physical phenomena without worrying about those aspects of Nature we haven't discovered yet. In the context of our toy model, this means that physicists who don't have access to energies above $M$ should still be able to come up with a theory that describes experimental results. 

Importantly, as long as $E^2/M^2$ is smaller than the precision with which one can measure the scattering amplitude $\mathcal M$, the theories \eqref{full theory} and \eqref{4pt tree effective theory} will yield predictions that are virtually indistinguishable. Hence, by Occam's razor, we might as well use the EFT \eqref{4pt tree effective theory}, because it provides a simpler and equally accurate explanation for the scattering process under consideration. The ratio $E/M$ controls how well our EFT approximates the full theory: it is our expansion parameter. At energies $E > M$, our EFT ceases to be useful and we need to resort to the full theory.

A remarkable feature of the effective action \eqref{4pt tree effective theory} is its \high{locality}: $S_{\rm eff}$ contains only products of $\Psi$ and its derivative evaluated at the same point. Locality follows from the fact that $\phi$ mediates an interaction between the fermions which at lowest order in perturbation theory is described by the Yukawa potential
\begin{align}
	V(r) = - \frac{y^2}{4 \pi} \frac{e^{-M r}}{r} \ .
\end{align}
The processes we are considering are such that the momentum exchanged by the fermions is $q \sim E \ll M$. Therefore, we are probing the Yukawa potential with a resolution $ \Delta r \sim 1/q \gg 1/M$ which is much larger than the range of our potential. In this limit, $V(r)$ reduces essentially to a contact interaction: fermions can only interact with each other if they are at the same point (within the limits imposed by the uncertainty principle).

\subsection{Matching tree level amplitudes}

The effective action \eqref{4pt tree effective theory} correctly reproduces the tree-level 4-point amplitude of fermions one would get from the theory \eqref{full theory} at lowest order in $E/M$. It is then natural to ask how far our effective theory approach can be pushed: can we also recover higher order corrections in $E/M$ to the 4-point function? What about higher-point functions? And loop diagrams? In this section we will address the first two questions, postponing the third one until Sec. \ref{sec: matching at loop level}.

Based on the discussion in the previous section, our EFT in Eq. \eqref{4pt tree effective theory} should contain not just a handful of terms, but in fact all possible terms one can build out of $\Psi$ and its derivatives while preserving all the symmetries. Each of these terms will enter the effective Lagrangian multiplied by an a priori arbitrary coupling constant, or \high{Wilson coefficient}. The infinite number of these coefficients reflects the fact that there are in principle an infinite number of theories that at low enough energies contain a single Dirac field $\Psi$ and have the same symmetries. The process of figuring out the precise values that the Wilson coefficients must take in order to reproduce the low-energy physics of the particular theory one is interested in---in our case, the toy model \eqref{full theory}---is known as \high{matching}. In practice, matching calculations are usually carried out at the level of individual amplitudes and up to the desired order in the loop expansion---similarly to what we did in the previous subsection for the 4-point function at tree level. It is however instructive to discuss how this procedure can be understood directly at the level of the generating functional. 

Scattering amplitudes can be extracted from time ordered correlation functions of fields by using the LSZ reduction formula~\cite{Srednicki:2007ab}. In turn, any time ordered correlator can be obtained from the generating functional
\begin{align} \label{toy model generating functional}
	Z [ J, \bar \eta, \eta] = \int D \phi D \bar \Psi D \Psi \, e^{i S [ \phi, \Psi] + i \int (J \phi + \bar \eta \Psi + \bar \Psi \eta)} ,
\end{align}
by taking a suitable number of derivatives with respect to the external currents $J, \bar \eta$ and $\eta$, and then setting them to zero. If we are only interested in scattering processes involving fermions, then all we need are correlation functions of $\Psi$ and $\bar \Psi$. We can therefore set $J=0$ in Eq. \eqref{toy model generating functional} from the get-go, in which case the path integral can be factored out as follows: 
\begin{align}
	Z [ J=0, \bar \eta, \eta] &= \int  D \bar \Psi D \Psi \, e^{i /\hbar\int (\bar \eta \Psi + \bar \Psi \eta)} \int D \phi \, e^{i S [ \phi, \Psi]/\hbar} \nonumber \\ 
	&\equiv \int  D \bar \Psi D \Psi \, e^{i /\hbar\int (\bar \eta \Psi + \bar \Psi \eta)} e^{i S_{\rm eff} [\Psi]/\hbar} , \label{path integral effective action}
\end{align}
where we have momentarily restored all factors of $\hbar$. From this perspective, the effective action for $\Psi$ is obtained by \high{integrating out} the scalar field. Performing the integral over $\phi$ exactly captures the contribution of internal scalar lines to Feynman diagrams with external fermions. This leads to an effective action that admits an expansion in powers of $\hbar$ of the form $S_{\rm eff} = S_{\rm eff}^{(0)} + \hbar S_{\rm eff}^{(1)} + \mathcal{O} (\hbar^2)$, with the pieces proportional to $\hbar$ coming from loops of scalars. The leading-order term in this expansion, $S_{\rm eff}^{(0)}$, describes an EFT that reproduces all tree-level amplitudes of fermions in the full theory \eqref{full theory}; the next-to-leading-order term, $S_{\rm eff}^{(1)}$, reproduces all 1-loop corrections; and so on. 

The tree-level effective action $S_{\rm eff}^{(0)}$ is particularly easy to calculate, as it can be obtained by performing a saddle point approximation of the functional integral over $\phi$, {\it i.e.}
\begin{align} \label{saddle point tree level effective action}
	e^{i S_{\rm eff}^{(0)}[\Psi]/\hbar}  = e^{i S [ \hat \phi, \Psi]/\hbar}, \qquad \quad \text{with $\hat \phi[\Psi]$ a solution of} \qquad \quad \frac{\delta S[ \hat \phi, \Psi]}{\delta \hat \phi} = 0 \ .
\end{align}
The solution $\hat \phi$ can be calculated perturbatively in the the expansion parameter $E/M$ of our EFT. The field $\Psi$ plays the role of an external field for the purposes of this calculation. Varying the action \eqref{full theory} with respect to $\phi$ we find the equation of motion
\begin{align}
	(\square -M^2)\phi + y \mathcal{J} - \frac{\lambda}{3!} \phi^3 = 0 ,
\end{align}
where we have streamlined our notation by defining $\mathcal{J} \equiv \bar  i \Psi \gamma^5 \Psi$. Postponing for now a detailed discussion about the systematics of the $E/M$ expansion, we can solve perturbatively this equation by expanding the solution in inverse powers of $M$:
\begin{align}
	\hat \phi = \frac{y}{M^2} \left(1 + \frac{\square}{M^2} + \frac{\square^2}{M^4} + \frac{\square^3}{M^6} \right) \mathcal{J} -\frac{\lambda y^3}{6 M^8} \mathcal{J}^3 +  \cdots \ .
\end{align}
Plugging this solution back into the action for our toy model, and using Eq. \eqref{saddle point tree level effective action}, we find 
\begin{align} \label{toy tree level effective action}
	S_{\rm eff}^{(0)} = \int d^4 x \left\{ \bar \Psi (i \gamma^\mu \d_\mu - m ) \Psi + \frac{y^2}{2 M^2} \mathcal{J} \left(1 + \frac{\square}{M^2} + \frac{\square^2}{M^4} + \frac{\square^3}{M^6} \right) \mathcal{J} - \frac{\lambda y^4}{4!M^8}\mathcal{J}^4 + \cdots \right\}.
\end{align}
This is an improved version of the effective action \eqref{4pt tree effective theory}, which now captures all tree level amplitudes, not just the 4-point one. It is clear that this action contains an infinite number of terms---in fact, all possible terms that are compatible with the symmetries. The Wilson coefficients in front of the various terms are very specific functions of the parameters $M, y, \lambda$ in the full theory. This particular choice of couplings ensures that the tree-level scattering amplitudes in our EFT are exactly the same as the ones we would have obtained using the full theory: we have successfully matched our EFT onto our toy model at tree-level. 

We have been able to perform the matching procedure explicitly because (1) we knew the full theory at energies $E >M$, and (2) this theory was weakly coupled. The assumption of weak coupling, in particular, allowed us to read off the relevant degrees of freedom directly from the Lagrangian, and to figure out that the expansion parameter at low energies should be $E /M$. It is often the case that one of the two conditions above is not satisfied. One then needs to rely on experiments, brilliant theoretical intuition, or a combination of both, to figure out what are the relevant symmetries and degrees of freedom at low energies.  

If the full theory is known and strongly coupled but the low-energy degrees of freedom are weakly coupled, as is the case for QCD, the matching procedure can be carried out numerically. If instead the full theory is unknown, as is the case for the Standard Model, one can still write down an effective action with arbitrary Wilson coefficients. Measuring these coefficients experimentally, or even just placing upper bounds on their values, can give us important clues about the full underlying theory. Thus,                                  EFTs remain a helpful tool to describe the physics one is interested in, and to parametrize systematically the physics one is neglecting.

One last comment. Matching is ultimately about ensuring that the EFT and full theory make the same exact predictions for physical observables, {\it e.g.} on-shell scattering amplitudes. Nevertheless, our generating functional approach showed that one doesn't necessarily have to compare observable quantities to read off the Wilson coefficients. You can match by comparing unphysical quantities that are easier to calculate ({\it e.g.} off-shell amplitudes with vanishing external momenta), and then use the matched EFT to derive physical predictions.

\subsection{Power counting and strong coupling scale}

At this point, some readers might feel a bit uneasy about actions that include an infinite number of terms. This is especially true if the Wilson coefficients are to be treated as arbitrary parameters. The plethora of models one encounters while studying physics are all described by finite Lagrangians with just a handful of parameters. How is one supposed to handle an infinite Lagrangian? An infinite number of measurements would seem necessary to determine all Wilson coefficients, in which case how can such a theory be useful, or even predictive? As we will now discuss, the answer to these questions lies in the third defining property of an EFT: its expansion parameter(s).

The 4-point amplitude in Eq. \eqref{4pt amplitude EFT} was the leading term in an expansion in powers of $E/M$. In fact, all physical predictions of our EFT should be expressed as a series in $E/M$. How far one should push this expansion depends on how precise our measurements can be: there is no point in including relative corrections of order $(E/M)^n$ if these contributions cannot be measured. EFTs are a very pragmatic business. 

A moment of reflection should be sufficient to realize that all the terms in the tree-level effective Lagrangian \eqref{toy tree level effective action} that are suppressed by more than two powers of $M$ either give rise to higher-point amplitudes, or contribute to the 4-point function at higher order in $E/M$. The interaction shown in Eq. \eqref{4pt amplitude EFT} is the only one that can give a $1/M^2$ contribution to the 4-point function. Thus, if an experiment is only sensitive to the 4-point function at lowest order in $E/M$, then there is only one parameter to be measured.

More in general, our EFT is a useful framework as long as the outcome of any low-energy experiment with finite precision is described by a finite number of parameters. Then, only a finite number of measurements is needed to determine the values of these parameters and thus specify the EFT for all practical purposes. Our EFT is also a predictive framework because, after its parameters have been fixed, say, by measuring some scattering amplitudes at a certain energy, it can be used to make predictions for other observables ({\it e.g.}, to predict the values of those same amplitudes at different energies). 

In practice, for any given experiment one must be able to determine at the outset which terms in the effective Lagrangian should be kept and which ones can be neglected. This can be achieved provided one can tell a priori at which order in $E/M$ each term will contribute to observable quantities. The process of figuring out how each term in the effective action scales with the expansion parameter(s) is known as \high{power counting}. Let's see how this works in practice for our toy~EFT.

Within the energy range in Eq.~\eqref{energy range}, the 4-momentum of each fermion involved in a scattering process scales with energy like $p^\mu \sim E$. This is also how each derivative acting on $\Psi$ should scale in this regime, {\it i.e.} $\d_\mu \sim E$. Invoking the uncertainty principle, we can then deduce that $x_\mu \sim 1/E$ and $d^4 x \sim 1/E^4$. Finally, for a weakly coupled effective theory it seems reasonable to assume that the kinetic term be the leading term in an $E/M$ expansion. This requirement fixes the scaling of $\Psi$:
\begin{align} \label	{scaling Psi toy model}
\int d^4 x \, \bar \Psi i \gamma^\mu \d_\mu \Psi \sim 1 \qquad\quad  \Longrightarrow \qquad\quad  \Psi \sim E^{3/2} .
\end{align}

Using these scaling rules, we can associate a definite power of $E/M$ to each operator in the Lagrangian. For example, the last term in Eq. \eqref{toy tree level effective action} scales as follows:
\begin{align}
	\int d^4 x \left\{ - \frac{\lambda y^4}{4!M^8}\mathcal{J}^4 \right\} \sim \left(\frac{E}{M} \right)^8.
\end{align}
In this simple example, the fact that the matched Wilson coefficient was proportional to $1/M^8$ already gave us a hint on how this term should scale with our expansion parameter. As we will see later on, figuring out the scaling of operators in the Lagrangian is often not so easy. The crucial point though is that each operator must scale like a definite power of the expansion parameter in order for the EFT to be well defined.

The scaling rules we have developed so far allow us to determine the relative importance of different terms in the Lagrangian. This sheds some light on why the Lagrangians you might have encountered so far contained only a handful of terms: for many applications, the expansion parameter is sufficiently small that considering only the most important terms is good enough. 

When $E \sim M$, instead, all interactions become as important as the kinetic term, and perturbation theory stops working. For this reason, the scale $M$ is usually referred to as the \high{strong coupling scale} of the EFT. More colloquially, many people will also call $M$ the \high{cutoff} of the EFT---a hint at the fact that this is also the scale at which loops of slow modes are cut off in Wilson's approach to the renormalization group. In what follows I will refrain from using this term since, as we'll discuss in the next section, a consistent power counting scheme at loop level requires a mass independent regularization scheme such as dimensional regularization.

Let's now discuss how we can use our scaling rules to figure out the leading order contribution of each term in the Lagrangian to observable quantities. In the case of scattering amplitudes, we can use the LSZ reduction formula~\cite{Srednicki:2007ab}, which relates the probability amplitude for a scattering process $i \to f$ involving $n$ fermions to an $n$-point function. For approximately massless fermions, the LSZ formula can be written very schematically as 
\begin{align} \label{LSZ formula}
	\langle f | i \rangle \equiv (2 \pi)^4 \delta^4 (p_f - p_i) \, i \mathcal{M}_n \sim \left\{ \prod_{j=1}^n \int d^4 x_j \, e^{\pm i p_j \cdot x_j} \slashed{\partial} \, u_j \right\} \langle T \Psi (x_1) \dots \Psi (x_n) \rangle .
\end{align}
We've been very liberal in writing this equation, since some of the $u_j$'s could be $\bar u_j$'s, $v_j$'s or $\bar v_j$'s, and some of the $\Psi$'s could be $\bar \Psi$'s. Furthermore, we have suppressed all polarization indices and haven't worried too much about the order in which the various factors appear. All these details won't matter for the purposes of figuring out how the amplitude $\mathcal M_n$ scales with $E/M$.  

We need however to find out how the mode functions $u_j$ scale with the energy. To this end, recall that the free operator $\Psi$ can be decomposed into creation and annihilation operators as follows:
\begin{align}
	\Psi (x) = \sum_\lambda \int \frac{ d^3 p}{(2 \pi)^3 2 E_{\vec p}} \left[ b_{\lambda, \vec p} \, u_\lambda (\vec p) e^{i \vec p \cdot x} + d^\dag_{\lambda, \vec p} v_\lambda (\vec p) e^{-i \vec p \cdot x} \right] ,
\end{align}
with $\lambda$ the polarization index and $[ b_{\lambda, \vec p},  b^\dag_{\lambda', \vec p'} ] = [ d_{\lambda, \vec p},  d^\dag_{\lambda', \vec p'} ] = 2 E_{\vec p} \, (2\pi)^3 \delta (\vec p - \vec p') \delta_{\lambda \lambda'}.$ These commutators determine the scaling of the creation and annihilation operators, and then we can use Eq. \eqref{scaling Psi toy model} to conclude that $u_j \sim E^{1/2}$.\footnote{This is also consistent with the fact that $u_\lambda (\vec p) \bar u_\lambda (\vec p) \sim \slashed{p}$.~\cite{Srednicki:2007ab}}

Finally, the $n$-point function on the righthand side of Eq. \eqref{LSZ formula} is given by
\begin{align} \label{expectation values}
	\langle T \Psi (x_1) \dots \Psi (x_n) \rangle = \frac{\int  D \bar \Psi D \Psi \, \Psi (x_1) \dots \Psi (x_n) e^{i S_{\rm eff} [\Psi]}}{\int  D \bar \Psi D \Psi \, e^{i S_{\rm eff} [\Psi]}} \ .
\end{align}
Combining this result with the LSZ formula and all the scaling rules we derived in this section, we can estimate how each term in the Lagrangian contributes to the amplitude $\mathcal M_n$. in perturbation theory. For example, the leading order perturbative contribution to a $n$-point function coming from a term $\Delta S$ in the effective action with $n$ fields will scale like $\Psi (x_1) \dots \Psi (x_n) \Delta S$. 

\begin{eBox}
{\bf Exercise 2.1:} Show that contribution of a quartic interaction $\Delta S = \int d^4 x \, \mathcal{O} (x)$ to the 4-point amplitude $\mathcal M_4$ of fermions scales like $\mathcal{O} (x)$.
\end{eBox}
\begin{eBox}
{\bf Exercise 2.2:} Determine how the 6-point amplitude $\mathcal M_6$ scales with energy at leading order in the effective theory in Eq. \eqref{toy tree level effective action}.
\end{eBox}

\subsection{Matching at loop level} \label{sec: matching at loop level}

We have argued so far that an EFT described by the action \eqref{toy tree level effective action} can reproduce all tree level scattering amplitudes calculated using the toy model \eqref{full theory} at low energies. We will now show how the matching procedure can be successfully carried out also at loop level. We will learn that loop diagrams built from the tree level effective action $S^{(0)}_{\rm eff}$ aren't sufficient to reproduce all 1-loop corrections in the full theory. This is fine, of course, because $S^{(0)}_{\rm eff}$ is just the tree-level part of the effective action. This apparent mismatch between loop corrections in the EFT and full theory is fixed by the 1-loop part of the effective action, $S^{(1)}_{\rm eff}$. In fact, comparing loop diagrams is often a more efficient way of calculating $S^{(1)}_{\rm eff}$ in practice, as opposed to calculating the path integral in Eq. \eqref{path integral effective action}.	 The advantage of this method becomes even more striking at higher loop order. To illustrate this procedure, we will focus for simplicity on the 1-loop corrections to the 2-point function of fermions, and we'll work at lowest order in the $E/M$ expansion.  This will suffice to introduce the main subtleties that arise when matching loop contributions.

Let's start by considering 1-loop corrections in the EFT. For our purposes, it will be sufficient to calculate 1PI diagrams rather than the full propagator. At lowest order in $E/M$, there is only one Feynman diagram that contributes to the 2-point function of fermions:
\vspace{.9cm}
\begin{align} \label{one loop toy EFT 1}
	\parbox[t][9.5mm][b]{20mm}{\begin{fmfgraph*}(50,60) 
	\fmfleft{i} \fmfright{o} 
	\fmf{fermion}{i,v,o} 
	\fmffreeze
	\fmf{fermion,tension=.5}{v,v}
	\end{fmfgraph*}} = - i \, \frac{y^2}{M^2} \int \frac{d^4 k}{(2 \pi)^4} \, \gamma^5 \, \frac{-i  (- \slashed{k} + m)}{k^2 + m^2 - i \epsilon} \, \gamma^5 = - \, \frac{y^2 m}{M^2} \int \frac{d^4 k}{(2 \pi)^4} \, \frac{1}{k^2 + m^2 - i \epsilon} ,
\end{align}
where in the last step we used $(\gamma^5) =1$, as well as the fact that the contribution linear in $k$ must vanish by symmetry. This diagram clearly has a quadratic divergence in the UV. On physical grounds, one might be tempted to regularize this divergence by cutting off the integral over $k$ at the scale $M$. After all, our EFT is only supposed to be valid at energies $E < M$, where our expansion parameter is smaller than one. Moreover, this approach would resonate with Wilson's perspective on the renormalization group, where low energy theories for coarse grained observables are obtained by integrating out ``fast'' Fourier modes~\cite{Wilson:1993dy}. And yet, this would be a bad idea. By doing so, the resulting factor of $M^2$ would cancel against the $1/M^2$ in the Wilson coefficient, giving a result that is independent of the UV scale $M$. This would directly contradict our power counting rules, according to which a quartic self-interaction should always be suppressed by a factor of $E^2 / M^2$.

The most common solution to this problem is to use instead dimensional regularization. This method presents multiple advantages: in a gauge theory, it respects gauge invariance; in conjunction with the modified minimal subtraction scheme (or $\overline{MS}$ scheme, for short), it greatly simplifies the renormalization process; and, as we we will now show, it doesn't spoil power counting. By switching to $d = 4 -\varepsilon$ dimensions\footnote{We are using a ``curly'' epsilon ($\varepsilon$) to quantify  deviations from $d=4$. This should not be confused with the ``script'' epsilon ($\epsilon$) that appears for instance in the denominator of Eq. \eqref{one loop toy EFT 1}.}, the righthand side of Eq. \eqref{one loop toy EFT 1} reduces to
\begin{align} \label{one loop toy EFT 2}
	- \, \frac{y^2 \tilde{\mu}^\varepsilon m}{M^2} \int \frac{d^{4-\varepsilon} k}{(2 \pi)^{4-\varepsilon}} \, \frac{1}{k^2 + m^2 - i \epsilon} =\frac{i y^2}{16 \pi^2} \frac{m^3}{M^2} \left[ \frac{2}{\varepsilon} + 1 + \log\left(\frac{\mu^2}{m^2}\right) + \mathcal{O}(\varepsilon) \right] ,
\end{align}
where we introduced the renormalization scale $\tilde \mu$ by replacing $y \to y \, \tilde{\mu}^{\varepsilon/2}$ to ensure that $y$ remains dimensionless away from $d=4$; moreover, we have simplified our notation by introducing as usual the related scale $\mu = \sqrt{4 \pi} e^{-\gamma} \tilde \mu$, with $\gamma = 0.55772...$ the Euler constant. The loop integral now must be proportional to the only scale that appears in the integrand: the fermion mass $m$. As a result, our one loop correction is now suppressed by a factor of $m^2 /M^2 = m^2 /E^2 \times E^2 / M^2$, as expected based on power counting.
\begin{eBox}
{\bf Exercise 2.3:} Derive the result in Eq. \eqref{one loop toy EFT 2}.
\end{eBox}

Let's now turn our attention to the full theory. In the EFT we calculated a 1PI diagram of fermions. Hence, we should now calculate all 1-loop diagrams that are 1PI with respect to fermions, but not necessarily with respect to the heavy scalar. There are therefore two Feynman diagrams that in principle we should consider at 1-loop. The first one is:
\vspace{3mm}

\begin{align}
	\parbox[t][11.5mm][b]{24mm}{\begin{fmfgraph*}(60,70) 
	\fmfleft{i} 
	\fmfright{o} 
	\fmftop{t}
	\fmf{fermion}{i,v1,o} 
	\fmffreeze
	\fmf{dashes,tension=.7}{v1,v2}
	\fmf{fermion,left,tension=.3}{v2,t,v2} 
	\end{fmfgraph*}} = (-y)^2 \gamma^5 \, \frac{-i}{M^2} \int \frac{d^4 k}{(2 \pi)^4} \,   \mbox{Tr} \left[ \gamma^5 \, \frac{-i  (- \slashed{k} + m)}{k^2 + m^2 - i \epsilon} \right] = 0 .  
\end{align}
This diagram vanishes because the tadpole diagram one would obtain by removing the external fermion lines must vanish to preserve parity. There is one more loop correction in the full theory, namely
\vspace{3mm}
\begin{align}
	\parbox[t][8mm][b]{30mm}{\begin{fmfgraph*}(80,50) 
	\fmfleft{i} \fmfright{o} 
	\fmf{fermion}{i,v1,v2,o} 
	\fmffreeze
	\fmf{dashes,left}{v1,v2}
	\end{fmfgraph*}} & = (-y)^2 \int \frac{d^4 k}{(2 \pi)^4} \,  \frac{(-i)^2 \gamma^5 (- \slashed{k} - \slashed{p} + m)\gamma^5}{[(k+p)^2 + m^2 - i \epsilon](k^2 + M^2 - i \epsilon)} \label{1 loop full theory 1} \\
	& = - y^2 \int_0^1 dx \int \frac{d^4 q}{(2 \pi)^4} \frac{\slashed{q} + (1-x) \slashed{p} +m}{(q^2 + D)^2} \ ,
\end{align}
where in the last step we have introduced the Feynman parameter $x$ to combine the denominators and have defined $D \equiv x(1-x) p^2 + x m^2 + (1-x)M^2$. In order to allow for a comparison with the EFT result, we are going to regularize once again our loop integral using dimensional regularization. This yields: 
\begin{align}
& - y^2 \tilde \mu^\varepsilon \int_0^1 dx \int \frac{d^{4-\varepsilon} q}{(2 \pi)^{4-\varepsilon}} \frac{\slashed{q} + (1-x) \slashed{p} +m}{(q^2 + D)^2} = \nonumber \\
& \qquad \qquad \qquad = \frac{i y^2}{16 \pi^2}\left\{ - \frac{1}{\varepsilon} (\slashed{p} + 2 m) + \int_0^1 dx \left[ (1-x) \slashed{p} + m \right] \log (D/\mu^2) + \mathcal{O}(\varepsilon)\right\} \ . \label{1 loop full theory 2}
\end{align}
Since our EFT calculation was valid only up to $\mathcal{O}(E^2/M^2)$, we should expand Eq. \eqref{1 loop full theory 2} in powers of $m^2/M^2$ and $p^2 / M^2$ before making any comparison. Keeping all the terms up to $\mathcal{O} (M^{-2})$, we find

\begin{align}
	\parbox[t][11.5mm][b]{30mm}{\begin{fmfgraph*}(80,70) 
	\fmfleft{i} \fmfright{o} 
	\fmf{fermion}{i,v1,v2,o} 
	\fmffreeze
	\fmf{dashes,left}{v1,v2}
	\end{fmfgraph*}} & = \frac{i y^2}{16 \pi^2} \bigg\{ - \frac{1}{\varepsilon} (\slashed{p} + 2 m) + \frac{\slashed{p}}{2} \left[ - \frac{1}{2} + \log \left( \frac{M^2}{\mu^2} \right) + \frac{m^2}{M^2} \right] \label{one loop full theory 3} \\
	&  \qquad   + m \left[ -1 + \log \left( \frac{M^2}{\mu^2} \right) + \frac{m^2}{M^2} \log \left( \frac{M^2}{m^2}\right) + \frac{p^2}{M^2}  \right] + \mathcal{O}(\varepsilon, p^3, M^{-4}) \bigg\} . \nonumber 
\end{align}
\begin{eBox}
{\bf Exercise 2.4:} Derive the result in Eq. \eqref{one loop full theory 3}.
\end{eBox}

Having calculated the 1-loop correction to the 2-point function of fermions both in the EFT, Eq.~\eqref{one loop toy EFT 2}, and in the full theory, Eq.~\eqref{one loop full theory 3}, we are now in a position to draw several conclusions:
\begin{enumerate}
	\item The EFT loop integral in Eq.~\eqref{one loop toy EFT 1}  can be obtained from the full theory loop integral in  Eq.~\eqref{1 loop full theory 1} by expanding the integrand in powers of $1/M$. At a pictorial level, this is encoded by the fact that the Feynman diagram in the EFT can be obtained from the one in the full theory by contracting the scalar propagator (dashed line) to a point.  
	\item Despite this connection, the final results in Eqs. \eqref{one loop toy EFT 2} and \eqref{one loop full theory 3} are different. This means that loop integration and series expansion are not interchangeable operations. Notice in particular that the EFT result misses all the terms that are non-analytic around $M =\infty$. This makes sense, since the very expansion in inverse powers of $M$ relies on the assumption of analyticity. 
	\item The UV divergent pieces, proportional to $1 /\varepsilon$, are also different in the EFT and the full theory. This implies that the mass parameters and fermion fields in the two theories have different anomalous dimensions, and thus should not be confused with each other. 
	\item On the contrary, the terms that are non-analytic around $m=0$ are identical. This must always be the case, and in practice can be used as a consistency check. As we will discuss in the next section, this cancellation is crucial for EFTs to be able to separate IR and UV scales and tame large logarithms.
	\item Last but not least, loop corrections in the EFT are automatically suppressed by the small ratio $m^2 / M^2$---see {\it e.g.} Eq. \eqref{one loop toy EFT 2}---unlike in the full theory. This ensures that only a finite number of loop diagrams need to be calculated at any given order in $E/M$. 
\end{enumerate}

From now on, we'll carry out the renormalization procedure both in the EFT and the full theory using the $\overline{MS}$ scheme. From a practical viewpoint this is tantamount to dropping all factors of $1/\varepsilon$ from the final results. The discrepancy between the renormalized 1-loop diagrams in both theories can then be eliminated by introducing an additional tree-level diagram in the EFT equal to
\begin{align}
	& \parbox[t][8mm][b]{30mm}{\begin{fmfgraph*}(78,50) 
	\fmfleft{i} \fmfright{o} 
	\fmf{fermion}{i,v,o} 
	\fmffreeze
	\fmfv{d.sh=circle,d.f=empty,d.si=.2w,l=$\!\!\!\!\! 1$}{v}
	\end{fmfgraph*}}  = \left[ \begin{array}{c} \text{1-loop} \\ \text{full theory}
	\end{array} \right]_{\overline{MS}} - \quad \left[ \begin{array}{c} \text{1-loop} \\ \text{EFT}
	\end{array} \right]_{\overline{MS}} \nonumber \\
	& =  \frac{i y^2}{16 \pi^2} \left\{ - \slashed{p} \left[ \frac{1}{4}-\frac{m^2}{M^2} - \log \left( \frac{M^2}{\mu^2} \right) \right] - m\left(1+\frac{m^2}{M^2} \right) \left[ 1 - \log \left(\frac{M^2}{\mu^2} \right) \right] + \frac{m}{M^2} \, p^2 \right\} . \label{toy matching correction}
\end{align}
This diagram arises from the 1-loop correction to the effective action, which therefore must have the form
\begin{align}
	S^{(1)}_{\rm eff} & = \int d^4 x \,  \frac{y^2}{16 \pi^2} \bigg\{ \left[ \frac{1}{4}-\frac{m^2}{M^2} - \log \left( \frac{M^2}{\mu^2} \right) \right] \bar \Psi i \gamma^\mu \partial_\mu \Psi \nonumber  \\
	&   \qquad \qquad \qquad \qquad \qquad - m\left(1+\frac{m^2}{M^2} \right) \left[ 1 - \log \left(\frac{M^2}{\mu^2} \right) \right] \bar \Psi \Psi - \frac{m}{M^2} \bar \Psi \square \Psi + \cdots \bigg\} ,  \label{toy one loop effective action}
\end{align}
where the dots stand for terms that are cubic and higher in $\Psi$. Notice that the first two terms can be combined with similar terms  in the tree-level effective action \eqref{toy tree level effective action}. The last term instead is new: it did not show up at tree-level, but it is generated at loop level in accordance with the principle that all terms allowed by symmetries belong in the effective action.  

One last comment: we have chosen to match off-shell---{\it i.e.}, for arbitrary values of $p^2$--- but we could have just as well worked on-shell---{\it i.e.} momenta such that $p^2 = - m^2$. This would have changed the last term in parentheses in Eq.~\eqref{toy matching correction} to $-m^3 / M^2$. Consequently, the last term in the 1-loop effective action \eqref{toy one loop effective action} would have been replaced by an additional contribution proportional to $\bar \Psi \Psi$. This new effective action is physically equivalent to the one we have derived. It can be obtained from ours via a (perturbative) field redefinition~\cite{Georgi:1991ch}---an operation that doesn't change the $S$-matrix~\cite{Weinberg:1995mt}. This field redefinition is tantamount to using the lowest order equations of motion to remove higher derivative terms. Incidentally, this shows that \high{some operators are redundant}, in the sense that their effect can be captured by modifying the Wilson coefficients of other operators. Matching on-shell automatically eliminates redundant operators. 

\begin{eBox}
{\bf Exercise 2.5:} Derive the field redefinition that connects the effective actions obtained by matching off- and on-shell.
\end{eBox}

To summarize, we've extracted the quadratic part of $S^{(1)}_{\rm eff}$ by comparing 1-loop corrections to 2-point amplitudes in the EFT and the full theory. Similarly, one could compare higher-loop and higher-point amplitudes in the two theories to calculate respectively higher-loop and nonlinear corrections to the effective action.

\subsection{Summing large logarithms}

So far we have argued that EFTs are convenient to use because they focus on the degrees of freedom that are relevant for the observables one is interested in. Nevertheless, in the particular case of our toy example this meant going from a Lagrangian with just a handful of terms to an effective Lagrangian that in principle includes an infinite number of terms. Although we have argued that only a finite number of terms will contribute to any calculation with a finite precision, some readers might still feel that the price to pay is just too high. To cheer up those readers, we will now discuss another reason why EFTs are the smart way of organizing calculations in theories with multiple scales. 

The introduction of a renormalization scale $\mu$ is an inevitable byproduct of the renormalization process. It is a priori an arbitrary scale, and therefore physical predictions cannot depend on it. To ensure this, the renormalized couplings must depend on $\mu$ in a way that is captured by the renormalization group (RG) equations. 

In perturbative calculations, it is helpful to choose $\mu$ in such a way as to avoid the appearance of large logarithms in the final result. Consider for instance the full theory 1-loop correction in Eq. \eqref{one loop full theory 3}. The part proportional to $\slashed{p}$ contains a term of the form $\frac{y^2}{16 \pi^2} \log (M/\mu)$. Similarly, the $n$-th loop correction will introduce terms of order $\left[ \frac{y^2}{16 \pi^2} \log (M/\mu) \right]^n$, and therefore a naive application of perturbation theory will converge more slowly if $\log (M /\mu)$ is large. This can be avoided by choosing $\mu \sim M$.\footnote{This requires one to know the value of the couplings at the scale $M$. If these couplings are only known at a different scale, one can always extrapolate their values at $\mu \sim M$ by solving the RG equations.} In theories with a large hierarchy of different scales, logarithms can also depend on ratios of UV and IR scales---see for instance the $\log (M^2 /m^2)$ in the second line of Eq.~\eqref{one loop full theory 3}---and thus large logarithms could seem unavoidable. We will now argue that this problem can be easily circumvented using an EFT at low energies.

Consider again the effective action we calculated in the last section. Adding together the tree-level and 1-loop contributions, the quadratic part reads
\begin{align}
	S^{(0)}_{\rm eff} + S^{(1)}_{\rm eff} & = \int d^4 x \,\left[ \mathcal{Z}(\mu) \bar \Psi i \gamma^\mu \partial_\mu \Psi -  \mathcal{M}(\mu) \bar \Psi \Psi - \mathcal{N}(\mu) \bar \Psi \square \Psi + \cdots \right] ,
\end{align}
where the dots stand once again for the non-linear part of the action, and we have introduced
\begin{subequations} \label{EFT wilson coefficients}
\begin{align}
	\mathcal{Z}(\mu) &= 1 + \frac{y^2}{16 \pi^2} \left[ \frac{1}{4}-\frac{m^2}{M^2} - \log \left( \frac{M^2}{\mu^2} \right) \right] , \label{Z 1-loop toy} \\
	\mathcal{M}(\mu) &= m \left\{ 1 + \frac{y^2}{16 \pi^2}\left(1+\frac{m^2}{M^2} \right) \left[ 1 - \log \left(\frac{M^2}{\mu^2} \right) \right] \right\} , \label{M 1-loop toy} \\
	\mathcal{N}(\mu) &= \frac{m}{M^2} \ . \label{N 1-loop toy}
\end{align}
\end{subequations}
We have argued above that this effective action exactly reproduces the 2-point function of the full theory at 1-loop up to $\mathcal{O}(E^2/M^2)$ included. And yet, something very interesting has happened. 

On the one hand, the Wilson coefficients in Eqs.~\eqref{EFT wilson coefficients} only depend on $\log (M^2 / \mu^2)$. Notice that this wouldn't have happened if the logarithms of the IR scale $m$ hadn't dropped out in the matching---a remarkable fact we have already emphasized in the previous section. Therefore, large logarithms in the effective action can be avoided altogether by carrying out the matching procedure at $\mu = M$. In other words, we should calculate the values of the Wilson coefficients in the EFT at the UV scale $M$. 

On the other hand, the EFT 1-loop correction in Eq.~\eqref{one loop toy EFT 2} only depends on $\log (m^2 / \mu^2)$. Thus, large logarithms in the 2-point function can be avoided by calculating the 1-loop correction at the IR scale $\mu = m$. This of course requires knowing the values of the Wilson coefficients at the IR scale $m$. These can be obtained by solving the RG equations to run the Wilson coefficients from $\mu = M$ (where their value is obtained from matching) down to $\mu = m$. 

To summarize, by switching from the full theory to an EFT at energies $E < M$ we are essentially decomposing large logarithms in the full theory in two parts: 
\begin{align}
	\log \left( \frac{M^2}{m^2} \right) = \log \left( \frac{M^2}{\mu^2} \right) - \log \left( \frac{m^2}{\mu^2} \right) \ . 
\end{align}
The first part gets absorbed in the Wilson coefficients of the EFT, while the second part is ``resummed'' by solving the RG equations. It is important to stress that it is the combination of \high{matching and running} that tames the large logarithms. RG running in the full theory wouldn't be sufficient to get the job done, because the RG group can only ``resum'' logarithms that depend on the renormalization scale $\mu$.

\subsection{A modern perspective on renormalizable theories} \label{sec: renormalizable theories}

Historically, a great emphasis has been placed by the particle physics community on renormalizable quantum field theories. As a matter of fact, the toy model with we started with in Eq.~\eqref{full theory} was renormalizable. This particular class of theories enjoys the special property that any UV divergence arising in a loop calculation can be absorbed into a redefinition of fields and a finite number of couplings. This was originally believed to be necessary in order for the theory to be predictive, so much so that renormalizability was often regarded as a fundamental requirement for particle physics models.  By now, we understand that EFTs are equally predictive within their regime of applicability, despite the fact that they include an infinite number of ``non-renormalizable'' interactions. Following our discussion on matching at loop level, we are in a position to understand the special status of renormalizable theories. 

The Wilson coefficients of our EFT depend on the UV scale $M$ in two ways: (1) powers of the ratio $m^2/M^2$, and (2) logarithms of the form of $\log (M^2/\mu^2)$. By evaluating the Wilson coefficients at the scale $\mu = M$, all the logarithms vanish, and we are left with the  inverse powers of $M$. These suppress what would be regarded as ``non-renormalizable'' (or \emph{irrelevant}) terms (see {\it e.g.} Eq.~\eqref{N 1-loop toy}), and also provide corrections to the coefficients of ``renormalizable'' (or \emph{relevant}) terms (see {\it e.g.} Eqs.~\eqref{Z 1-loop toy} and~\eqref{M 1-loop toy}).  Then, our EFT reduces to a (in our case, free) renormalizable theory when we take the formal limit $M \to \infty$ while keeping $m$ and $y$ fixed. 

More in general, renormalizable theories arise in the limit where the UV scale is taken to be infinitely larger than all IR scales, and the UV physics decouples completely. The bare couplings of a renormalizable theory are then just a finite subset of Wilson coefficients evaluated at $M = \infty$. The divergences arising in loop calculations can now all be absorbed in a redefinition of the bare couplings and fields, because any other divergence would be suppressed by powers of $1/M \to 0$. (I am obviously being cavalier about the order of the limits $M \to \infty$ and $\varepsilon \to 0$ in order to get the main point across).

\subsection{Naturalness} \label{sec: naturalness}

It is instructive to consider our toy model in Eq. \eqref{full theory} in a different region of parameter space, where the fermion is much heavier than the scalar, {\it i.e.} $m \gg M$. We can then follow the same logic we have developed so far in this section and integrate out the fermion to obtain an EFT for the scalar valid at energies $E \ll m$.

 It is easy to see that the tree-level effective action $S^{(0)}_{\rm eff}$ is equal to the action in Eq. \eqref{full theory} with $\Psi = 0$. To calculate the quadratic part of the 1-loop effective action $S^{(1)}_{\rm eff}$, we compare the 1-loop correction to the scalar propagator obtained using the tree-level EFT and the full theory. In dimensional regularization, the EFT yields
\vspace{.7cm} 
\begin{align}
	\parbox[t][9.5mm][b]{20mm}{\begin{fmfgraph*}(50,60) 
	\fmfleft{i} \fmfright{o} 
	\fmf{dashes}{i,v,o} 
	\fmffreeze
	\fmf{dashes,tension=.5}{v,v}
	\end{fmfgraph*}}	= - \frac{i \lambda \tilde{\mu}^\varepsilon}{2} \int \frac{d^{4-\varepsilon} k}{(2 \pi)^{4-\varepsilon}} \frac{-i}{k^2 + M^2 - i \epsilon} = \frac{i \lambda}{32 \pi^2} M^2 \left[ \frac{2}{\varepsilon} + 1 + \log\left(\frac{\mu^2}{M^2}\right) + \mathcal{O}(\varepsilon) \right] . \label{toy EFT scalar loop diagram}
\end{align}

This same diagram appears also in the full theory, supplemented by a second one with a fermion running in the loop:
\begin{align}
	\parbox[t][11.5mm][b]{24mm}{\begin{fmfgraph*}(60,70) 
	\fmfleft{i} 
	\fmfright{o} 
	\fmf{dashes}{i,v1} 
	\fmf{dashes}{v2,o} 
	\fmf{fermion,left,tension=.3}{v1,v2,v1} 
	\end{fmfgraph*}} \!\! = \,\, - (- y)^2 \int \frac{d^4 k}{(2 \pi)^4}\text{Tr} \left[ \frac{-i(- \slashed{k} - \slashed{p} + m)}{(k+p)^2 + m^2 - i \epsilon} \gamma^5 \frac{-i(- \slashed{k} +m)}{k^2 + m^2 - i \epsilon}\gamma^5 \right] . 
\end{align}
A word of caution: the matrix $\gamma^5$ is not well defined away from $d=4$. Thus, we need to eliminate it using $(\gamma^5)^2 = 1$ and $\gamma^5 \slashed{p} \gamma^5 = - \slashed{p}$ before switching to $d = 4 - \varepsilon$. Taking this precaution, and expanding the final result in powers of $p^2 / m^2$, we find
\begin{align}
	\parbox[t][11.5mm][b]{24mm}{\begin{fmfgraph*}(60,70) 
	\fmfleft{i} 
	\fmfright{o} 
	\fmf{dashes}{i,v1} 
	\fmf{dashes}{v2,o} 
	\fmf{fermion,left,tension=.3}{v1,v2,v1} 
	\end{fmfgraph*}} \!\! = \,\, - \frac{i y^2}{4 \pi^2} \left\{ \frac{1}{\varepsilon} (p^2 + 2 m^2) +\frac{p^2}{2} \log \left( \frac{\mu^2}{m^2} \right) + m^2 \left[ 1 + \log \left( \frac{\mu^2}{m^2} \right) \right] +\mathcal{O}(\varepsilon, m^{-2} )\right\} .  \label{toy EFT loop of fermions} 
\end{align}

The diagram in Eq. \eqref{toy EFT scalar loop diagram} contributes equally to the EFT and the full theory, and thus drops out when comparing 1-loop corrections in the two theories. This implies that the 1-loop corrections to the effective action must yield a contribution equal to the diagram in Eq. \eqref{toy EFT loop of fermions} in the $\overline{MS}$ scheme:   
\begin{align}
	\parbox[t][8mm][b]{30mm}{\begin{fmfgraph*}(78,50) 
	\fmfleft{i} \fmfright{o} 
	\fmf{dashes}{i,v,o} 
	\fmffreeze
	\fmfv{d.sh=circle,d.f=empty,d.si=.2w,l=$\!\!\!\!\! 1$}{v}
	\end{fmfgraph*}} =  - \frac{i y^2}{4 \pi^2} \left\{\frac{p^2}{2} \log \left( \frac{\mu^2}{m^2} \right) + m^2 \left[ 1 + \log \left( \frac{\mu^2}{m^2} \right) \right] +\mathcal{O}(\varepsilon, m^{-2} )\right\} .
\end{align}
From this result, we can easily infer the form of $S^{(1)}_{\rm eff}$ at quadratic order in $\phi$, as was done for instance in Sec. \ref{sec: matching at loop level}. Adding up this result and the tree-level one, we find that the quadratic part of the effective action at 1-loop must read
\begin{align}
	S^{(0)}_{\rm eff} + S^{(1)}_{\rm eff} = \int d^4 x \left\{ - \tfrac{1}{2} \mathcal{Z} (\mu) \partial_\mu  \phi \, \partial^\mu \phi - \tfrac{1}{2} \mathcal{M}^2 (\mu) \phi^2 + \dots \right\} \ ,
\end{align}
where the dots stand for terms with higher derivatives or higher powers of $\phi$, and we have defined the Wilson coefficients
\begin{subequations}
\begin{align}
	\mathcal{Z} (\mu) &= 1 + \frac{y^2}{8 \pi^2} \log \left( \frac{\mu^2}{m^2} \right)  , \\
	\mathcal{M}^2 (\mu) &= M^2 + \frac{y^2}{4 \pi^2} m^2 \left[ 1 + \log \left( \frac{\mu^2}{m^2} \right) \right] \label{1 loop correction mass scalar} .
\end{align}
\end{subequations}

There is a very important qualitative difference between the mass squared coefficient $\mathcal{M}^2 (\mu)$ we just calculated for the scalar and the one we obtained for the fermion by integrating out the scalar---see Eq. \eqref{M 1-loop toy}. The 1-loop correction to the fermion mass parameter was proportional to the tree-level fermion mass $m$, and therefore $\mathcal{M}(\mu) \sim m$ for values of $\mu$ such that there is no large logarithm. In other words, the size of the mass parameter $\mathcal{M}(\mu)$ in the EFT for the fermion was determined by the IR scale $m$ rather than the UV scale $M$. As a consequence, one can freely choose the fermion mass to be arbitrarily smaller than the strong coupling scale of the EFT for any fixed value of the coupling $y$.

The 1-loop correction to the scalar mass $M$ is instead proportional to the mass $m$ of the heavy fermion, {\it i.e.} it is determined by the UV scale rather than the IR one. This means that we should expect $\mathcal{M}^2(\mu)$ to be no smaller than its 1-loop contribution, {\it i.e.}
\begin{align} \label{fine tuning 1}
	\mathcal{M}^2(\mu) \gtrsim \frac{y^2}{4 \pi^2} \, m^2 , 
\end{align}
for $\mu \sim \mathcal{O}(m)$. Unlike in the previous case, we now have a lower bound on $\mathcal{M}^2(m)$ for any fixed value of~$y$. One could circumvent this bound by arranging a delicate \high{fine-tuning} of the tree-level and loop contributions at a particular scale  $\mu$. However, such fine-tuning would not survive  if we changed the renormalization scale by a relative factor of $\mathcal{O}(1)$, and is thus regarded as an ``unnatural'' possibility. The ``natural'' expectation, instead,  is  that the scalar mass be in the neighborhood of the strong coupling scale $m$, within a window of size determined by  the coupling $y$. When applied to realistic theories of Nature, this expectation is based on the sentiment that the properties of our universe should be fairly generic. 

We should stress that the sensitivity of the scalar mass to the UV scale $m$ is not an artifact of switching to an EFT at low energies. A similar result can be inferred using exclusively the full theory. In the $\overline{MS}$ renormalization scheme, the physical mass of the scalar  doesn't coincide with the renormalized mass parameter $M$, but instead is defined as the location of the pole of the scalar propagator. Denoting with $i \Pi (k^2)$ the sum of all 1PI loop corrections, the physical mass is determined by the relation $M_{\rm ph}^2 = M^2 - \Pi (-M_{\rm ph}^2)$. At 1-loop, this yields
\begin{align}
	M_{\rm ph}^2 = M^2 \! \left\{ 1 - \frac{\lambda}{32 \pi^2} \left[ 1 + \log \left( \frac{\mu^2}{M^2} \right) \right] - \frac{y^2}{8 \pi^2} \log \left( \frac{\mu^2}{M^2} \right) \! \right\} + \frac{y^2}{4 \pi^2} \, m^2 \!\left[ 1 + \log \left( \frac{\mu^2}{m^2} \right) \right] \! , 
\end{align}
from which we conclude that a fine-tuning would be required for $M_{\rm ph}^2$ to be much smaller than $\frac{y^2}{4 \pi^2} \, m^2 \simeq \frac{y^2}{4 \pi^2} \, m_{\rm ph}^2$.

The issue of the \high{UV sensitivity} of a scalar mass is often discussed by regulating loops in the full theory with a hard cutoff $\Lambda$, and noticing  that the scalar mass receives a correction that diverges quadratically with the cutoff. In the case of fermions, instead, the mass only depends logarithmically on $\Lambda$. Therefore, as the argument goes, scalar masses are more sensitive to UV physics. There is of course a kernel of truth behind this statement, but as a student I always found it confusing, since a cutoff regularization is just a trick we use to make sense of intermediate steps when calculating loops. Eventually, any trace of $\Lambda$ gets absorbed in the renormalized fields and couplings, so why would it matter if the divergence is quadratic or logarithmic? The point is that imposing a hard cutoff in momentum space can be thought of as a (admittedly, highly unphysical) modification of the UV physics that kicks in at the scale $\Lambda$. From this perspective, the quadratic divergence can once again be interpreted as a sign that the natural size of a scalar mass is set by the ``UV physics''. In dimensional regularization we only deal with physical mass scales, and this makes it clear that what ultimately matters is not how loops diverge, but rather how the physical mass of a scalar depends on physical UV scales.
 
The requirement that there be no unexplained fine-tuning can also be expressed in a different way. Clearly, for the bound in Eq.~\eqref{fine tuning 1} to be satisfied it is sufficient to have
\begin{align} \label{t Hooft naturalness}
	M^2 \gtrsim \frac{y^2}{4 \pi^2} \, m^2 .
\end{align}
Based on Eq.~\eqref{1 loop correction mass scalar}, this requirement means that the 1-loop correction to the mass parameter in the EFT should not be larger than its tree-level value at the UV scale $\mu = m$. This observation allows us to introduce a slightly more precise notion of naturalness: the Wilson coefficients of an EFT are considered natural in a technical sense, or technically natural for short, if their magnitude is not altered significantly by loop corrections. When phrased this way, the requirement of \high{technical naturalness} is tantamount to demanding that the Wilson coefficients admit a well defined perturbative expansion. 

We have seen that the toy model in Eq. \eqref{full theory} would be technically natural if $m^2 \ll M^2$, but not if $M^2 \ll \frac{y^2}{4 \pi^2} \, m^2$. It all boils down to the fact that loop corrections to the fermion mass $m$ are proportional to $m$ itself, whereas loop corrections to the scalar mass squared $M^2$ are proportional to $m^2$ rather than $M^2$. What is the origin of this difference? It turns out that our toy model acquires an additional discrete symmetry when $m = 0$, namely
\begin{align} \label{some sort of chiral symmetry}
	\Psi \to \gamma^5 \Psi \ , \qquad \qquad \qquad \phi \to - \phi \ .
\end{align}
Barring anomalies, which are not an issue in this model, loop corrections are supposed to preserve the symmetries of the  theory. For this reason, corrections to the mass parameter can be non-zero only if $m$ is already non-zero at tree-level. In other words, these corrections must vanish when $m \to 0$ to preserve the symmetry \eqref{some sort of chiral symmetry} in this limit. More broadly, it is natural for any dimensionless parameter (such as, in our case, $m/M$) to be much smaller than one if setting it to zero enhances the symmetry of the model. In this case, we say that the smallness of such parameter is ``protected'' by the additional symmetry. This particular form of the naturalness criterion is known as \high{'t Hooft naturalness}~\cite{tHooft:1979bh}. It is a slightly more stringent requirement than our notion of technical naturalness, since there exist theories in which certain Wilson coefficients do not receive {\it any} loop corrections without being protected by any symmetry.\footnote{I find the distinction between 't Hooft and technical naturalness a useful one, albeit it is not common in the literature. The reader should be aware that what I am calling 't Hooft naturalness is often referred to as technical naturalness.} A particularly striking example of this is provided by Galileon theories~\cite{Goon:2016ihr}.

Over the last few decades, naturalness has become one of the guiding principles in our search for physics beyond the Standard Model (SM). Given the important role that this concept has played in shaping the research agenda of the particle physics community, I would be remiss if I didn't address briefly its broader implications. Before I do so, however, a disclaimer is in order: the following discussion reflects my own viewpoint on the subject. You are encouraged to develop your own perspective by consulting a variety of sources. To get you started, I have provided a few references in the following section. 

I think it is important to stress that the naturalness criterion in Eq.~\eqref{t Hooft naturalness} is specific to the particular toy model we have been considering: it is a constraint on the parameters of the action in Eq.~\eqref{full theory}. If we were to replace the heavy fermion with a different spectrum of heavy particles---{\it i.e.} choose a different \high{UV completion} for the scalar EFT---this constraint would take a different form. Thus, in order to establish whether the scalar mass is natural, one needs to ({\it i}\,) know the correct UV completion of the scalar EFT, and ({\it ii}\,) measure experimentally its parameters, {\it e.g.} by studying scattering processes involving both the light scalar and the heavy particles. This suggests that naturalness should be regarded as a constraint that a low-energy EFT imposes on its possible UV completions.\footnote{This should be contrasted with constraints that one can impose on low-energy EFTs by making some assumptions about the UV completion, {\it e.g.} connection between spin and statistics~\cite{Weinberg:1995mt} and positivity constraints on the Wilson coefficients~\cite{Adams:2006sv} in theories with a Lorentz-invariant UV completion, constraints on symmetries and particle content of EFTs coming from string theory~\cite{Vafa:2005ui}, etc ...} 

The two most famous ``naturalness problems'' in particle physics are associated with the two UV sensitive parameters of the SM coupled to General Relativity (GR): the Higgs mass parameter, and the cosmological constant. The UV sensitivity of the Higgs mass is known as the \high{hierarchy problem}, and it is essentially the same phenomenon we encountered in our toy scalar EFT. The hierarchy problem certainly makes it more challenging to write down models of physics above the weak scale that are not finely tuned. Whether or not this turns out to be an actual problem---by which I mean, an aspect of Nature we don't understand---will depend on what the correct UV completion of the SM happens to be. 

For now, we should keep an open mind and search for new physics above the weak scale, remaining aware of the fact that none of the open problems in particle physics ({\it e.g.} baryogenesis, dark matter, neutrino masses, strong CP problem, ...) necessarily requires a solution in the UV (even though some of such solutions are particularly compelling). The strongest argument in favor of new UV physics is in my opinion the fact that GR is an effective theory that ceases to be valid at the Planck scale. Based on our experience with EFTs, we would expect new degrees of freedom to show up at or below this scale and give rise to a more complete description of gravitational interactions. It is also reasonable to expect this new physics to provide large quantum corrections to the Higgs mass parameter. Unfortunately, our present understanding of gravity at the Planck scale doesn't allow us to perform a detailed matching calculation and explicitly verify this expectation. 

The \high{cosmological constant problem} poses  a much more serious challenge to our understanding of naturalness and its role in particle physics, for at least two reasons. First and foremost, unlike the hierarchy problem, it doesn't hinge upon the existence of new UV physics. Integrating out any SM particle with mass $m$ gives rise to a 1-loop correction to the cosmological constant of the form $\Delta \Lambda =\mathcal{O}(1) \frac{m^4}{16 \pi^2 \mpl^2}$. Even the electron would yield a contribution of $\mathcal{O}(10^{-52}) \, \text{GeV}^2$ which is much larger than the observed value of $\mathcal{O}(10^{-84}) \, \text{GeV}^2$. Second, the only known symmetry that can protect the cosmological constant from quantum corrections is supersymmetry. Unfortunately there is so far no experimental evidence for supersymmetry, which means that, at best, it must be broken at a scale $M_{\rm susy}$ that is higher than the weak scale. In this scenario, the cosmological constant would receive corrections that are generically of order  $\Delta \Lambda =\mathcal{O}(1) \frac{M_{\rm susy}^4}{16 \pi^2 \mpl^2} \gtrsim \mathcal{O}(10^{-34}) \, \text{GeV}^2$, which is again much larger than the observed value.  This state of affair has led many to believe that a solution to the cosmological constant problem might point towards a new organizing principle in the fundamental laws of nature.  Interesting possibilities include anthropic considerations~\cite{Weinberg:1987dv} and cosmological relaxation mechanisms~\cite{Alberte:2016izw,Graham:2019bfu}.
\begin{eBox}
{\bf Exercise 2.6:} By convention, the cosmological constant enters the effective action as $S \supset \int \sqrt{-g} \left( - \mpl^2 \Lambda \right)$. Calculate the 1-loop matching correction to the cosmological constant generated when integrating out the Higgs particle. Work at lowest order in the SM couplings. Compare your result with the vacuum expectation value of the stress energy tensor of the free Higgs field.
\end{eBox}

\subsection{Additional resources}

There are many excellent reviews on EFT methods, {\it e.g.} ~\cite{Georgi:1994qn,Manohar:1996cq,Pich:1998xt,Rothstein:2003mp,Kaplan:2005es,Burgess:2007pt,Skiba:2010xn,Gripaios:2015qya,Petrov:2016azi,Manohar:2018aog,Cohen:2019wxr}. In particular, a nice example of a matching calculation at 1-loop implemented at the level of the generating functional can be found in~\cite{Burgess:2007pt}. For a very pedagogical discussion of matching and running, see~\cite{Cohen:2019wxr}. For interesting discussions on naturalness, see~\cite{Dine:2015xga,Giudice:2017pzm}. For a provocative take on the hierarchy problem, see instead the appendix of~\cite{Manohar:2018aog}.
The classic reference about the cosmological constant problem is~\cite{Weinberg:1988cp}.

\newpage

\section{EFTs for Nambu-Goldstone modes}

We have argued that symmetries are one of the defining ingredients of any EFT: any term that is invariant should be included in the effective action. In practice, this is fairly easy to achieve when dealing with fields that transform linearly under the relevant symmetries, {\it i.e.} according to transformation rules of the form $\varphi^i \to M^i{}_j \varphi^j$ for some constant matrix $M$.\footnote{One can of course always complicate the transformation rules at will by performing an arbitrarily involved field redefinition. Steven Weinberg's ``Third Law of Progress in Theoretical Physics'' addresses this possibility: {\it ``You may use any degrees of freedom you like to describe a physical system, but if you use the wrong ones, you'll be sorry''}~\cite{Weinberg:1981qq}.} Then it is just a matter of contracting indices in an invariant way.  Spontaneously broken symmetries, however, are realized non-linearly on the ensuing Goldstone modes. In this case, writing down all possible invariant operators is not always straightforward. In the first part of this chapter, we will introduce non-linearly realized symmetries in the context of Chiral Perturbation Theory (CPT)---the EFT that describes the lightest QCD bound states. This EFT will also feature additional subtleties compared to the ones we discussed in the previous chapter. In the second part of this chapter, we will introduce the coset construction---a general technique to write down effective actions that are invariant under non-linearly realized symmetries. Along the way, we will also learn how to handle EFTs with approximate or anomalous symmetries.

\subsection{Chiral perturbation theory} \label{sec: CPT}

Imagine we are interested in studying QCD bound states that are lighter than proton and neutron. We might as well integrate out all SM particles with mass at or above the GeV scale and work the resulting effective theory. The remaining quarks---$u$, $d$ and $s$---are described at low energies by the effective action
\begin{align} \label{SM action below 1 GeV}
	S = \int d^4 x \left\{ \sum_{j=u,d,s} \bar \Psi_j (i \gamma^\mu D_\mu - m_j ) \Psi_j - \frac{1}{4} F_{\mu\nu} F^{\mu\nu} - \frac{1}{2} \mbox{Tr} \, G_{\mu\nu} G^{\mu\nu} + \, \dots \right\} , 
\end{align}
with $D_\mu \Psi_j = (\partial_\mu - i q_j A_\mu - i g \, G_\mu ) \Psi_j$, $A_\mu$ and $G_\mu$ the photon and gluon fields, and $F_{\mu\nu}$ and $G_{\mu\nu}$ their respective field strengths. Color indices won't play any role in what follows and will be suppressed throughout. The dots in the Lagrangian above stand for all sorts of ``non-renormalizable'' terms, which are suppressed at energies below the GeV. These include in particular interactions between quarks and light leptons.

The mass of all three remaining quarks are smaller than the only other scale in the problem, {\it i.e.} the QCD scale $\Lambda_{\rm QCD}  \simeq 300$ MeV associated with confinement. It seems therefore a reasonable approximation to study bound states by neglecting quark masses at a first pass. Admittedly, this approximation works better for the $u$ and $d$ quarks, whose masses are of $\mathcal{O}(1)$ MeV, than it does for the $s$ quark, since $m_s \simeq 100$ MeV. Still, with this caveat in mind, let's press on and double down by neglecting also the electromagnetic coupling. We will reintroduce the effect of finite masses and electromagnetism in the next section. In this limit, it is convenient to break up the quark fields into their left-handed and right-handed components, defined by
\begin{align}
	\Psi^j_L = \tfrac{1}{2} (1 - \gamma^5) \Psi^j, \qquad \qquad \quad \Psi^j_R = \tfrac{1}{2} (1 + \gamma^5) \Psi^j, 
\end{align}
so that the action above reduces to
\begin{align} \label{approximate SM action below 1 GeV}
	S \simeq \!\int d^4x \!\!\! \sum_{j=u,d,s} \!\! \left[ \bar \Psi_L^j i \gamma^\mu (\partial_\mu - i g \, G_\mu ) \Psi_L^j + \bar \Psi_R^j i \gamma^\mu (\partial_\mu - i g \, G_\mu ) \Psi_R^j \right] - \frac{1}{2} \mbox{Tr} \, G_{\mu\nu} G^{\mu\nu} + \, \dots .
\end{align}

This action is invariant under several global symmetry. First, it is invariant under conformal transformations. This follows from the fact that there is no mass scale in Eq. \eqref{approximate SM action below 1 GeV}. Nevertheless, a preferred mass scale is generated dynamically, because quantum effects make the QCD coupling run with energy and interactions become strong at the scale $\Lambda_{QCD}$. In other words, conformal invariance is anomalous: it is broken by quantum effects and therefore is not a symmetry of the full theory.

The action \eqref{approximate SM action below 1 GeV} is also invariant under separate $U(3)$ transformations of the left-handed and right-handed quarks:
\begin{align}
	(\Psi_L^i)' = L^i{}_j \Psi_L^j, \qquad \qquad \quad (\Psi_R^i)' = R^i{}_j \Psi_R^j \ .
\end{align}
It is helpful to decompose this $U(3)_L \times U(3)_R$ symmetry group into the product $SU(3)_L \times SU(3)_R \times U(1)_V \times U(1)_A$, where $U(1)_V$ corresponds to global rephasings of all the quark fields, while $U(1)_A$ transforms left-handed and right-handed quarks with opposite phases, {\it i.e.} 
\begin{align}
	\Psi_L' = e^{i \alpha} \Psi_L, \qquad \qquad \quad \Psi_R' = e^{-i \alpha} \Psi_R \ .
\end{align}
Each of these subgroups enjoys a different status. The \high{axial symmetry} $U(1)_A$ is anomalous, and therefore is not an actual symmetry of the quantum theory. More on this in Sec.~\ref{sec: chiral anomalies}. Moreover, strong interactions cause the operator $\bar \Psi^i_R \Psi_L^j$ to develop a non-trivial vacuum expectation value~(vev) of the form
\begin{align} \label{chiral symmetry breaking}
	\langle \bar \Psi^i_R \Psi_L^j \rangle = v^3 \delta^{ij} ,
\end{align}
with $v \sim \Lambda_{\rm QCD}$. This vev breaks spontaneously the \high{chiral symmetry} $SU(3)_L \times SU(3)_R$ down to its diagonal subgroup $SU(3)_V$, such that left- and right-handed fermions transform in lockstep.\footnote{There is a very general argument based on anomalies that implies that $SU(3)_L \times SU(3)_R$ must be spontaneously broken down to some unspecified subgroup~\cite{tHooft:1979bh}.} (This vev would also break spontaneously the $U(1)_A$ and conformal symmetries, if it wasn't for the fact that they are not  actual symmetries to begin with.) As a result, Goldstone's theorem dictates that the spectrum of this theory must contain (at least) eight massless bound states, {\it i.e.} one Nambu-Goldstone (NG) mode for each broken symmetry generator. Finally, the vev in Eq. \eqref{chiral symmetry breaking} preserves the \high{baryon symmetry} $U(1)_V$.\footnote{To be precise, baryon symmetry is also anomalous when weak interactions are taken into account. It is only the difference between baryon and lepton number that is exactly conserved. However, the lepton symmetry acts trivially on quarks and gluons, and therefore the action \eqref{SM action below 1 GeV} appears invariant under baryon symmetry alone. This means that baryon number is conserved by strong interactions.}. This means that bound states of quarks and gluons must have a definite charge under $U(1)_V$.

In what follows, we will derive an EFT for the NG modes associated with the spontaneous breaking of chiral symmetry. This EFT is known as \high{Chiral Perturbation Theory}, and its expansion parameter will once again be a ratio of the form $E /\Lambda$, with $\Lambda$ some UV scale to be determined. Compared to the toy model we considered in the previous chapter, this EFT presents two novel features. First, the action \eqref{approximate SM action below 1 GeV} doesn't contain any mass scale, as we already pointed out. For this reason, it is not immediately obvious what the strong coupling scale $\Lambda$ should be. Notice in particular that it doesn't have to coincide with the symmetry breaking scale $v$---although in practice it won't be too far from it. Second, the NG fields do not appear directly in the UV action \eqref{approximate SM action below 1 GeV}. Therefore, we cannot derive our EFT simply by integrating out some heavy fields as we did in the previous chapter: strong interactions prevent us from carrying out an explicit matching calculation. We will therefore need to be guided by particle content and symmetries alone, and treat the EFT coefficients as free parameters to be fixed by experiments.  

In order to figure out how symmetries act on the NG fields, it is convenient to think of NG excitations as broken transformations of the vev that are ``mildly'' modulated in space and time. A constant broken transformation would, by definition, change the vev \eqref{chiral symmetry breaking} to a different but physically equivalent vev; by contrast a NG excitation smoothly changes the vacuum at each point in spacetime. This is often expressed by saying that NG modes interpolate between different vacua. Based on this, we act on the vev \eqref{chiral symmetry breaking} with a local $SU(3)_L \times SU(3)_R$ transformation such that $L(x) = R^\dag (x) = e^{i \pi^a(x) Q_a}$, with $\pi^a(x)$ the eight NG fields, and the $Q_a$'s the generators of $SU(3)$. The result is
\begin{align}
	\langle \bar \Psi^i_R \Psi_L^j \rangle \to v^3 U^{ji} , \qquad \quad \text{with} \qquad \quad U (x) = L(x) \cdot R^\dag (x) = e^{2 i \pi^a(x) Q_a} .
\end{align}
Hence, NG excitations correspond to those particular local deformations of the $3 \times 3$ complex matrix $\langle \bar \Psi^i_R \Psi_L^j \rangle$ that can be described by an $SU(3)$ transformation $U(x)$. Notice that this transformation preserves the overall ``size'' of the matrix, {\it i.e.} its determinant. Now, a generic $SU(3)_L \times SU(3)_R$ transformation is such that $L(x) \to \tilde L \cdot L(x)$ and $R(x) \to \tilde R \cdot R(x)$, and therefore 
\begin{align} \label{chiral transformations U}
	U \to U' = \tilde L \cdot  U \cdot \tilde R^\dag .
\end{align}
Similarly, it is fairly easy to see that the $\pi^a$'a must carry zero baryonic charge, because the generators of $SU(3)_L \times SU(3)_R$ commute with the generator of $U(1)_V$.

It is not difficult to write down an effective action for $U$ that is invariant under chiral transformations of the form \eqref{chiral transformations U}, because these transformations are linear. We will need to use both $U$ and $U^\dag$, as well as some derivatives, since any product of the form $U^\dag U$ is trivially invariant because $U$ is unitary. A moment of reflection leads to the conclusion that there is only one invariant term we can write down with two derivatives, and therefore the effective action for the our NG bosons takes the form
\begin{align} \label{CPT action}
	S = \int d^4 x \left\{ - \frac{f^2}{4} \mbox{Tr} \, ( \partial_\mu U^\dag \partial^\mu U ) + \dots \right\} ,
\end{align}
where the factor of $1/4$ has been introduced for later convenience, and the dots stand for all terms with more than two derivatives. This EFT displays many remarkable features:
\begin{itemize}
	\item The effective action depends on a single parameter $f$ at lowest order in the derivative expansion, despite the fact that it contains an infinite number of terms when expanded in powers of the NG fields. Up to quartic order, it reads 
\begin{align} \label{CPT expanded action}
	S = \int d^4 x \, f^2 \left\{ - \frac{1}{2} \partial_\mu \pi_a \partial^\mu \pi^a + \frac{1}{6} f_{abc} f_{de}{}^c \pi^b \pi^e \partial_\mu \pi^a \partial^\mu \pi^d  + \dots \right\} ,
\end{align}
where $f_{abc}$ are the structure constants of $SU(3)$, which can be found for instance in~\cite{Scherer:2002tk}. To derive this result, we used the fact that the normalization of the $SU(3)$ generators in the fundamental representation is such that $\mbox{Tr} (Q_a Q_b) = \frac{1}{2}\delta_{ab}$. 
	\item Each term in Eq. \eqref{CPT expanded action} is manifestly invariant under linear $SU(3)$ transformations of the NG fields: this is the unbroken symmetry, $SU(3)_V$. 
	\item The relative coefficients of the quadratic, quartic, and higher order terms are completely nailed by symmetry. This  is a consequence of the spontaneously broken part of $SU(3)_L \times SU(3)_R$, which is \high{realized non-linearly} on the $\pi$'s. Consider for instance a broken transformation such that $\tilde L = e^{2 i \epsilon^a Q_a}$ and $\tilde R = \mathbf{1}$. According to Eq. \eqref{chiral transformations U}, the NG fields must change in such a way that 
	\begin{align} \label{nonlinearly realized symmetries}
		e^{2 i \pi_a' Q^a}  = e^{2 i \epsilon_b Q^b} e^{2 i \pi_a Q^a} = e^{2 i (\pi_a + \epsilon_a - f_{abc} \epsilon^b \pi^c + \cdots ) Q^a } ,
	\end{align}
	where in the last step we used the Baker-Campbell-Hausdorff formula to combine the exponentials, and the dots stand for an infinite number of terms that are higher order in $\pi$, $\epsilon$, or both. The relative tuning of quadratic, quartic, and higher order terms is necessary for the effective action \eqref{CPT expanded action} to be invariant under these very complicated transformation rules.  
	\item Incidentally, this shows why it's much easier to work with $U$, which transforms linearly as shown in Eq. \eqref{chiral transformations U}, rather than directly with the $\pi$'s. It would have been much more difficult to come up with the action \eqref{CPT expanded action} starting from the transformation rules for the NG fields.
	\item Since $U_{ij}(x)$ was a $SU(3)$ matrix, the action \eqref{CPT expanded action} was obtained by calculating traces of products of generators in the fundamental representation. Had we worked with a different representation ({\it e.g.}, the adjoint one), the relative coefficients between quadratic, quartic, etc... terms would have remained the same. The only difference would have been the overall normalization, which can always be reabsorbed in the coupling $f$.\footnote{This is essentially because $\mbox{Tr} (T_a T_b) = C \delta_{ab}$, with $C$ some coefficient that depends on the representation.} Thus, the representation of the operator breaking the symmetry only affects the values of the Wilson coefficients. The allowed non-linear structures, instead, are completely determined by the \high{symmetry breaking pattern}, {\it i.e.} by which symmetry group is broken down to which subgroup.
\end{itemize}

By convention, the canonical normalization for a scalar field is such that the coefficient in front of the kinetic term is equal to $-1/2$ (with a mostly plus metric signature). Therefore, the canonically normalized NG fields are $\hat \pi^a = f \pi^a$. When expressed in terms of the $\hat \pi$'s, our effective action has the schematic form
\begin{align}
	S = \int \left\{ \sum_{n=0}^\infty c_n (\partial \hat \pi)^2 \! \left( \frac{\hat \pi}{f} \right)^{2n} + \dots \right\} , 
\end{align}
with $c_n$ some coefficients determined by symmetries. Thus, additional powers of $\hat \pi$ in the Lagrangian are suppressed by the scale $f$. It is worth emphasizing that this is \emph{not} equal to the strong coupling scale $\Lambda$ of the EFT, which is instead the scale that suppresses additional derivatives in the Lagrangian, and at which perturbation theory breaks down. In other words, observable quantities such as $n$-point amplitudes will admit an expansion in powers of $E /\Lambda$, not $E/f$, and the effective action takes the schematic form 
\begin{align}\label{full action CPT}
	S = \int d^4 x \, \Lambda^2 f^2 \mathcal{L} (U, U^\dag,\partial_\mu / \Lambda).
\end{align}

We can derive an upper bound on the strong coupling scale $\Lambda$ by invoking the naturalness criterion we discussed in the previous chapter. To this end, we will compare the size of tree-level and their 1-loop contributions the 4-point function. At tree-level, the 4-point function scales like $E^2 / f^2$, as one can easily see from the action in Eq. \eqref{CPT expanded action}. The 1-loop correction built out of two tree-level quartic vertices can be easily estimated in dimensional regularization: each vertex comes with a $1/f^2$ factor; each derivative contributes an external momentum of $\mathcal{O}(E)$, since there are no other scales around; finally, we have the ubiquitous ``loop factor'' $1/(4\pi)^2$. Putting it all together, we find schematically
\begin{align}
	\parbox[t][6.5mm][b]{24mm}{\begin{fmfgraph*}(60,40) 
	\fmfleft{i1,i2} 
	\fmfright{o1,o2} 
	\fmf{vanilla}{i1,v1,i2} 
	\fmf{vanilla}{o1,v2,o2} 
	\fmf{vanilla,left,tension=.3}{v1,v2,v1} 
	\end{fmfgraph*}} \!\!\!\!\! \sim \,\, \frac{1}{(4 \pi)^2} \times \left( \frac{1}{f^2} \right)^2 \times E^4 = \frac{E^2}{f^2} \times \frac{E^2}{(4 \pi f)^2} \ .
\end{align}
This 1-loop diagram is suppressed compared the tree-level result by a factor of $E^2 / (4 \pi f)^2$. This suggests that perturbation theory will stop working at a scale $\Lambda \sim 4 \pi f$, which is an order of magnitude larger than $f$. Notice that 4-derivative operators in the effective action will yield a tree-level contribution to the 4-point function that is also proportional to $E^4$. Therefore, it is possible that perturbation theory breaks down at energies smaller than $4 \pi f$ if these operators have Wilson coefficients larger than $1/(4\pi)^2$. On the contrary, having a strong coupling scale much larger than $4 \pi f$ would require a delicate fine-tuning of the Wilson coefficients of higher-derivative terms, and is therefore considered an unnatural possibility. The statement that the strong coupling scale of this EFT should not be larger than $4 \pi f$ is usually regarded as indication that new physics should appear at or below this scale. 

Incidentally, the separation between $f$ and the strong coupling scale is needed to ensure that chiral perturbation theory is actually useful. A comparison with experiments shows that $f \simeq 93$ MeV,\footnote{Notice that $f$ is sometimes used in the literature to denote $\sqrt{2} \times 93 $ MeV $\simeq 130$ MeV.} which is smaller than the mass of the lightest meson: if $f$ was the strong coupling scale, we would be in trouble!

\subsection{Approximate symmetries and spurions}

Particle physics experiments have of course never discovered eight Goldstone bosons. This is because chiral symmetry is only an \high{approximate symmetry} of the action \eqref{SM action below 1 GeV}: it is explicitly broken by the quark mass terms and the electromagnetic interactions.\footnote{Weak interactions also break explicitly chiral symmetry but, in the effective action \eqref{SM action below 1 GeV}, do so via higher-dimensional operators whose effects are suppressed at low energies.} Approximate symmetries introduce additional expansion parameters in the EFT. Depending on the sources of explicit symmetry breaking, these can be either dimensionless couplings ({\it e.g.} quark charges), or ratios between the explicit symmetry breaking scales ({\it e.g.} quark masses) and the strong coupling scale. These new expansion parameters can be taken into account in a systematic way by resorting to a nifty trick: pretending that the symmetry breaking parameters are fictitious fields, usually referred to as \high{spurions}.

Let's see how this strategy works in practice in Chiral Perturbation Theory, and consider first the quark masses. We start by adding to the quark action in Eq. \eqref{approximate SM action below 1 GeV} the following term
\begin{align}
	\Delta S = \int d^4 x \left\{- \bar \Psi^j_R M_{jk} \Psi_L^k + \mbox{h.c.} \,  \right\} ,
\end{align}
where $M$ is the spurion field. This term is invariant under $SU(3)_L\times SU(3)_R$ (and, of course, under $U(1)_V$) provided the spurion transforms as follows:
\begin{align}
	 M \to \tilde R \cdot M \cdot \tilde L^\dag .
\end{align}
The term we just introduced reduces precisely to the mass terms we dropped in going from the action in Eq. \eqref{SM action below 1 GeV} to the one in Eq. \eqref{approximate SM action below 1 GeV} when the spurion takes the following value
\begin{align} \label{value M spurion}
	 M_{jk} = \mbox{diag} (m_u, m_d, m_s) .
\end{align}
Our strategy will consist of extending the effective action for the NG fields in Eq. \eqref{full action CPT} to include all possible couplings with the spurions that respect chiral symmetry. Chiral symmetry is a good approximation to the extent that the quark masses are small compared to the strong coupling scale. Therefore, we will treat $M/ \Lambda$ as an additional expansion parameter which controls the size of explicit symmetry breaking, so that our effective action now reads, 
\begin{align} \label{spurion action CPT}
	S = \int d^4 x \, \Lambda^2 f^2 \mathcal{L} (U, U^\dag,\partial_\mu / \Lambda, M/\Lambda) , 
\end{align}
where the Lagrangian contains in principle all possible operators that are invariant under chiral symmetry. After replacing the spurion $M$  with its value in Eq. \eqref{value M spurion}, the action \eqref{spurion action CPT} captures all possible ways in which non-zero quark masses modify the dynamics of the NG fields.

At lowest order in $M/\Lambda$, the spurion couples to the NG fields via the following operator in the effective Lagrangian:
\begin{align} \label{CPT breaking mass}
	\Delta S = \int d^4 x \, f^2 \Lambda^2 \, \left\{ \frac{y}{2} \, \mbox{Tr} \left( \frac{M}{\Lambda} \, U \right) + \mbox{h.c.} \,\right\} , 
\end{align}
with $y$ some coupling assumed to be of $\mathcal{O}(1)$ on naturalness grounds. Expanding this term up to quadratic order in the $\pi$'s, we find that our NG fields are now endowed with mass terms. More precisely, it turns out that the eight NG modes arrange themselves into four pairs, three of which have masses
\begin{subequations}\label{mass mesons quark masses 1}
\begin{align} 
	(\pi^+, \pi^-): \qquad \qquad  m^2 &= y \, \Lambda (m_u+m_d), \\
    (K^+, K^-): \qquad \qquad m^2 &= y \, \Lambda (m_u+m_s), \\
    (K^0, \bar K^0): \qquad \qquad m^2 &= y \, \Lambda (m_d+m_s),
\end{align}
\end{subequations}
while the fourth one has a non-diagonal mass matrix of the form 
\begin{align} \label{mass mesons quark masses 2}
	(\pi^0, \eta): \qquad \qquad m^2 = y \, \Lambda \begin{pmatrix}
		m_u +m_d & m_u-m_d \\
		m_u-m_d & \tfrac{1}{3}(m_u+m_d +4m_s)                                 
	\end{pmatrix} . 
\end{align}
I have added in parentheses the mesons with which these mass eigenstates should be identified. 
\begin{eBox}
{\bf Exercise 3.1:} Derive the results in Eqs. \eqref{mass mesons quark masses 1} and \eqref{mass mesons quark masses 2}. {\bf Hint:}  the generators of $SU(3)$ in the fundamental representation are given by $Q_a = \tfrac{1}{2} \lambda_a$, where the $\lambda$'s are the eight Gell Mann matrices. These matrices satisfy the following properties:
\begin{align}
	\mbox{Tr} (\lambda_a) = 0, \qquad \quad \mbox{Tr} (\lambda_a \lambda_b) = 2 \, \delta_{ab}, \qquad \quad \{\lambda_a, \lambda_b\} =\tfrac{4}{3} \delta_{ab} \mathbf{1} + 2 d_{abc} \lambda^c . 
\end{align}
Explicit expressions for the Gell Mann matrices and the totally symmetric coefficients $d_{abc}$ can be found for instance in~\cite{Scherer:2002tk}.
\end{eBox}
The results in Eqs. \eqref{mass mesons quark masses 1} and \eqref{mass mesons quark masses 2} are non-trivial predictions of Chiral Perturbation Theory: a priori, one could have imagined that each NG field would develop a different mass once chiral symmetry was broken! 

We can further refine our discussion of explicit breaking due to quark masses by introducing an additional expansion parameter, namely $(m_d - m_u)/\Lambda$. At lowest order in this expansion, $m_d \simeq m_u$ and the  spurion value in Eq. \eqref{value M spurion} preserves a $SU(2)$ subgroup of $SU(3)_V$: this is known as \emph{isospin symmetry}. In this limit, $\eta$, $(K^+, K^0)$, $(K^-, \bar K^0)$, $(\pi^0, \pi^+,\pi^-)$, form one spin 0, two spin 1/2 (conjugate to each other), and one spin 1 representations of isospin, and their masses satisfy the \emph{Gell-Mann--Okubo relation}:
\begin{align}
	m_\pi^2 + 3 m^2_\eta = 4 m^2_K.
\end{align}

This is of course not end of the story, since the chiral symmetry is also broken explicitly by electromagnetic interactions. This breaking can be described once again by introducing some fictitious fields, although there are some differences compared to the case of quark masses we just considered, so it is worth discussing this in detail. Reintroducing electromagnetic interactions means adding the following term to the quark action in Eq. \eqref{approximate SM action below 1 GeV}:
\begin{align}
	\Delta S = \int d^4 x \left\{ e A_\mu \, \bar \Psi_L^j \gamma^\mu Q^L_{jk} \Psi^k_L +  e A_\mu \, \bar \Psi_R^j \gamma^\mu Q^R_{jk} \Psi^k_R \right\} ,
\end{align}
with %
\begin{align}
	Q_{ij}^L = Q_{ij}^R = \text{diag} \left(\tfrac{2}{3}, - \tfrac{1}{3}, - \tfrac{1}{3} \right) . 
\end{align}
This provides us with a new expansion parameter: the dimensionless electromagnetic coupling, $e$. We will now treat the symmetry generators $Q^L$ and $Q^R$ as spurion fields, and exploit the fact that $\Delta S$ remains invariant under chiral transformations if these spurions transform as
\begin{align} \label{QL QR defs}
	Q^L \to \tilde L \cdot Q^L \cdot \tilde L^\dag , \qquad \qquad Q^R \to \tilde R \cdot Q^R \cdot  \tilde R^\dag \ .   
\end{align}
Following the same strategy we used for the quark masses, we then modify the effective action for chiral perturbation theory by including all possible couplings between the NG fields, the gauge field $A_\mu$, and the spurions $Q^{L,R}$ that preserve gauge invariance and chiral symmetry. 

Gauge invariance can be easily enforced by replacing ordinary derivative $\partial_\mu U$ with covariant ones, defined by $D_\mu U = \partial_\mu U - i e A_\mu Q^L U + i e  A_\mu Q^R U$. We can also write down interactions between NG fields and spurions that do not involve the gauge field.  Notice that each spurion must appear with a factor of ``$e$'', since the quark action is invariant under $e \to \lambda e, \, Q^{L,R} \to Q^{L,R} / \lambda$. At lowest order in $e$, there is only one operator that is allowed by the symmetries, namely
\begin{align} \label{EM spurion masses}
	\Delta S = \int d^4 x \, f^2 \Lambda^2 \, \left\{ \frac{e^2 y'}{2} \, \mbox{Tr} \left( Q^L U Q^R U^\dag \right) + \mbox{h.c.}  \right\} , 
\end{align}
with $y'$ another arbitrary coupling. When expanded in powers of $\pi$'s, this operator generates an infinite number of self-interactions that, unlike the ones in Eq. \eqref{CPT expanded action}, do not involve any derivative. In particular, at quadratic order one finds that Eq. \eqref{EM spurion masses} yields the following corrections to the meson masses:
\begin{align} \label{Dashen}
	(\pi^+, \pi^-): \qquad \qquad  m^2 &\to y \, \Lambda (m_u+m_d) +  y' e^2 f^2, \\
    (K^+, K^-): \qquad \qquad m^2  &\to y \, \Lambda (m_u+m_s) +  y' e^2 f^2 ,
\end{align}
while the other masses are not affected. The fact that the fields $\pi^\pm$ and $K^\pm$ receive exactly the same correction at lowest order is another non-trivial result that goes under the name of \emph{Dashen's theorem}. Neglecting isospin breaking, this correction is entirely responsible for the mass splitting between, say, $\pi^0$ and $\pi^\pm$.

\begin{eBox}
{\bf Exercise 3.2:} Derive the result in Eq. \eqref{Dashen}. {\bf Hint:} think about which generators of $SU(3)_{L,R}$ commute with $Q^{L,R}$. You might want to use again the explicit form of the Gell-Mann matrices given in~\cite{Scherer:2002tk}.
\end{eBox}

\subsection{Naturalness and pseudo-Nambu-Goldstone modes}

The discussion in the previous section provides a different viewpoint on the issues of naturalness of scalar masses and symmetry protection. Take for instance the mass of the $\pi^\pm$ mesons, which using $\Lambda \simeq 4 \pi f$ can be written as 
\begin{align} \label{naturalness mass PNGB}
	m^2 = \left[ y \, \frac{m_u +m_d}{\Lambda} + y' \frac{e^2}{(4\pi)^2} \right] \Lambda^2 .
\end{align}
At face value, this doesn't look too different from the natural value of the scalar mass squared in the toy model of the previous chapter: it has the same form, namely (small parameters) $\times$ (strong coupling scale)$^2$. There is however an important qualitative difference. In the toy model of the previous chapter, the small parameter was the coupling $y^2$ which determined the strength of interactions between the IR degrees of freedom (the scalar) and the UV ones (the fermion). This quantity has a well defined meaning only in the context of the full theory: it is an unknown quantity from the perspective of the low energy EFT. 

The small parameters that appear in Eq. \eqref{naturalness mass PNGB}, instead, have a well defined meaning from a purely low-energy perspective: they are expansion parameters of the EFT. These can be measured experimentally without knowing anything about the physics above the strong coupling scale, and must be small \emph{by definition} in order for the EFT to be a valid perturbative framework. In other words, the masses of the NG fields are guaranteed to be parametrically smaller than the strong coupling scale within the regime of validity of the EFT.

Something similar happened in the EFT for light fermions discussed in the previous chapter. In fact, we could have expressed the natural value of the fermion mass as
\begin{align} 
	m = \left( \frac{m}{M} \right) \times M ,
\end{align}
with $M$ the strong coupling scale, and $m/M$ the expansion parameter that controls the explicit breaking of the discrete symmetry in Eq. \eqref{some sort of chiral symmetry}. This expression is now very similar to the one for the scalar mass in Eq. \eqref{naturalness mass PNGB}. The fermion mass was protected by the discrete symmetry \eqref{some sort of chiral symmetry}, the   meson masses are protected by the non-linearly realized symmetries \eqref{nonlinearly realized symmetries}. Massive scalars whose mass is protected by symmetries realized non-linearly are called \high{pseudo-Nambu-Goldstone bosons} (PNGBs). Presently, we only know of one other way of protecting scalar masses: imposing invariance under supersymmetry, so that scalar masses can inherit the symmetry protection of fermion masses.

\subsection{Coset construction}

In Sec. \ref{sec: CPT}, we pointed out that the form of the effective action \eqref{CPT action} is independent of the representation of the operator that spontaneously breaks chiral symmetry. Knowing that this operator was $\bar \Psi^i_R \Psi_L^j$ allowed us to guess that $U=e^{2 i \pi^a Q_a}$ was a particularly convenient quantity to work with. However, the only thing that ultimately mattered was the symmetry breaking pattern, {\it i.e.} what was the full symmetry group $G$ of the EFT, and what subgroup $H$ was realized nonlinearly. This of course resonates with the overarching philosophy behind EFTs: the symmetry breaking mechanism is a UV detail, and we shouldn't need to know what physics looks like in the UV to describe what's going on in the IR. This suggests that there should be a way to formulate effective theories of NG bosons based on the symmetry breaking pattern alone: such an approach goes under the name of \high{coset construction}.\footnote{This name comes from the fact that, when a symmetry group $G$ is spontaneously broken down to a subgroup $H$, the broken transformations belong to the {\it coset space} $G/H$.} It was first developed by Callan, Coleman, Wess and Zumino~\cite{Callan:1969sn} for spontaneously broken internal symmetries, and later extended to space-time symmetries by Volkov~\cite{Volkov:1973vd} and Ogievetsky~\cite{ogievetsky:1974ab}. In this section, we are going to review the basic ideas behind the coset construction in a way that applies both to internal and space-time symmetries.  

Consider an arbitrary symmetry breaking pattern of the form $G \to H$. We will start by grouping the generators of $G$ into three categories:
\begin{enumerate}
	\item the generators of \emph{unbroken translations}: $P_\alpha$
	\item the generators of all other \emph{unbroken symmetries}: $T_A$
	\item the generators of \emph{broken symmetries}: $X_a$
\end{enumerate}
This classification is always possible, albeit it is not unique since one can alway redefine the broken generators as $X_a \to X_a + c_{a A}T^A+ c_{a \alpha}P^\alpha$, with $c_{a A}$ and $c_{a \alpha}$ some arbitary coefficients.

In chiral perturbation theory, for example, all translations are unbroken; the $T_A$'s are the generators of $SU(3)_V \times U(1)_V$; the broken generators are $X_a = 2 Q_a$.\footnote{To be more precise, if $Q^{L,R}_a$ are the generators of $SU(3)_{L,R}$, then broken generators we worked with are $Q_L^a-Q_R^a$. The quantity $U_{ij}$ transforms in the ${\mbf 3}$ ($\bar {\mbf  3}$) representation of $SU(3)_L$ ($SU(3)_R$), in which case $Q_L^a= - Q_R^a = Q^a$ with $Q^a= \tfrac{1}{2} \lambda^a$ and $\lambda^a$ the Gell-Mann matrices.} With this notation, the building block $U_{ij}$ we used in chiral perturbation theory was just a particular representation of the abstract group element $g = e^{i \pi^a(x) X_a}$. A natural guess would be to try using this quantity for any symmetry breaking pattern. For reasons that will become more clear in a moment, however, it is actually better to work instead with the following quantity:
\begin{align} \label{coset}
	g  = e^{i x^\alpha P_\alpha} e^{i \pi^a(x) X_a} \ . 
\end{align}
The element $g$ is usually referred to as a \high{coset parametrization}. The extra factor of $e^{i x^\alpha P_\alpha}$, describing a translation from the origin of our coordinate system to the point $x^\alpha$ at which the NG fields $\pi^a(x)$ are evaluated, will ensure that the $\pi$'s transform as expected under spatial translations. This is especially non-trivial in the presence of spontaneously broken space-time symmetries. But we are getting ahead of ourselves: we still need to define how \emph{any} symmetry acts on the NG fields, let alone translations. 

Using the group structure, we define the action of a symmetry transformation $\tilde g$ on the quantity $g$ as the product of the two group elements, {\it i.e.} $g \to \tilde g \cdot g$. What does this imply for the coordinates and the NG fields? In general the group element $\tilde g \cdot g$ will not have the particular form in Eq. \eqref{coset}, since it will also depend on unbroken generators. However, this element can always be decomposed uniquely into the product of a term that looks like $g$ (albeit with different values of $x^\alpha$ and $\pi^a$) and an element $h$ of the unbroken group $H$ (which in general will depend on the $\pi$'s and on $\tilde g$):
\begin{align} \label{coset transformation rule}
	\tilde g  \cdot g (x, \pi) = g (\tilde x, \tilde \pi) \cdot h(\pi, \tilde g) .
\end{align}
This equation defines the transformation rules of coordinates and NG fields, in the sense that $(x^\alpha, \pi^a(x) ) \stackrel{\tilde g}{\longrightarrow}  (\tilde x^\alpha, \tilde \pi^a(\tilde x) )$. These transformation rules are in general highly non-linear. The explicit form of $\tilde x$, $\tilde \pi$, and $h(\pi, \tilde g)$ can in principle be calculated  using the algebra of the symmetry generators. In practice, though, one seldom needs this information. 

\begin{eBox}
{\bf Exercise 3.3:} Use Eq. \eqref{coset transformation rule} to show that the NG fields of spontaneously broken internal symmetries transform like scalars under Poincar\'e transformations.
\end{eBox}

The need for the factor $h(\pi, \tilde g)$ in Eq. \eqref{coset transformation rule} might not seem obvious based on our discussion of chiral perturbation theory. This is because, in that example, we worked in a particular representation where the generators of $SU(3)_L$ and $SU(3)_R$ were the same up to a sign. As a result, $U = e^{2 i \pi^a Q_a}$ was the most general parametrization of an $SU(3)$ element, and therefore it kept the same form no matter which $SU(3)$ transformation acted on it. In other words, in this particular representation one could choose to absorb $h(\pi, \tilde g)$ in the transformation rules of the NG fields---and that's exactly what we did in Eq. \eqref{nonlinearly realized symmetries}. This is however not possible for a generic representation, and the simple strategy we followed to write down the effective action for chiral perturbation theory will not work. We will have to be more clever.

Our plan will be to derive some fundamental building blocks that depend on the NG fields and  have relatively simple transformation rules. NG fields are derivatively coupled, meaning that each term in their Lagrangian contains some derivatives. We therefore start by taking a derivative of $g$, or even better, by introducing the \high{Maurer-Cartan form} $ dx^\mu g^{-1} \partial_\mu g $. This quantity is particularly convenient because it is an element of the algebra of $G$, and therefore can be expressed as a linear combination of all the generators, which we can always cast in the following form: 
\begin{align} \label{MC form}
	dx^\mu  g^{-1} \partial_\mu g = i\, dx^\mu \left(  e_\mu{}^\alpha P_\alpha + e_\mu{}^\alpha \nabla_\alpha \pi^a X_a + C_\mu^A T_A \right) .
\end{align}
The coefficients $e_\mu{}^\alpha$, $\nabla_\alpha \pi^a$, and $C_\mu^A$ are non-linear functions of the Goldstone $\pi$'s, and are the main building blocks that we'll use to build the effective action for NG fields. Their explicit form can be calculated using the algebra of the group $G$. We will not need to know the exact commutation relations in what follows, but we will assume that 
\begin{subequations} \label{assumptions re commutators}
\begin{align}
	[T_A, X_a] = i f_{Aab} X^b , \label{[T,X]} \\
	[T_A, P_\alpha] = i f_{A\alpha\beta} P^\beta ,
\end{align}
\end{subequations}
meaning that the generators of translations and broken transformations form a (possibly reducible) representation of the unbroken group $H$. This requirement is not very restrictive; for example, Eq. \eqref{[T,X]} is satisfied by any compact, internal symmetry group, as well as in many other physically interesting situations.
\begin{eBox}
{\bf Exercise 3.4:} For compact groups, one can always choose the generators in such a way that the structure constants are totally antisymmetric~\cite{Weinberg:1996kr}. Use this result together with the fact that the $T_A$'s span a subgroup to prove Eq. \eqref{[T,X]}. {\bf Hint:} this amounts to showing that the righthand side of \eqref{[T,X]} cannot depend on the $T_A$'s. 
\end{eBox}

Using now the transformation rule \eqref{coset transformation rule} and the commutation relations \eqref{assumptions re commutators}, it is easy to show that the coefficients of the Maurer-Cartan form must transform as follows under the action of $\tilde g$:
\begin{subequations}
\begin{align}
	 e_\mu{}^\alpha &\to h(\pi, \tilde g)^\alpha{}_\beta \, e_\mu{}^\beta , \label{coset vierbein transformation} \\
	 \nabla_\alpha \pi^a &\to  h(\pi, \tilde g)^a{}_b \, \nabla_\beta \pi^b \,  h^{-1}(\pi, \tilde g)^\beta{}_\alpha ,\\
	  C_\mu^A T_A &\to h (\pi, \tilde g) \, C_\mu^A T_A \, h^{-1} (\pi, \tilde g) - i \,  h(\pi, \tilde g) \partial_\mu h^{-1} (\pi, \tilde g) , \label{coset connection transformation}
\end{align}
\end{subequations}
where $h(\pi, \tilde g)^\alpha{}_\beta$ and $ h(\pi, \tilde g)^a{}_b$ are different representations of the abstract group element $h(\pi, \tilde g)$. These transformation rules are not as complicated as they might appear at first sight. They are highly non-linear in the NG fields, but the non-linearity is ``packaged'' inside $h(\pi, \tilde g), e_\mu{}^\alpha, \nabla_\alpha \pi^a$ and $C_\mu^A$. The generic group element $\tilde g$ only enters through the unbroken element $h(\pi, \tilde g)$, which in turn acts on the coefficients of the Maurer-Cartan form in fairly simple ways, suggesting the role that each of these building blocks will play.  

The coefficients $e_\mu{}^\alpha$ are known as the \high{coset vierbein}, because their transformation properties resemble those of the vierbein in General Relativity under local Lorentz transformations~\cite{Carroll:2004st}. If the $X$'s are generators of internal symmetries, the coset vierbein is just a Kronecker delta and $h(\pi, \tilde g)$ can only be a global Lorentz transformation (or Galilei transformation, in non-relativistic systems). Things are more interesting when some space-time symmetries are spontaneously broken, in which case $e_\mu{}^\alpha$ and $h(\pi, \tilde g)^\alpha{}_\beta$ acquire a non-trivial dependence on the NG fields. The latter however remains a tensor with unit determinant, and therefore the measure of integration $d^4 x \det e$ is invariant under the transformations of the form \eqref{coset vierbein transformation}. As a bonus, this measure is also invariant under arbitrary redefinitions of the coordinates $x^\mu$. This allows us in principle to work not just in the Cartesian coordinates implicitly used in Eq. \eqref{coset}, but in fact in arbitrary coordinate systems.

The quantities $\nabla_\alpha \pi^a$ are known as \high{coset covariant derivatives}. Despite the notation, we shouldn't think of these covariant derivatives as the action of some derivative operator $\nabla_\alpha$ on the fields $\pi^a$. These are instead particularly clever non-linear combinations of the NG fields of the form
\begin{align}
	\nabla_\alpha \pi^a = \partial_\alpha \pi^a + \mathcal{O}(\pi^2). 
\end{align}
The non-linear terms ensure that $\nabla_\alpha \pi^a$ transforms in a linear representation of $h(\pi, \tilde g)$---unlike the $\partial_\alpha \pi^a $, which by itself would transform in a much more complicated way. It is very easy to build invariant operators out of quantities that transform linearly: we just need to contract all the indices in the appropriate way. The amazing property of the coset covariant derivatives is that contractions that are manifestly invariant under unbroken transformations $h$, are also guaranteed to be secretly invariant under broken transformations! 

Finally, we will call $C_\mu^A$ the \high{coset connection}, because it transforms exactly like a gauge field under $h(\pi, \tilde g)$ transformations. We can use it to define a covariant derivative
\begin{align}
	D_\alpha = (e^{-1})_\alpha{}^\mu (\partial_\mu + i C_\mu^A T_A), 
\end{align}
which preserves the linearity of the transformation rules when acting on $\nabla_\nu \pi^a$---or, for that matter, on any other field that transforms at $\Psi^i \to h(\pi, \tilde g)^i{}_j \Psi^j$. 

Putting it all together, we have found that effective action for the NG fields associated with any symmetry breaking patter $G \to H$ takes the form
\begin{align} \label{coset effective action}
	S = \int d^4 x \det e \, \mathcal{L} (\nabla_\alpha \pi^a, D_\alpha) ,
\end{align}
where the Lagrangian $\mathcal{L}$ contains all possible operators that are manifestly invariant under the unbroken transformations. The main building blocks of this action are defined in Eq. \eqref{MC form}, and can be calculated using only the algebra of the symmetry generators. This procedure is in principle straightforward, but as usual the devil is in the details, so let's spell out a few of them:
\begin{itemize} 
	\item We have already mentioned that spontaneously broken generators are only defined up to the addition unbroken generators. Different choices of broken generators correspond to field redefinitions of the NG fields. This freedom can always be leveraged to simplify the calculations of the Maurer-Cartan form. For example, in chiral perturbation theory it is especially convenient to work with the broken generators $X_a = Q_a^L = \tfrac{1}{2} (Q_a^L - Q_a^R) + \tfrac{1}{2} (Q_a^L + Q_a^R)$, because this choice makes the coset connection vanish. 
	\item It is usually convenient to break up the broken generators $X_a$ into irreducible representations $X_a^{(j)}$ of the unbroken symmetries, and to use a coset parametrization given by a product of several factors, {\it i.e.} $e^{i \pi^a(x) X_a^{(1)}} e^{i \pi^b(x) X_b^{(2)}} \dots$. These exponentials should be ordered in such a way that the $X^{(1)}$'s form a representation of the symmetries generated by the $X^{(2)}$'s ({\it i.e.} $[ X_a^{(1)},  X_b^{(2)} ] = i f_{abc}  X^c_{(1)}$), and so on. Factorizing the coset parametrization in this way amounts once again to a field redefinition of the $\pi$'s. From this perspective, chiral perturbation theory is a particularly simple example because the broken generators form an irreducible representation of $SU(3)_V$. 
	\item A couple of identities you might find helpful when performing explicit calculations are:
	\begin{subequations}
	\begin{align}
	e^{-Y} X e^Y &= X + [X,Y] +\frac{1}{2!}  [[X,Y],Y] +\frac{1}{3!}  [[[X,Y],Y] ,Y] + \dots ,\\
	e^{- i\pi^a X_a} \d_\mu  \, e^{i \pi^a X_a} &=  i \, \partial_\mu \pi^b \int_0^1 ds \, e^{-i (1-s)\pi^a X_a} X_b \, e^{i (1-s)\pi^c X_c} \\
	&= i \, \d_\mu \pi^a \l(X_a + \frac{i }{2!} \pi^b [X_a, X_b] + \frac{i^2}{3!}  \pi^b \pi^c [ \, [X_a, X_b], X_c] + \dots \r) .
	\end{align}
	\end{subequations}
	Calculating the first few terms in these expansions is usually sufficient to notice a pattern and re-sum the entire series.
	\item The effective action \eqref{coset effective action} can be easily generalized by including additional matter fields that transform under irreducible representations of the unbroken symmetries. Interaction terms that are manifestly invariant under all unbroken symmetries will also preserve the broken ones as long as the matter fields transform as $\Psi \to h(\pi, \tilde g) \,\Psi$. In chiral perturbation theory, one can describe in this way processes that involve both mesons and light leptons.
	\item Finally, one can also take into introduce gauge fields by promoting the ordinary derivative appearing in the Maurer-Cartan form to a gauge covariant derivative:
	\begin{align}
		d x^\mu g^{-1} \partial_\mu g \to d x^\mu g^{-1} (\partial_\mu - i e A_\mu^i Q_i) g,  
	\end{align}
	where the $Q_i$'s can be a combination of broken and unbroken generators. Our coset building blocks will now depend not only on the NG fields, but also on the gauge fields $ A_\mu^i$. For any gauge generator that is broken, one can always set to zero the corresponding NG field in the coset parametrization by working in unitary gauge.
\end{itemize}
\begin{eBox}
{\bf Exercise 3.5:} Rederive the effective action for chiral perturbation theory in Eq. \eqref{CPT action} using the coset construction.
\end{eBox}

\subsection{Inverse Higgs constraints}

The textbook version of Goldstone's theorem, asserting the existence of one NG mode for each broken generator, only applies to internal symmetries. When space-time symmetries are spontaneously broken, there is often a mismatch between the number of broken generators and that of NG modes. Think for instance of a relativistic 4D brane in 5 dimensions, whose embedding is defined by the condition $X^5 = 0$. This constraint breaks spontaneously translations in the 5-direction (generated by $P_5$) as well as Lorentz transformations that mix the 5-direction with directions along the brane (generated by $J_{\alpha 5}$). And yet, this system has a single massless mode corresponding to local deformations of the brane in the normal direction. 

How does the coset construction capture this mismatch between NG modes and broken generators?  The answer is that, when space-time symmetries are spontaneously broken, it is possible to impose additional local constraints that preserve all the symmetries and can be solved to express some of the NG modes in terms of others. These are known as \high{inverse Higgs constraints}, and amount to setting to zero one or more multiplets (under $H$) of coset covariant derivatives. The rule of thumb to determine which constraints can be imposed is that, if  $X_a^{(1)}$ and $X_a^{(2)}$ are two multiplets of broken generators such that
\begin{align} \label{IH rule of thumb}
	[P_\alpha, X_a^{(1)}] \supset i f_{\alpha a b} X^b_{(2)} ,
\end{align}
then we can set $f^{\alpha a b} \nabla_\alpha \pi_b^{(2)} = 0$ to eliminate the $\pi_a^{(1)}$'s from the effective theory. Notice that the righthand side of Eq. \eqref{IH rule of thumb} can be non-zero only if the $X_a^{(1)}$'s generate space-time symmetries. It is easy to see that the our rule of thumb is indeed satisfied by the 4D brane we mentioned above, since $[P_\alpha, J_{5 \beta}] = i \eta_{\alpha \beta} P_5$. 

What is the rationale behind the criterion in Eq. \eqref{IH rule of thumb}? Whenever this condition is satisfied, one can show that 
\begin{align} \label{IH constraint}
	f^{\alpha a b} \nabla_\alpha \pi_b^{(2)} =  f^{\alpha a b} \partial_\alpha \pi_b^{(2)} + \kappa \, \pi^a_{(1)} +\dots , 
\end{align}
where $\kappa$ is some constant and the dots stand for terms non-linear in the NG fields. The inverse Higgs constraint can then be solved order-by-order in the $\pi$'s to obtain a local expression for $\pi^a_{(1)}$ in terms of $\pi_b^{(2)}$, their derivatives, and possibly other NG fields.

By imposing all possible inverse Higgs constraints, one obtains the most minimal field content needed to realize a given symmetry breaking pattern. We can choose not to impose some inverse Higgs constraints, but then Eq. \eqref{IH constraint} shows that operators of the form $(f^{\alpha a b} \nabla_\alpha \pi_b^{(2)})^2$ will  give rise to a mass term for the $\pi^a_{(1)}$'s. Thus, those NG modes that could be removed by inverse Higgs constraints are always massive. Keeping them in the effective theory is possible, but it's akin to including additional matter fields. This can be seen explicitly by performing a local field redefinition to trade the $\pi^a_{(1)}$'s for $\Psi^a = f^{\alpha a b} \nabla_\alpha \pi_b^{(2)}$, which transform exactly like matter fields, {\it i.e.} $\Psi^a \to h(\pi, \tilde g)^a{}_b \Psi^b$.
\begin{eBox}
{\bf Exercise 3.6:} Consider a relativistic point particle at rest. What space-time symmetries does it break? Use the coset construction and impose all possible inverse Higgs constraints to derive the usual world-line action for a relativistic point particle.  
\end{eBox}

\subsection{Wess-Zumino-Witten terms}

Let's turn our attention back to chiral perturbation theory for a moment. Remarkably, all the terms in the effective action \eqref{CPT action} are invariant under the symmetry $U \leftrightarrow U^\dag$, which is also equivalent to $\pi^a \to - \pi^a$. Even those terms that explicitly break chiral symmetry, such as those in Eqs. \eqref{CPT breaking mass} and \eqref{EM spurion masses}, preserve this discrete symmetry. This symmetry forbids scattering processes involving an odd number of mesons. Unfortunately, this is in conflict by experiments, where processes such as $K^+ K^- \to \pi^+ \pi^0 \pi^-$ are observed.   

This apparent contradiction is due to the fact that all the operators we have considered so far are exactly invariant under chiral symmetry\footnote{Assuming that the spurions transform appropriately.}---as opposed to invariant up to a total derivative. The coset construction makes this very clear: by contracting indices in a way that preserves the unbroken symmetries, we can only build operators that are exactly invariant. However, terms that are invariant only up to a total derivative also belong in the effective action. These terms are usually referred to as \high{Wess-Zumino-Witten} (WZW) terms in the context of effective theories for NG modes. 

In the case of chiral perturbation theory, there happens to be only one WZW term. In order to write it down, we will resort to a clever trick. Following~\cite{Witten:1983tw}, we will extend spacetime to 5 dimensions, in such a way that ordinary 4D space-time is the boundary of this 5D region. We will denote with $X^M$ the 5D coordinates. Consider now the following 5D action: 
\begin{align}\label{WZW pi 5D}
\!\!\!	S_{WZW} =\! \int \! d^5 X \!\left\{ - \frac{2i \lambda}{15 \pi^2}\epsilon^{MNPQR} \mbox{Tr} \left[ U^\dag (\partial_M U)  U^\dag (\partial_N U) U^\dag (\partial_P U) U^\dag (\partial_Q U) U^\dag (\partial_R U) \right] \right\} ,
\end{align}
where the definition of the coupling $\lambda$ has been chosen for later convenience.  It is exactly invariant under chiral transformations, and its integrand has a special property: when expanded in powers of NG fields, each order is a total derivative. This means that, despite the appearances, this is actually a boundary term, {\it i.e.} an integral over 4D space-time. At lowest order in the canonically normalized NG fields, it reads

\begin{align} \label{WZW pi}
	S_{WZW} = \int d^4 x \, \frac{64 \lambda}{15 \pi^2 f^5} \, \epsilon^{\mu\nu\lambda\rho} \hat \pi^a \partial_\mu \hat  \pi^b \partial_\nu \hat  \pi^c \partial_\lambda \hat  \pi^d \partial_\rho \hat  \pi^e \, \mbox{Tr} (Q_a Q_b Q_c Q_d Q_e) + \dots ,
\end{align}
and therefore it violates the $\pi^a \to - \pi^a$ symmetry. At lowest order, the broken chiral transformations act on the NG fields as constant shifts: $\pi^a \to \pi^a + \epsilon^a + \dots$. It is easy to check that the leading order term in Eq. \eqref{WZW pi} is in fact invariant up to a total derivative under these transformations. The quintic coupling in Eq. \eqref{WZW pi} accounts precisely for processes such as $K^+ K^- \to \pi^+ \pi^0 \pi^-$.

WZW terms enjoy very special non-renormalization properties, in that they do not get renormalized by operators that are exactly invariant. To prove this, imagine promoting the Wilson coefficients of all the exactly invariant terms to spurion fields. These terms will remain invariant under chiral symmetry provided the spurions do not transform. However, if the couplings in front of WZW terms were to receive quantum corrections proportional to the spurions, they would effectively become functions of coordinates, spoiling invariance up to a total derivative. We conclude therefore that WZW terms cannot get renormalized by exactly invariant operators. 

In the particular case of chiral perturbation theory, one could have of course reached the same conclusion by simply invoking the fact that loops built out of exactly invariant operators should preserve the $\pi^a \to - \pi^a$ symmetry, and thus cannot give rise to the WZW term in Eq.~\eqref{WZW pi}. In fact, this term is even more robust, in that doesn't get renormalized at all: it is protected for topological reasons, which require $\lambda$ to be an integer number~\cite{Witten:1983tw}.\footnote{I should stress that this is a special property of the WZW term in Eq. \eqref{WZW pi 5D}. It follows from the fact that, in this case, the homotopy group $\pi_4 (G/H)$ is non-trivial. WZW terms do not necessarily have Wilson coefficients that are quantized.}

The procedure we followed to introduce the WZW term for chiral perturbation theory might have seemed very \emph{ad hoc}. However, it is just a particular example of a very general method, based once again on the coset construction. To conclude this section, we will briefly review the main idea, pointing out along the way a connection with a beautiful mathematical structure. Unfortunately, we will only be able to scratch the surface. 

Let's forget for a moment that the NG fields are functions of the space-time coordinates, and treat them as independent quantities. This allows us to build exactly invariant 5-forms by wedging together the coefficients of the Maurer-Cartan from $g^{-1} d g$, each of which is a 1-form, and contracting all the indices appropriately.  In order to build WZW terms, we are interested in those 5-forms $\alpha$ that are exact, {\it i.e.} such that $\alpha = d \beta$ with $\beta$ some 4-form. These 5-forms should be built \emph{before} imposing any inverse Higgs constraint. The form $\beta$ does not need to be exactly invariant, but it can at most shift by a total derivative under a symmetry transformation since $\alpha$ must remain invariant:
\begin{align}
	\beta \to \beta + d \gamma . 
\end{align}
If $\beta$ does shift by a total derivative, then its integral over 4D space-time is a WZW term. 

According to this definition, $\beta$ is only defined up to the addition of exactly invariant terms. How many truly independent WZW terms are there for a given symmetry breaking pattern $G \to H$?  It turns out that there is a one-to-one correspondence between the number of WZW terms and the elements of the relative Lie algebra cohomology of $G/H$. See \cite{Goon:2012dy} and references therein for many more details.
\begin{eBox}
{\bf Exercise 3.7:} The standard kinetic term for a non-relativistic point particle is invariant under boosts only up to a total derivative. Using the algebra of the Galilei group (which you can find at the end of Sec. 2.4 of~\cite{Weinberg:1995mt}), carry out the coset construction and follow the procedure outlined above to write the kinetic term as a WZW term. {\bf Hint:} The EFT for a point particle is a (0+1)-dimensional field theory. Therefore, WZW terms are integrals of exact 2-forms. 
\end{eBox}

\subsection{Chiral anomalies} \label{sec: chiral anomalies}

The WZW term in Eq. \eqref{WZW pi} has deep connections with anomalies, which unfortunately we will not be able to explore in full detail. Still, I cannot resist the temptation of saying at least a few words about chiral anomalies. One can introduce the coupling with electromagnetism by supplementing the terms in Eq. \eqref{WZW pi} with 
\begin{align} \label{Delta S WZW}
	\Delta S_{WZW} &= \int d^4 x \bigg\{ - \frac{e\lambda }{48\pi^2}  \epsilon^{\mu\nu\lambda\rho} A_\mu \mbox{Tr} \big[ Q_L (\partial_\nu U ) U^\dag (\partial_\lambda U ) U^\dag(\partial_\rho U ) U^\dag  \\
	& \qquad \qquad \qquad \qquad \qquad \qquad \qquad \qquad \qquad \qquad + Q_R U^\dag (\partial_\nu U) U^\dag (\partial_\lambda U) U^\dag (\partial_\rho U)\big] \nonumber \\
	& \quad +\frac{i e^2\lambda }{24\pi^2}\epsilon^{\mu\nu\lambda\rho} (\partial_\mu A_\nu) A_\lambda \mbox{Tr} \big[ Q_L^2 (\partial_\rho U)U^\dag + Q_R^2 U^\dag (\partial_\rho U) + Q_L U Q_R U^\dag (\partial_\rho U)U^\dag \big] \bigg\} , \nonumber
\end{align}
with $Q^L$ and $Q^R$ defined in Eq. \eqref{QL QR defs}. Notice that in this case gauging is not as straightforward as usual because the WZW term is not exactly invariant. When expanded in powers of the NG fields, the righthand side of Eq. \eqref{Delta S WZW} contains a term of the form 
\begin{align}
	\Delta S_{WZW} \supset \int d^4 x \, \frac{\lambda e^2 }{48 \pi^2 f} \, \hat \pi^0 \epsilon^{\mu\nu\lambda\rho} F_{\mu\nu} F_{\lambda \rho} 
\end{align}
with $\hat \pi^0$  the canonically normalized field that describes neutral pions. This interaction is responsible for the decay process $\pi^0 \to \gamma \gamma$, which in turn is a consequence of the fact that the non-linearly realized symmetry associated with $\pi^0$ is anomalous. To see this, we can calculate the Noether current $J^\mu_{3, A}$ for this symmetry\footnote{The subscript $A$ on the current stands for ``axial'', meaning that $J^\mu_{3, A} = J^\mu_{3, L} - J^\mu_{3, R}$; the subscript ``3'' stands for the fact that the symmetry generator associated with $\pi^0$ is $Q_3 = \tfrac{1}{2} \lambda_3$.} following the standard procedure. We find that its divergence doesn't vanish even in the limit of zero quark masses:
\begin{align} \label{chiral anomaly}
	\partial_\mu J^\mu_{3, A} = - \frac{\lambda e^2 }{48 \pi^2} \, \epsilon^{\mu\nu\lambda\rho} F_{\mu\nu} F_{\lambda \rho} . 
\end{align}
The righthand side arises precisely from the term in Eq. \eqref{Delta S WZW}. This result is an exact relation between operators that are well defined at all energies. As such, it can be derived not only using chiral perturbation theory, but also from a one-loop calculation in QCD, where one finds that $\lambda$ is equal to the number of colors, {\it i.e.} $\lambda = 3$.

It is worth stressing that, in the absence of explicit symmetry breaking corrections, $\Delta S_{WZW}$ is the \emph{only} term that prevents the divergence of $J^\mu_{3, A}$ from vanishing: in other words, it is the only term that breaks the associated symmetry. From the perspective of QCD, this is a manifestation of the fact that anomalies are 1-loop exact, meaning that the righthand side of  \eqref{chiral anomaly} doesn't receive contributions from higher-loops~\cite{Harvey:2005it}. By contrast, when a symmetry is explicitly broken, the effective action contains an infinite number of symmetry-breaking operators. Anomalies are therefore a very special way of breaking symmetries.

The reader might  remember that we invoked another anomaly at the beginning of this chapter to argue away the axial symmetry $U(1)_A$, and therefore the existence of a corresponding NG mode. This could seem at odds with the fact that $J^\mu_{3, A}$ is also anomalous, and yet its spontaneous breaking gives rise to the NG field we identify with the $\pi^0$ meson. The difference between these two anomalies lies in the fact that, had we done a 1-loop calculation in QCD, we would have found that
\begin{align}
	\partial_\mu J^\mu_A = - \frac{g^2}{16 \pi^2} \epsilon^{\mu\nu\lambda\rho} \mbox{Tr} (G_{\mu\nu} G_{\lambda \rho} ) .
\end{align}
Unlike in Eq. \eqref{chiral anomaly}, the anomaly in the conservation of $J^\mu_A$ is now proportional to the field strength of strong interactions. This is once again a relation among operators, and therefore should hold at all scales. However, at low energies all quarks are confined, and there is no trace of QCD; therefore, there shouldn't be any trace of the $U(1)_A$ symmetry either.

\subsection{Gauge anomalies}

Since we have opened up the pandora's box of anomalies, I would like to also say a few words about anomalous gauge symmetries. The standard lore is that gauge anomalies should be avoided, otherwise the resulting gauge theory doesn't make sense. In fact, absence of gauge anomalies is often used as a constraint on the particle content of EFTs. And yet, I am now going to argue that anomalous gauge theories are just fine from an EFT perspective as long as the gauge boson is massive~\cite{Preskill:1990fr}. 

Let's consider a concrete example, namely a $U(1)$ gauge theory with $N$ left-handed Weyl fermions described by the action
\begin{align}
	S = \int d^4 x \left\{ - \frac{1}{4} F_{\mu\nu} F^{\mu\nu} + \sum_{j=1}^N  i \, \psi_j^\dag  \bar \sigma^\mu (\partial_\mu - i e q_j A_\mu ) \psi_j \right\} .
\end{align}
Without going too much into details---which you can find for instance in~\cite{Srednicki:2007ab}---it turns out that the the $U(1)$ symmetry is anomalous unless the following condition is satisfied: 
\begin{align} \label{anomaly cancellation condition}
	\sum_j q_j^3 = 0 .
\end{align}
This combination of charges is usually referred to as the \emph{anomaly coefficient}.  
Let's assume therefore that such requirement is met, and now add to the theory a complex scalar field with the following interactions:
\begin{align}
	\Delta S = \int d^4 x \left\{ - |(\partial_\mu - i e  A_\mu )\Phi |^2 - \frac{\lambda}{2} ( |\Phi|^2 - v^2 )^2 - (y \, \Phi^\dag \psi_N \cdot \psi_N + \text{h.c.} ) \right\} .
\end{align}
The Yukawa interaction between $\Phi$ and $\psi_N$ is gauge invariant provided $q_N=\tfrac{1}{2}$; moreover, it is technically natural for all other Yukawa interactions to be absent as long as the charges of the remaining Weyl fermions do not have this value. Expanding around the vev $\langle\Phi\rangle = v$, we find that the gauge field  , the scalar, and the $N$-th fermion acquire masses  $m_A = 2 e v, m_\Phi = \sqrt{2 \lambda} v$ and $m_N = 2 y v$ respectively. If $\lambda, y \gg e$, we can choose to integrate out the scalar fluctuation and the $N$-th fermion and work with a low-energy EFT that only includes the (massive) gauge boson and $N-1$ fermions. By construction, this EFT doesn't satisfy Eq. \eqref{anomaly cancellation condition}, and yet we know that this EFT must be fine because it originated from an anomaly-free, renormalizable theory. In this construction, we violated \eqref{anomaly cancellation condition} at low energies at the expense of the gauge bosons becoming massive. This is actually a general result: since the masslessness of gauge bosons is enforced by the gauge symmetry, it make sense for $A_\mu$ to have a mass if the $U(1)$ is anomalous and hence not an actual symmetry.

\begin{figure}
	\begin{center}
	 	\includegraphics[scale=.25]{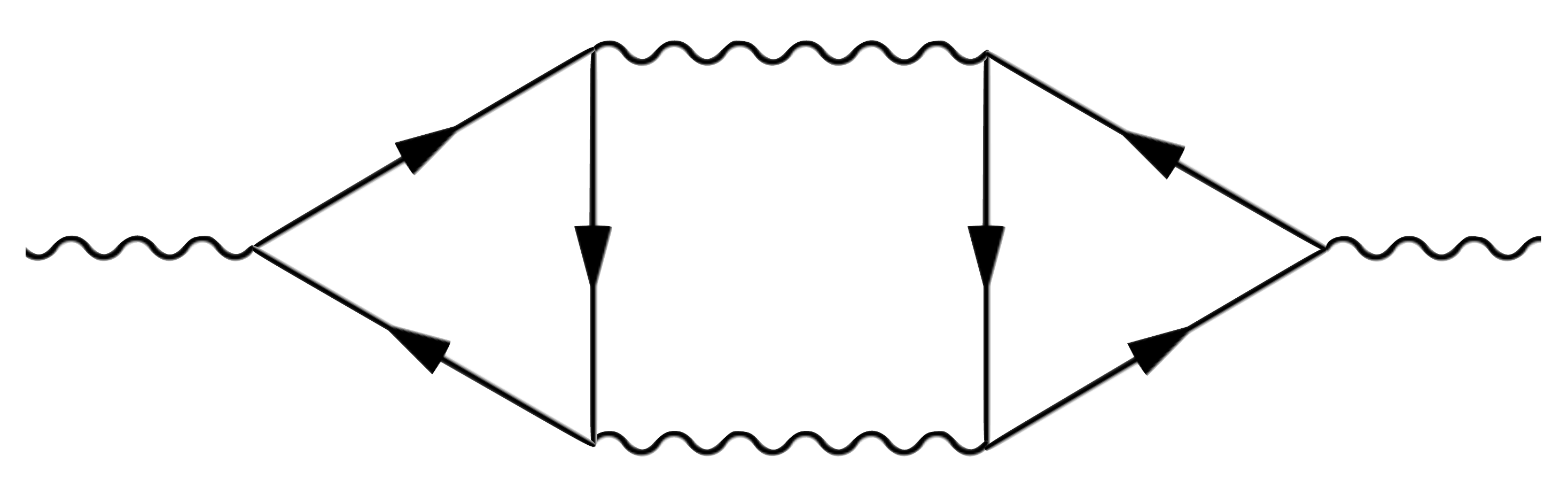}
	\end{center}
\caption{\small Feynman diagram contributing to the gauge boson mass.} \label{anomalous gauge boson mass}
\end{figure}

We can even derive a lower bound on the gauge boson mass by resorting to a naturalness argument. Imagine starting with a massless gauge boson at tree level, and consider the Feynman diagram shown in Fig. \ref{anomalous gauge boson mass}, with the $N$-th quark running in the triangular loops. A reader well-versed in anomalies will recognize that this graph is equal to two copies of a triangular fermion diagram---the hallmark of chiral anomalies---glued together by two photon lines. This 3-loop diagram is UV divergent, but there is no way to regulate it in such a way that gauge invariance is preserved. In order to match this diagram in the low-energy EFT one needs to add, among others, a mass term for the gauge boson with 
\begin{align} \label{m_A}
	m_A^2 \sim \left[\frac{1}{(4 \pi)^2}\right]^3 \times (e q_N)^6 \times (2 y v)^2 . 
\end{align}
The origin of the three factors on the right-hand side should be clear: one loop factor $1/(4 \pi)^2$ for each loop, one factor of $e q_N$ for each vertex, and the fermion mass $2 y v$ is the only mass scale appearing in this diagram since we assumed the gauge boson mass to be zero at tree level. We can also rewrite Eq. \eqref{m_A} in terms of quantities that are defined purely in the context of the low-energy EFT:
\begin{align} \label{m_A 2}
	m_A^2 \sim \left(\frac{e}{4 \pi}\right)^6 \times \left(\sum_{j=1}^{N-1} q_j^3 \right)^2 \times \Lambda^2 ,
\end{align}
with $\Lambda$ the strong coupling scale of the EFT. \emph{A posteriori}, we could have actually guessed this result based on the following observations:
\begin{enumerate}
	\item the strong coupling scale of the EFT is the only scale in the problem;
	\item physical quantities can only depend on the square of $e$, because the Lagrangian is invariant under $e \to -e, A_\mu \to - A_\mu$;
	\item it makes sense for the mass to be proportional to the anomaly coefficient, since this is the quantity that controls whether the $U(1)$ is actually a symmetry;
	\item each factor of $e$ should come with a factor of $q_j$, since the Lagrangian is invariant under $e \to -e, q_j \to - q_j$.
\end{enumerate}
The expression in Eq. \eqref{m_A 2} is actually quite general and, on naturalness grounds, provides a lower bound on the mass of an abelian gauge field in an EFT with anomalous fermion content.

\subsection{Additional resources}

More in-depth reviews on chiral perturbation theory are~\cite{Ecker:1994gg,Pich:1995bw,Scherer:2002tk}. General EFTs for Goldstone modes are the focus of the reviews~\cite{Burgess:1998ku,Pich:2018ltt}. You can read much more about anomalies in~\cite{Harvey:2005it,Bilal:2008qx}, and about anomalous Yang-Mills theories in \cite{Preskill:1990fr}.

\newpage

\section{EFTs for non-relativistic fermions}

\subsection{Non-relativistic fermions in vacuum} \label{sec: NR fermions in vacuum}

Power counting was particularly easy in the toy model of Sec. \ref{sec: toy model} because the fermions were relativistic, {\it i.e.} they had energies $E \gg m c^2$.\footnote{In this section we will reintroduce all factors of $c$; this means in particular that $\d_0 = \d_t / c$.} As a result, each component of a partial derivative acting on $\Psi$ scaled the same way, namely $\d_\mu \sim E /c$. This is however no longer true for scattering processes involving non-relativistic fermions, which have a momentum $p \ll mc$. In this case,
\begin{align}
	\d_\mu \Psi \sim i \, (- m c, \, \vec p \, ) \, \Psi,
\end{align}
and the zero-th component is now much larger than the spatial components. To be clear, the fact that different derivatives scale differently is not necessarily an issue---as we will see, this happens in many EFTs. The problem here is that the zero-th component doesn't scale at all with the momentum (or equivalently, the energy) of the fermion. For this reason, higher time derivatives of $\Psi$ are not necessarily more suppressed at low energies, and therefore operators in the Lagrangian are not clearly organized in powers of momentum. Fortunately, this can be easily remedied by introducing a new field $\tilde \Psi$ via the following \high{rephasing}:
\begin{align} \label{Psi tilde def}
	\Psi (t, \vec x)  \equiv e^{- i m c^2 t} \tilde \Psi (t, \vec x) 
\end{align}
Derivatives of $\tilde \Psi$ now scale like 
\begin{align} \label{Psi tilde derivatives}
	\d_\mu  \tilde \Psi = \d_\mu \left( e^{i m c^2 t} \Psi \right) \sim i \, (- \frac{p^2}{2mc}, \, \vec p \, ) \, \tilde \Psi
\end{align}
and therefore operators with additional derivatives are more suppressed for small momenta (compared to the strong coupling scale of the effective theory).  

It is instructive to implement the field redefinition directly at the level of the Lagrangian. Take for instance the action in Eq. \eqref{4pt tree effective theory} with $(\bar  \Psi \gamma^5 \Psi)^2 \to (\bar  \Psi  \Psi)^2$. We can write the action directly in terms of $\tilde \Psi$ with all the powers of $c$ and gamma matrices spelled out as follows: 
\begin{align} \label{4pt tree effective action with c}
	S_{\rm eff} = \int dt \, d^3 x \, c \left\{ \tilde \Psi^\dag \l[ \frac{ i \d_t}{c}  +  i \gamma^0 \gamma^i \d_i +  (1 - \gamma^0) m c  \r] \tilde \Psi - \frac{y^2}{2 M^2c} (\tilde \Psi^\dag \gamma^0 \tilde \Psi )^2\right\} \ .
\end{align}
The factor of $1/c$ in front of the quartic interaction can always be reabsorbed by redefining the coupling constant, but has been introduced for later convenience. According to the power counting rule \eqref{Psi tilde derivatives} for the derivatives of $\tilde \Psi$, the dominant term in square brackets is the one proportional to $m c$. This suggests that we introduce the projection matrices\footnote{It is easy to check that $P_\pm$ indeed satisfy the defining properties of projectors ($P_\pm^2 = P_\pm$). Moreover, they are orthogonal ($P_+ P_- = P_- P_+ = 0$) and form a complete set ($P_+ + P_- = 1$).} $P_\pm \equiv \tfrac{1}{2} (1 \pm \gamma^0)$ to decompose $\tilde \Psi$ as a sum of $\tilde \Psi_\pm \equiv P_\pm \tilde \Psi$. Because the leading quadratic term in the Lagrangian is $mc^2  \tilde \Psi_+^\dag \tilde \Psi_+$, we can integrate out $\tilde \Psi_+$ by working perturbatively in $p/mc$---the expansion parameter of our non-relativistic EFT---and obtain an effective Lagrangian for $\tilde \Psi_-$ alone. Notice that, since for non-relativistic particles we have $p = m v$, our expansion parameter can also be expressed as $v /c$. 
\begin{eBox}
{\bf Exercise 4.1:} integrate out $\tilde \Psi_+$ at tree-level and derive an effective action for $\tilde \Psi_-$ at lowest order in $p/mc$. 
\end{eBox}

Physically, we can integrate out half of the modes in the non-relativistic regime because low-energy scattering events involving only, say, particles, are unable to produce anti-particles by conservation of energy. We can make this more explicit by working with the Dirac representation of the Dirac matrices~\cite{Zee:2003mt}, in which $\gamma^0$ is diagonal:\footnote{This basis is more convenient in the non-relativistic limit compared to the Weyl basis used in~\cite{Srednicki:2007ab}, in which $\gamma^5$ is diagonal. Spinors in the two basis are connected by a linear transformation $\Psi_{\rm Dirac} = S \, \Psi_{\rm Weyl}$ with $S = \frac{1}{\sqrt{2}} \left( \begin{smallmatrix} 1 & 1 \\ -1 & 1 \end{smallmatrix} \right)$.}  
\begin{align} \label{2-spinor def}
	\gamma^0 = \begin{pmatrix}
		1 &  0 \\ 0 & -1 
	\end{pmatrix} \qquad \Longrightarrow \qquad \tilde \Psi_- = \begin{pmatrix}
		0 \\ \psi 
	\end{pmatrix} .
\end{align}
Then, the leading order Lagrangian for the 2-component spinor $\psi$ is  
\be \label{S psi}
S_{\rm eff} = \int dt \, d^3 x \left\{ \psi^\dag \l( i \d_t  + \frac{\nabla^2}{2 m} \r) \psi  - \frac{y^2}{2 M^2} (\psi^\dag \psi )^2 \right\} . 
\ee

It is now time to do some more power counting. From the scaling of derivatives in Eq. \eqref{Psi tilde derivatives} we infer that $d t d^3 x \sim 1/p^5$. Notice that, unlike the examples discussed in the previous two sections, the scaling with momentum can no longer be inferred from dimensional analysis in natural units. The scaling of $\psi$ is determined as usual by the requirement that the free action doesn't scale, which implies $\psi \sim p^{3/2}$. Therefore, we see that the quartic interaction we have kept scales like $\sim p$ and is therefore irrelevant at small momenta.\footnote{You can use Eq. \eqref{expectation values} and the non-relativistic version of the LSZ equation, in which $\slashed{\partial} \, u_j \to i \d_t  + \frac{\nabla^2}{2 m}$, to convince yourself that this scaling is consistent with the fact that the 4-point amplitude is a constant at lowest order in perturbation theory.}  In fact, \emph{any} interaction involving just fermions in vacuum turns out to be irrelevant. We will see in a moment that this is not the case for fermions at finite density. Before, though, let's discuss how the symmetries of our effective action \eqref{S psi} relate to those of the relativistic action we started from.

\subsection{Contracted and accidental symmetries}

Our relativistic Lagrangian is invariant under Lorentz transformations $x^{\mu \, \prime} = \Lambda^\mu{}_\nu x^\nu$ provided $\Psi$ transforms as follows (in the Dirac representation):
\begin{align} \label{trans rule relativistic spinor}
	\Psi' (x') = 
		\frac{1}{2}\begin{pmatrix}
		 e^{\frac{i}{2} \vec \sigma \cdot (\vec \alpha + i \vec \beta)} + e^{\frac{i}{2} \vec \sigma \cdot (\vec \alpha - i \vec \beta)} & e^{\frac{i}{2} \vec \sigma \cdot (\vec \alpha + i \vec \beta)} - e^{\frac{i}{2} \vec \sigma \cdot (\vec \alpha - i \vec \beta)} \\ e^{\frac{i}{2} \vec \sigma \cdot (\vec \alpha + i \vec \beta)} - e^{\frac{i}{2} \vec \sigma \cdot (\vec \alpha - i \vec \beta)} & e^{\frac{i}{2} \vec \sigma \cdot (\vec \alpha + i \vec \beta)} + e^{\frac{i}{2} \vec \sigma \cdot (\vec \alpha - i \vec \beta)}
		\end{pmatrix} \Psi (x) ,
\end{align}
where the $\alpha$'s parametrize rotations and the vector $\vec \beta \equiv \vec v / c$ parametrizes Lorentz boosts. At lowest order in a $p/mc$ expansion, Eq. \eqref{trans rule relativistic spinor} implies the following transformation rule for the 2-component spinor $\psi$:
\begin{align}
	\psi (t, \vec x^{\, \prime}) = e^{i m (v^2 t/2 - \vec v \cdot \vec x)} e^{i \vec \sigma \cdot \vec \alpha/2} \psi (t, \vec x) ,
\end{align}
while coordinates transform as usual under rotations and Galilean boosts.
\begin{eBox}
{\bf Exercise 4.2:} show that the differential operator $D \equiv i \d_t  + \nabla^2/(2 m)$ is a covariant derivative with respect to boosts, in the sense that $D' \psi' (t, \vec x^{\, \prime}) = e^{i m (v^2 t/2 - \vec v \cdot \vec x)} D \psi (t, \vec x)$ with $\vec x^{\, \prime} = \vec x - \vec v \, t$.
\end{eBox}

By using this result, we can see immediately that the non-relativistic Lagrangian \eqref{S psi} is invariant under rotations as well as Galilean boosts. This is of course just the $c \to \infty$ limit of Lorentz invariance. For lack of a better term,  I will call this a \high{contracted symmetry}, since the algebra of Galilean transformations follows from a contraction of the Lorentz algebra~\cite{Weinberg:1995mt}.

There is also an internal $U(2) = SU(2) \times U(1)$ symmetry transformation that acts on $\psi$ without changing the coordinates. The $U(1)$ subgroup is just a remnant of the $U(1)$ symmetry of the relativistic action---except that a low energies it is the number of particles, rather than particles minus antiparticles, that is conserved by kinematics. The $SU(2)$ subgroup instead doesn't have a relativistic counterpart, and is a manifestation of the well known fact that spin and coordinates can rotate independently in a non-relativistic theory. 

This additional $SU(2)$ is good example of \high{accidental symmetry}. As the name suggests, these are symmetries that arise accidentally when the particle content and fundamental symmetries of an EFT are so constraining that one can only write down invariant operators up to some order in the EFT expansion. Accidental symmetries are eventually broken by higher order corrections in the EFT. Thus, one can in principle distinguish between accidental and fundamental symmetries by testing the EFT with high enough accuracy. Some examples of accidental symmetries in the Standard Model are the ones associated with baryon and lepton number conservation, or the custodial symmetry of the Higgs sector.

Looking back at the lowest-order effective action \eqref{S psi}, we now  realize that we could have \emph{almost} guessed its form based on symmetry arguments alone. Invariance under spatial rotations requires the spin index of each $\psi$ to be contracted with an index carried by a $\psi^\dag$, while invariance under Galilean boosts requires us to use the covariant derivative $D$. Granted, at first order in derivatives we could have also added a term of the form $(D \psi)^\dag \psi$, but it is easy to show that this is equivalent to $\psi^\dag D \psi$ up to integrations by parts. More puzzling instead is the absence  of a quadratic term $\psi^\dag \psi$. This term would be compatible with the all the symmetries we've discussed so far, and its absence doesn't lead to a symmetry enhancement. In light of our previous discussions on naturalness, you might now be worried that our effective action \eqref{S psi} is fine-tuned. In fact, it is easy to see that such a term doesn't get renormalized, and therefore it is self-consistent to set it to zero. 

Consider for instance the following loop correction, which would renormalize the mass term of a relativistic scalar:\\

\begin{align}
	\parbox[t][10mm][b]{20mm}{\begin{fmfgraph*}(50,60) 
	\fmfleft{i} \fmfright{o} 
	\fmf{fermion}{i,v,o} 
	\fmffreeze
	\fmf{fermion,tension=.5}{v,v}
	\end{fmfgraph*}} = - i \frac{y^2}{M^2} \int \frac{dE d^3 p}{(2 \pi)^3} \frac{i}{E -p^2/2m + i \epsilon} = - i \frac{y^2}{M^2} \int \frac{d Q^2 d^3 p}{(2 \pi)^3} \frac{i}{Q^2 -p^2 + i \epsilon } . 
\end{align}
This integral is UV divergent, but after redefining the integration variable $E$ it doesn't involve any length scale, and therefore vanishes in dimensional regularization. A similar argument applies also to higher loop corrections to the 2-point function. After absorbing the factor of $2m$ in the energies, the only scales in the loop integrals are the external energy and momentum. This means that, in dimensional regularizations, loops will only renormalize quadratic operators with derivatives. 

The fact that the operator $\psi^\dag \psi$ doesn't get renormalized admits an intuitive explanation. Using standard arguments to calculate the $U(1)$ Noether current from the action \eqref{S psi}, one finds that operator $\psi^\dag \psi$ is the density of $U(1)$ charge. Thus, adding a term of the form $\Delta L = \mu_{\rm nr} \psi^\dag \psi$ corresponds to turning on a source for the charge density---{\it i.e.}, a chemical potential. This is something that an experimenter should be able to do at will, without having to worry about loop corrections!

\subsection{Non-relativistic fermions at finite density}

Let's now consider a state with a finite density of fermions. This can be achieved by turning on a finite chemical potential $\mu_{\rm nr} \ll m c^2$, and thus introducing another scale in the problem. At zero temperature, all the states with energies less than or equal to the Fermi energy $E_F \equiv \mu_{\rm nr}$ will be occupied, while all the other ones will be empty. This condition defines a closed surface in momentum space---the \emph{Fermi surface}---separating occupied from unoccupied momentum states. In the case of free fermions, this is just a sphere with radius $p_F = \sqrt{2 m E_F}$. In the presence of interactions this sphere gets deformed and the Fermi surface can   display more complicated shapes.  

The theory of \high{Fermi liquids} is based on the assumption that the relevant collective excitations at zero temperature and finite density remain fermions. This makes intuitive sense in the weak coupling regime, where one can visualize these excitations as fermions that are ``dressed'' by interactions\footnote{By this I mean that interactions will change the dispersion relation of one-particle states near the Fermi surface.} and excited slightly above the Fermi surface. Remarkably, though, this assumption proves to be empirically correct also in many strongly coupled systems. These fermionic excitations are often referred to as quasi-particles, to distinguish them from the actual fermions that make up the system.

I should point out that fermionic quasi-particles are not the only option. For example, fermions could pair up into bosons which then condense at low energies, as is the case in Helium 3 or in ordinary superconductors at low temperatures. We will briefly discuss this possibility at the end of this section. Some strongly coupled systems of fermions can even lack quasi-particle excitations altogether, in which case they are simply referred to as \emph{non-Fermi liquids}. In what follows, we will restrict our attention to Fermi liquids, and derive an EFT description of their fermionic quasi-particles.

Quasi-particles that are slightly excited above the Fermi surface are fermions whose energy differs slightly from $E_F$. This should be reminiscent of our previous discussion of non-relativistic fermions, whose energy deviated slightly from $m c^2$. The difference here is that particles on the Fermi surface have also finite momenta ${\vec p}_F$. To handle that, we will generalize the procedure followed in Sec. \ref{sec: NR fermions in vacuum} and parametrize our non-relativistic fermionic field $\psi$ as follows:  
\begin{align} \label{Psi tilde def}
	\psi (t, \vec x) \equiv e^{- i E_{F} t} \sum_{{\vec p}_F} e^{i {\vec p}_F \cdot \vec x} \, \psi_{{\vec p}_F} (t, \vec x) .
\end{align}
Thus, by extracting a phase that depends not only on $E_F$ but also on $\vec p_F$, we have effectively traded a single field $\psi$ for a collection of fields $\psi_{{\vec p}_F}$ labeled by all possible momenta on the Fermi surface. By construction, these fields describe excitations such that $\d_t \psi_{{\vec p}_F} \sim \epsilon \,\psi_{{\vec p}_F}$ with $\epsilon \ll E_F$. Therefore, the natural expansion parameter of our EFT is the ratio $\epsilon / E_F$. 

Let's now turn to the business of writing down an effective action for the $\psi_{{\vec p}_F}$'s. For strongly interacting systems we cannot derive our EFT directly from the underlying theory, as we did for instance for the non-relativistic model discussed in Sec. \ref{sec: NR fermions in vacuum}. For this reason, we will rely instead on symmetries as our guideline. Starting from the transformation properties of $\psi$, it is easy to show that the fields $\psi_{{\vec p}_F}$ must transform under space-time symmetries as~follows:
\begin{subequations} \label{sp symmetries psi p_F}
\begin{align}
	\text{time translations:} \qquad \psi_{{\vec p}_F}' (t + c, \vec x) &= e^{i E_F c} \,\psi_{{\vec p}_F} (t, \vec x) \\
	\text{spatial translations:} \qquad \psi_{{\vec p}_F}' (t, \vec x + \vec c) &= e^{-i \vec p_F \cdot \vec c} \,\psi_{{\vec p}_F} (t, \vec x) \\
	\text{spatial rotations:} \qquad  \psi_{R \cdot {\vec p}_F}' (t, R \cdot \vec x ) &= M(R) \,\psi_{{\vec p}_F} (t, \vec x) \\
	\text{Galilean boosts:} \qquad  \psi_{{\vec p}_F- m \vec v}' (t, \vec x - \vec v t) &= \psi_{{\vec p}_F} (t, \vec x) , \label{boost transf rule}
\end{align}
\end{subequations}
where $M(R)$ is the spin $1/2$ representation of the rotation $R$. 
\begin{eBox}
{\bf Exercise 4.4:} use the algebra of the Galilei group, and in particular the commutators
\begin{align}
	[ H, K_j ] = i P_j, \qquad \qquad [P_i, K_j ] = i M \delta_{ij} , \qquad \qquad [M, K_j ] = 0, 
\end{align}
to argue that, under a boost transformation such that $\vec x^{\, \prime} = \vec x - \vec v \, t$, the Fermi momentum and energy change as follows:
\begin{align} 
	\vec p_F^{\, \prime} = \vec p_F - m \, \vec v, \qquad \quad E_F' = E_F - \vec p_F \cdot \vec v + \frac{m v^2}{2}. \label{pF EF transformation rules}
\end{align}
Use then these results to derive the transformation rule \eqref{boost transf rule}. Finally, convince yourself that Galilean boosts change the position of the Fermi surface in momentum space but do not alter its shape.
\end{eBox}
To the extent that relativistic corrections are negligible, one should also impose the  accidental symmetry under spin rotations:
\begin{align} \label{int symmetry psi p_F}
	\text{spin rotations:} \qquad \psi_{{\vec p}_F}' (t, \vec x) &= U \,\psi_{{\vec p}_F} (t, \vec x) .
\end{align}

Which of the above symmetries we eventually impose depends on the physical system under consideration. For instance, in the case of electrons in a metal, the underlying lattice spontaneously breaks boosts and rotations. These symmetries are non-linearly realized by the phonons, but appear to be explicitly broken if couplings with phonons are not taken into consideration. In what follows, we will consider the simplest possible scenario, namely one in which there are only fermions and there is no underlying lattice. In this case, our effective action should be invariant under all the transformations discussed above, and the Fermi surface remains a sphere.  

Taken all together, the transformation rules \eqref{sp symmetries psi p_F} and \eqref{int symmetry psi p_F} place strong constraints on the form of our effective action. In particular, invariance under time translations imply that each $\psi_{{\vec p}_F}$ should appear together with a $\psi_{{\vec p}_F^{\, \prime}}^\dag$. Spatial translations require instead that the sum of all the labels of the $\psi_{{\vec p}_F}$'s minus those of the $\psi_{{\vec p}_F}^\dag$'s add up to zero. Finally, spin rotations is ensured provided all spinor indices are contracted.

The transformation properties of the $\psi_{\vec p_F}$'s under boosts require us to introduce covariant derivatives that are invariant under boosts. To this end, is helpful to define at each point on the Fermi surface the normal~vector
\begin{align}
	\vec v_F \equiv \frac{\d E_F}{\d \vec p_F},
\end{align}
which is usually referred to as \emph{Fermi velocity}. It is easy to see that this vector shifts as expected under boosts, {\it i.e.} $\vec v_F^{\, \prime} = \vec v_F - \vec v$, and therefore that the differential operator $\mathcal D \equiv \d_t + \vec v_F \cdot \vec \nabla$ is invariant. At lowest order in derivatives there is only one kinetic term that we can write down while preserving all the symmetries, which is
\begin{align} \label{S kin fermi liquids}
	S_{\rm kin} = \sum_{\vec p_F} \int dt \, d^3 x \, \psi_{\vec p_F}^\dag i \mathcal D \, \psi_{\vec p_F} \ , 
\end{align}
where the spinor indices are understood to be contracted. 

\begin{figure}
	\begin{center}
		\includegraphics[scale=0.4]{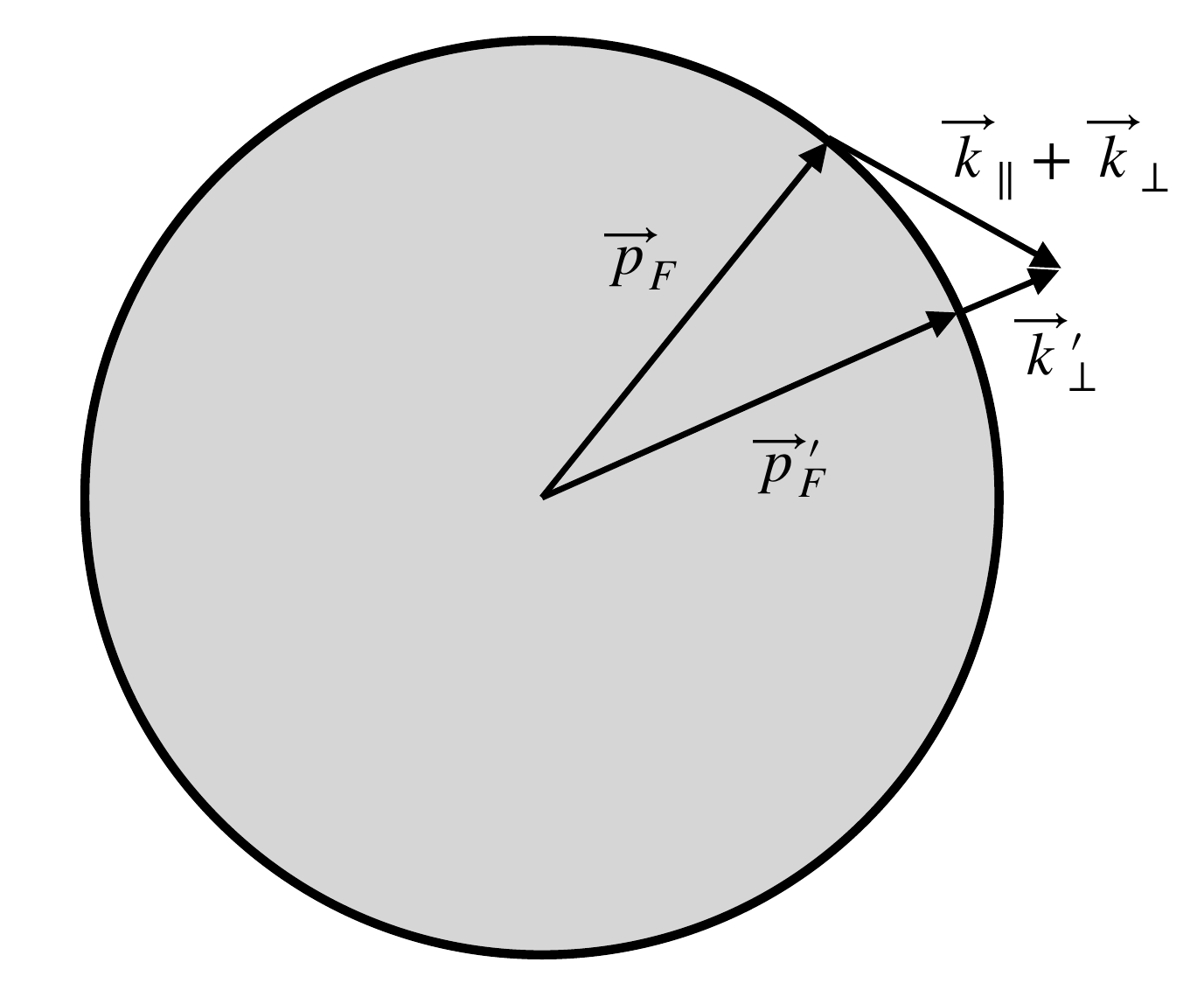}
	\end{center}
	\vspace{-.6cm}
	\caption{\small Momenta parallel and perpendicular to the Fermi surface.} \label{fig: Fermi surface}
\end{figure}

Now that we have determined the leading kinetic term, we are in a position to derive the power counting rules of our EFT. First, in order for the invariant differential operator to scale homogeneously we must have 
\begin{align} \label{fermi liquids power counting rule 1}
	\vec v_F \cdot \vec \nabla \sim \d_t \sim \epsilon.
\end{align}
The kinetic term determines the scaling of only one component of spatial gradients---the one perpendicular to the Fermi surface. This is unlike all other models we have considered so far. To understand why that's the case, we can decompose the momentum $\vec k$ of each Fourier mode $\psi_{{\vec p}_F} (t , \vec k)$ along directions parallel and perpendicular to the Fermi surface~at~$\vec p_F$:
\begin{align}
	\vec k = \hat v_F (\hat v_F \cdot \vec k ) + \hat v_F \times ( \vec k \times \hat v_F) \equiv \vec k_\perp + \vec k_\parallel \ .
\end{align}
Then, as shown in Fig \ref{fig: Fermi surface}, one can always redefine the field label to absorb any component of the momentum parallel to the Fermi surface. To be more precise, we have the following relation between the Fourier modes of fields at different points on the Fermi sphere:
\begin{align}
	\psi_{\vec p_F} (t, \vec k_\parallel + \vec k_\perp) = \psi_{\vec p_F^{\, \prime}} (t, \vec k_\perp^{\, \prime}), 
\end{align}
with
\begin{align}
	k_\perp' = | \vec p_F + \vec k_\parallel + \vec k_\perp|, \qquad \qquad \hat k_\perp' = \hat p_F' = \frac{\vec p_F + \vec k_\parallel + \vec k_\perp}{| \vec p_F + \vec k_\parallel + \vec k_\perp|} \ .
\end{align}
We can therefore always assume that our fields $\psi_{\vec p_F}$ only carry momentum in the direction perpendicular to the Fermi surface. The integration measure and our fields must scale therefore like
\begin{align} \label{fermi liquids power counting rule 2}
	dt\,  d^3 x \sim d t \, d x_\perp d^2 x_\parallel \sim \epsilon^{-2}, \qquad \qquad \psi_{\vec p_F} \sim \epsilon^{1/2}, 
\end{align}
to ensure that the kinetic term \eqref{S kin fermi liquids} is of $\mathcal{O}(1)$ in the derivative expansion.

Armed with the power counting rules \eqref{fermi liquids power counting rule 1} and \eqref{fermi liquids power counting rule 2}, we are now in a position to discuss interactions among quasi-particles. First, notice that there is in principle one relevant quadratic term that we could write down, namely
\begin{align} \label{quadratic int fermi liquids}
	-  \sum_{\vec p_F} \int dt \, d^3 x \, \Delta \, \psi_{\vec p_F}^\dag  \psi_{\vec p_F} \sim \epsilon^{-1} , 
\end{align}
where $\Delta$ is a parameter that must be independent of $\vec p_F$ to preserve boost invariance. This term would modify the dispersion relation of quasi-particles by adding an energy gap, {\it i.e.} $\epsilon \to \vec v_F \cdot \vec k + \Delta$. In other words, this operator shifts the energy of all quasi-particles by a constant amount. Such shift can always be absorbed into a redefinition of the Fermi surface, and that is why the quadratic operator \eqref{quadratic int fermi liquids} can always be removed by re-phasing our fields: $\psi_{\vec p_F}' \equiv e^{i \Delta t} \psi_{\vec p_F}$.

More interesting is the fact that there exist some quartic marginal interactions that have the schematic form
\begin{align} \label{fermi liquids marginal interactions}
	S_{\rm int} = \sum_{\vec p_{F, i}}{\vphantom{\sum}}' \int dt \, d^3 x \,  g \,  \psi_{- \vec p_{F,3}}^\dag   \psi_{\vec p_{F,1}}  \, \psi_{- \vec p_{F,4}}^\dag \psi_{\vec p_{F,2}} \sim 1 \ ,
\end{align}
where the prime on the sum is a reminder of the constraint $\vec p_{F,1} +   \vec p_{F,2} + \vec p_{F,3} +\vec p_{F,4} = 0$ imposed by translational invariance. Moreover, the coefficient $g$ carries an index structure to ensure that all spinor indices are contracted in a $SU(2)$-invariant way, and can depend on the labels $\vec p_{F, i}$ in the most general way consistent with rotational and boost invariance. 
\begin{figure}[t]
	\begin{center}
		\includegraphics[scale=0.4]{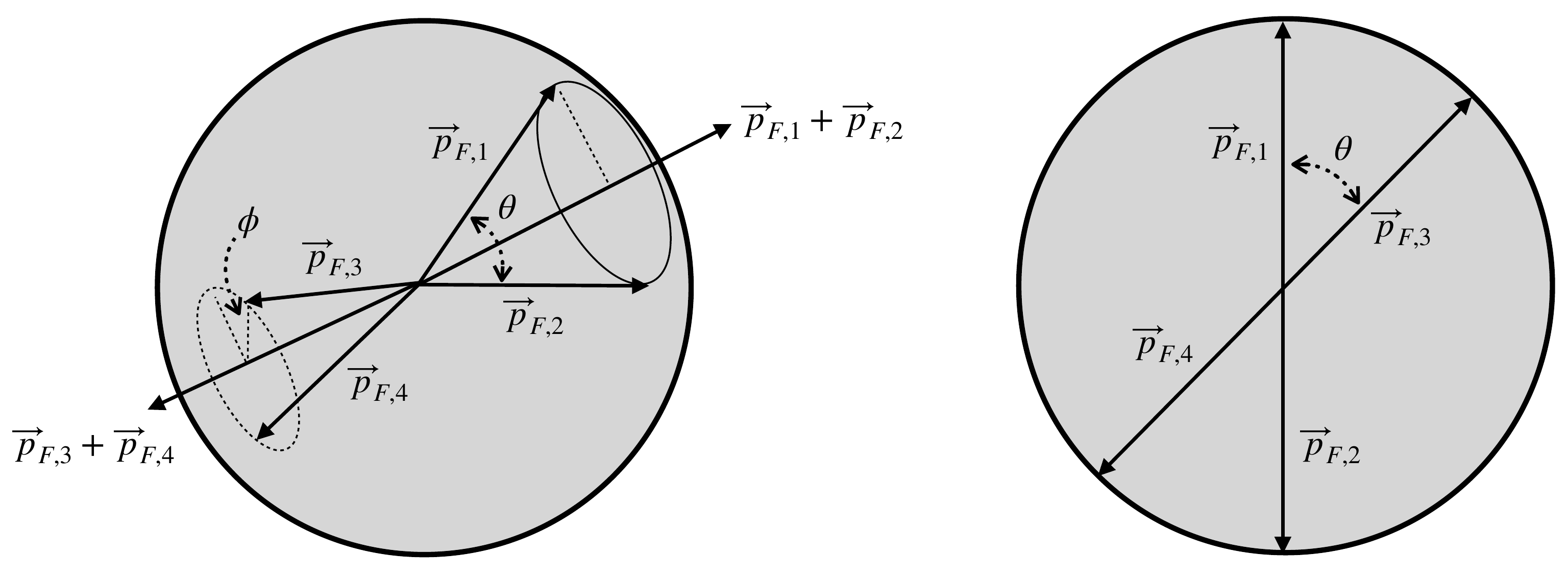}
	\end{center}
	\vspace{-.6cm}
	\caption{\small Interactions that conserve momentum on the Fermi surface.} \label{fig: scattering Fermi surface}
\end{figure}

Let's first discuss the momentum dependence. The requirement of translational invariance is particularly stringent given that all the $\vec p_{F,i}$'s must simultaneously be on the Fermi surface. To see this, let's choose $\vec p_{F,1}$ and $\vec p_{F,2}$ along two arbitrary directions in momentum space. These two vectors will lie on a cone with axis in the direction of $\vec p_{F,1} + \vec p_{F,2}$, as shown in the right panel of Fig. \ref{fig: scattering Fermi surface}. Conservation of momentum requires $\vec p_{F,3}$ and $\vec p_{F,4}$ to lie on a cone with the same axis but opposite orientation with respect to the origin. Up to a global rotation, this configuration is fully described by just two angles, {\it i.e.} the angle $\theta$ between $\vec p_{F,1}$ and $\vec p_{F,2}$, and the angle $\phi$ between the planes spanned by $\vec p_{F,1}, \vec p_{F,2}$ and $\vec p_{F,3}, \vec p_{F,4}$. Following~\cite{Shankar:1993pf}, we will refer to this configuration of momenta as the \emph{zero-sound channel}.

This channel becomes ill-defined in the special case where $p_{F,1} = - \vec p_{F,2}$. Momentum conservation then implies that $p_{F,3} = - \vec p_{F,4}$. These ``back-to-back'' interactions are characterized by a single angle $\theta$, as shown in the right panel of Fig \ref{fig: scattering Fermi surface}. We will refer to this type of interactions as the \emph{BCS channel}.
\begin{eBox}
{\bf Exercise 4.5:} Express the angles $(\theta,\phi)$ of the zero-sound channel and the angle $\theta$ of the BCS channel in terms of the labels $\vec p_{F,i}$.
\end{eBox}
Let's now restrict our attention to the BCS channel, for which the quartic interaction takes the form
\begin{align}
	S_{\rm BCS} = \sum_{\vec p,\vec p^{\, \prime}} \int dt \, d^3 x \,  g_{abcd} (\theta)  \psi_{\vec p^{\, \prime}}^{a \dag}   \psi_{\vec p}^b  \, \psi_{- \vec p^{\, \prime}}^{c\dag} \psi_{- \vec p}^d \ ,
\end{align}
with $\theta$ the angle between $\vec p$ and $\vec p^{\, \prime}$. It is easy to check that  the only two independent index structures that are allowed by the $SU(2)$ spin symmetry are 
\begin{align} \label{BCS channel tensor structure}
	g^{abcd} (\theta) &= g_1 (\theta) \, \delta^{ab} \delta^{cd} + g_2	 (\theta) \, \vec \sigma^{ab} \cdot \vec \sigma^{cd} \ . 
\end{align}
The functions $g_i(\theta)$ can be further decomposed into Legendre polynomials, effectively introducing an infinite number of coupling constants $g_{i, \ell}$: 
\begin{align}
	g_i (\theta) = \sum_{\ell} g_{i,\ell} P_\ell (\cos \theta) \ .
\end{align}
Taken at face value, this seems to defy one of the standard lores about EFTs, namely that only a finite number of coupling constants should appear in the action at any given order in the expansion parameter. This is because the standard lore is based on the implicit assumption that the EFT contains a finite number of fields. In our case, we have effectively traded the $\psi$ for an infinite number of fields $\psi_{\vec p}$. Notice however that predictivity is not lost: for any fixed value of $\theta$, the coupling $g_i (\theta)$ describes processes with arbitrary kinematical configurations of the ``soft'' momenta $\vec k$.

\begin{eBox}
{\bf Exercise 4.6:} Use the results derived in Exercise 4.5 to show that the angles in the zero-sound channel transform like $(\theta, \phi) \to (\theta, \pi+ \phi)$ under a parity transformation. Then, show that the only possible tensor structures in the zero-sound channel are also of the form \eqref{BCS channel tensor structure} if the theory is invariant under parity.   
\end{eBox}

In summary, we have seen that having a finite density of fermions drastically changes the power counting rules compared to the vacuum case and allows for the existence of marginal interactions of the form \eqref{fermi liquids marginal interactions}. These interactions can be divided into two channels---zero-sound and BCS---and are in principle described by an infinite number of couplings. As usual, quantum corrections can make these interactions either relevant or irrelevant, depending on the anomalous dimension of the couplings. It turns out that attractive interactions become relevant whereas repulsive interactions become irrelevant~\cite{Shankar:1993pf}. 

Whether our couplings are positive or negative for any given system can only be determined by experiments or by matching onto an underlying UV theory (whose couplings have been determined by other experiments). On general grounds, though, we can expect the couplings that turn relevant to give rise to remarkable phenomena at low energies. For example, in Helium 3 couplings in the BCS channel with $\ell =1$ are marginally relevant, prompting Helium atoms to pair up and condense at low temperatures, giving rise to superfluidity. In metals, instead, it is the BCS channel with $\ell =0$ which grows relevant and is responsible for ordinary superconductivity. A detailed discussion of these phenomena goes well beyond the scope of these lectures, but the interested reader can find some useful resources listed at the end of this section. Overall, I hope to have conveyed the idea that, like any EFT, Fermi liquid theory relies on a particular choice of degrees of freedom, expansion parameter and symmetries. Lots of interesting physics follows from just these few assumptions.

\subsection{Non-relativistic QED}

We will now consider a slightly more involved example, namely non-relativistic QED (NRQED). This is an EFT where non-relativistic fermions in vacuum experience long range electromagnetic interactions rather than the contact interaction in Eq. \eqref{S psi}. Unlike the EFTs we have discussed until now, NRQED features more than one relevant kinematic region, and as a result power counting becomes considerably more complicated. NRQED can be used for instance to study bound states made of equal-mass constituents, with the goal of calculating observable quantities in a systematic expansion in powers of the velocity $v$.\footnote{In this section we will work again in units where $c=1$.} It is in fact easy to argue that $v$ must be small in a bound state by combining the virial theorem and the uncertainty principle:
\begin{align}
	\frac{m v^2}{2} \sim \frac{\alpha}{r} \lesssim \alpha \, m v \qquad \Longrightarrow \qquad v \lesssim \alpha \ll 1. 
\end{align}
It is worth stressing that the motivation behind NRQED is different compared to other EFTs we have encountered so far in these lectures. Here we are not integrating out particles that are too heavy to be produced at low-energies, nor are we preoccupied with modeling the low-energy consequences of strongly coupled or unknown physics in the UV. Instead, we are interested in using a well understood, perturbative theory like QED to calculate some observable quantities in an expansion in powers of the small velocity $v$. The most efficient way to achieve this is to switch to an EFT such as NRQED, where power counting in $v$ is implemented directly a the level of the action.

Our starting point is the usual action for QED,
\begin{align}
	S = \int d^4 x \left\{ \bar \Psi [i \gamma^\mu (\partial_\mu - i e A_\mu) - m ] \Psi - \frac{1}{4} F_{\mu\nu} F^{\mu\nu} - \frac{1}{2\xi} (\partial_\mu A^\mu)^2 \right\},
\end{align}
where we have already added explicitly a gauge fixing term. In what follows we will work in the Feynman gauge, where $\xi = 1$. At energies much smaller than $m$, we can rephase the fermion field and integrate out the anti-particles exactly as we did in Sec. \ref{sec: NR fermions in vacuum}. Denoting $A^\mu = (\Phi, \vec A)$, we obtain the following tree-level effective action:
\begin{align} \label{NRQED action}
	S_{\rm eff}^{(0)} = \int dt \, d^3 x \left\{  \psi^\dag \left( i \partial_t  -e \Phi + \frac{\vec{D}^2}{2 m} + \frac{e}{2m} \vec B \cdot \vec \sigma \right) \psi  - \frac{1}{2} \partial_\mu A_\nu \partial^\mu A^\nu + \mathcal{O}(1/m^2) \right\} ,
\end{align}
with the 2-component spinor $\psi$ defined in Eq. \eqref{2-spinor def}, the electric and magnetic fields are defined as usual in terms of the scalar and vector potentials~\cite{Jackson:1998nia}, and $\vec D = \vec \nabla - i e \vec A$. This action is the analogue of the one in Eq. \eqref{S psi}, with the contact interaction replaced by one mediated by the photon. 
\begin{eBox}
{\bf Exercise 4.7:} Derive the tree-level effective action in Eq. \eqref{NRQED action}.
\end{eBox}
Notice that the term linear in $\vec B$ breaks the accidental $SU(2)$ spin symmetry and spatial rotations down to their diagonal subgroup. This term has the form $\vec \mu \cdot \vec B$, with  $\vec \mu = 2 \mu_B \vec S$ the density of magnetic moment, $\mu_B = q / (2 m c)$ the Bohr magneton, and $\vec S = \frac{1}{2}\tilde \Psi_-^\dag \sigma^i \tilde \Psi_-$ the spin density: it is (minus) the familiar potential energy of a magnetic dipole density in a magnetic~field. 

\begin{figure}
	\begin{center}
		\includegraphics[scale=0.3]{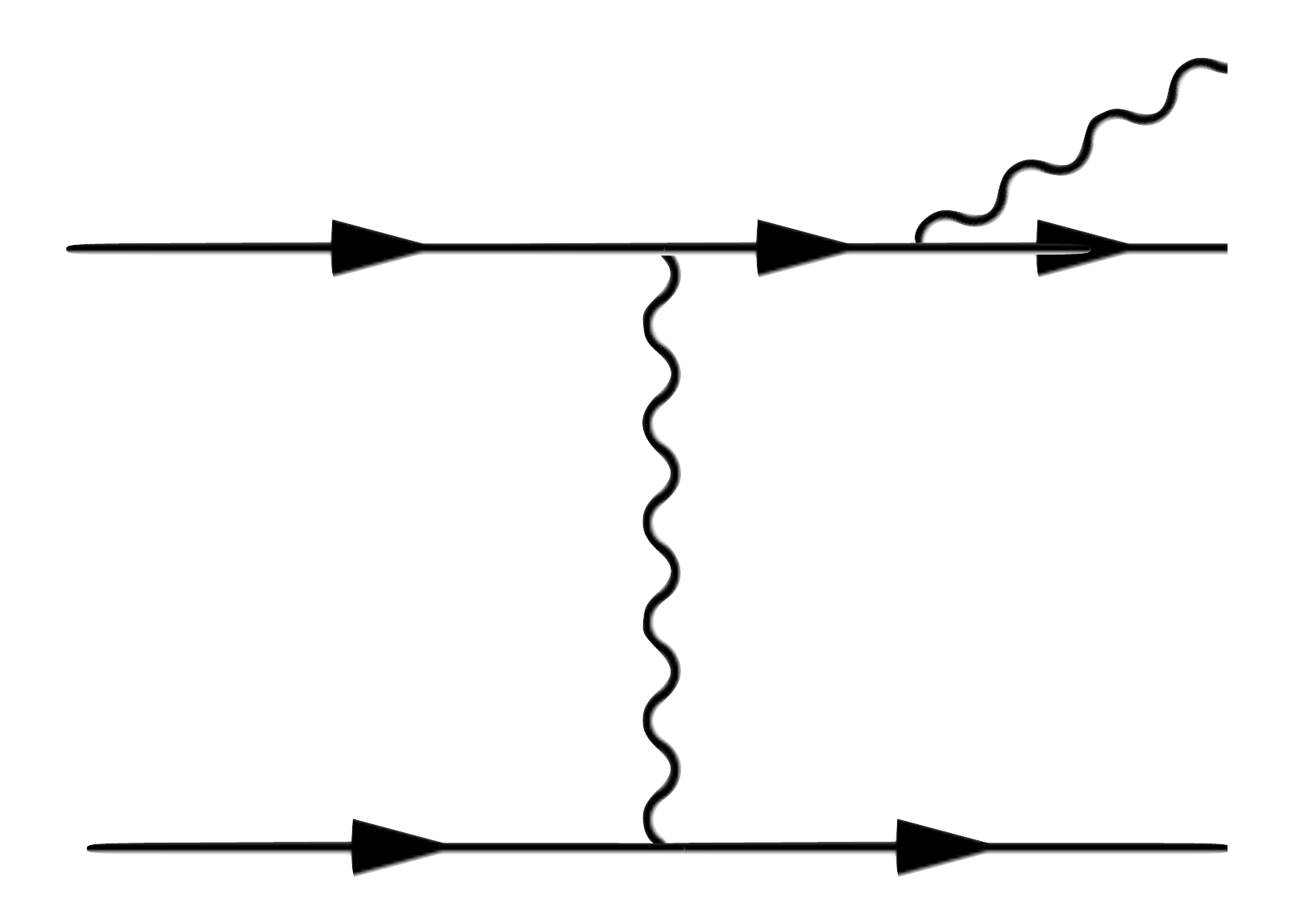}
	\end{center}
	\vspace{-.6cm}
	\caption{\small Sample tree-level diagram in QED.} \label{fig:NRQED}
\end{figure}

Gauge invariance demands that each derivative in the action \eqref{NRQED action} appears together with a gauge potential. For this reason, it is no longer clear that expanding the effective action in inverse powers of $m$ is equivalent to expanding in powers of $p / m = v$. For this to be the case, the potentials would also need to scale like $v$. Checking this requires in turn knowing how time and space derivatives act on the potentials, because then the scaling rule for the fields will follow as usual by demanding that the kinetic be a quantity of $\mathcal{O}(1)$. To this end, let's consider the diagram in Fig. \ref{fig:NRQED}. All external fermion lines have energy and momentum of order $(mv^2, mv)$. Thus, the energy and momentum exchanged in the scattering process, which is carried by the internal photon line, must generically be of order $(mv^2, mv)$ as well. We will call this kinematic region the \emph{potential} region. Consider now the photon emitted in this process: by conservation of energy it can at most have energy of $\mathcal{O}(mv^2)$. Since the corresponding external line must be on-shell, it must therefore carry energy and momentum of order $(mv^2, mv^2)$. We will refer to this kinematic region as the \emph{ultrasoft} region.\footnote{This is also known as the \emph{radiation} region.} This shows that the momentum carried by $A^\mu$ depends on the role that the photon line is playing in a Feynman diagram.  Taken at face value, this result seems to prevent us from assigning a definite in $v$ to the gauge potential. We will show in a moment how to address this problem and implement a well defined power counting in $v$. 

It turns out that the potential and ultrasoft kinematic regions are not the end of the story. To see this, we'll need to first make a little detour and discuss one more EFT technique, known as the \high{method of regions}. We will first introduce it using a simple example, following~\cite{Griesshammer:1997wz}. Consider the following one-dimensional integral,
\begin{align} \label{toy integral regions}
	I(a, b) = \int_{- \infty}^{+\infty}  dk \, \frac{1}{k^2 - a^2+ i \epsilon}\, \frac{1}{k^2 - b^2 + i \epsilon},
\end{align}
whose structure is reminiscent of certain 1-loop integrals in relativistic field theories. We will be interested in a situation where the location of the two poles are widely separated, {\it i.e.} $ a / b \ll 1$. Then, $I(a, b)$ can be calculated as an expansion in powers of the small parameter $a / b$. It seems reasonable to assume that the dominant contributions to this integral will come from the regions around the poles. Therefore we are going to approximate the full integral as a sum of two integrals centered around $k= \pm a$ and $k=\pm b$.
\begin{align}
	I(a, b) \simeq  \left[ \int_{k^2 \sim a^2} dk + \int_{k^2 \sim b^2} dk \right] \, \frac{1}{k^2 - a^2+ i \epsilon}\, \frac{1}{k^2 - b^2 + i \epsilon}. 
\end{align}
In the first region we can expand the second term in the integrand in powers of $k^2/b^2$, whereas in the second one we can expand the first term in powers of $a^2/k^2$, to obtain 
\begin{align} \label{toy 2 regions}
	I(a, b) \simeq \sum_{n=0}^\infty \left[ -\frac{1}{b^2}\int_{k^2 \sim a^2} dk  \, \frac{1}{k^2 - a^2+ i \epsilon} \frac{k^{2n}}{b^{2n}}  + \int_{k^2 \sim b^2} \frac{dk}{k^2} \, \frac{1}{k^2 - b^2 + i \epsilon} \frac{a^{2n}}{k^{2n}}  \right]. 
\end{align}
The exact values of these two integrals will depend in general on how we choose the regions around $k= \pm a$ and $k=\pm b$---something we've been purposefully vague about until now. What we are going to do is extend these two regions to the entire real axis. At first this might seem like a very bad idea, since far away from the two poles the series expansions we have used do not seem justified. Moreover, the first integral is UV divergent for $n \neq 0$, whereas the original integral in Eq. \eqref{toy integral regions} was not. We can easily address this, however, by using dimensional regularization. Then, the two integrals over the entire real axis can be calculated exactly to obtain 
\begin{align}
	I(a, b) \simeq \frac{i \pi}{a} \frac{1}{b^2 - a^2} - \frac{i \pi}{b} \frac{1}{b^2 - a^2} = \frac{i \pi}{ab (a+b)}. 
\end{align}
Something remarkable has happened: despite our seemingly questionable intermediate step, our final result turns out to be not just a good approximation, but in fact the \emph{exact} value of $I(a,b)$! (This can be checked by calculating directly the integral in Eq. \eqref{toy integral regions} using standard contour integration methods.) The reason why were able to get away with extending the integrals to the entire real axis is that, in regions far away from each pole, all scales in the integrand can be neglected, and scale-less integrals vanish in dimensional regularization.

To summarize, we have shown that the integral in Eq. \eqref{toy integral regions} can be calculated by identifying all the poles in the integrand, and then adding up the contributions that arise when expanding around each pole. This strategy is known as the method of regions, and it is a very helpful tool when applied to loop integrals in EFTs. In our simple example we have summed the series and obtained an exact result to give you some confidence that this method actually works. In an EFT, however, we are more interested in the fact that the integrals in \eqref{toy 2 regions} are explicitly organized in an expansion in the small parameter $a / b$. 

In fact, if we think of the two factors in the integrand in Eq. \eqref{toy integral regions} as the propagators of two different particles, the expression \eqref{toy 2 regions} lends itself to a very interesting interpretation. The first integral is equal to a loop made of one propagator $1/(k^2 - a^2 + i \epsilon)$, $n$ static propagators $1/b^2$, and $n$ field insertions $k^2$. These are propagators and quadratic vertices that involve fields with momentum in the region $k^2 \sim a^2$. The second integral admits an analogous interpretation, with the relevant region being $k^2 \sim b^2$. The lesson is that, in order to organize $I(a,b)$ as an expansion in powers of $a/b$, we need to formally introduce a copy of each field for each kinematic region they are involved with, as well as additional interactions among all these copies. 

\begin{figure}
	\begin{center}
		\includegraphics[scale=0.3]{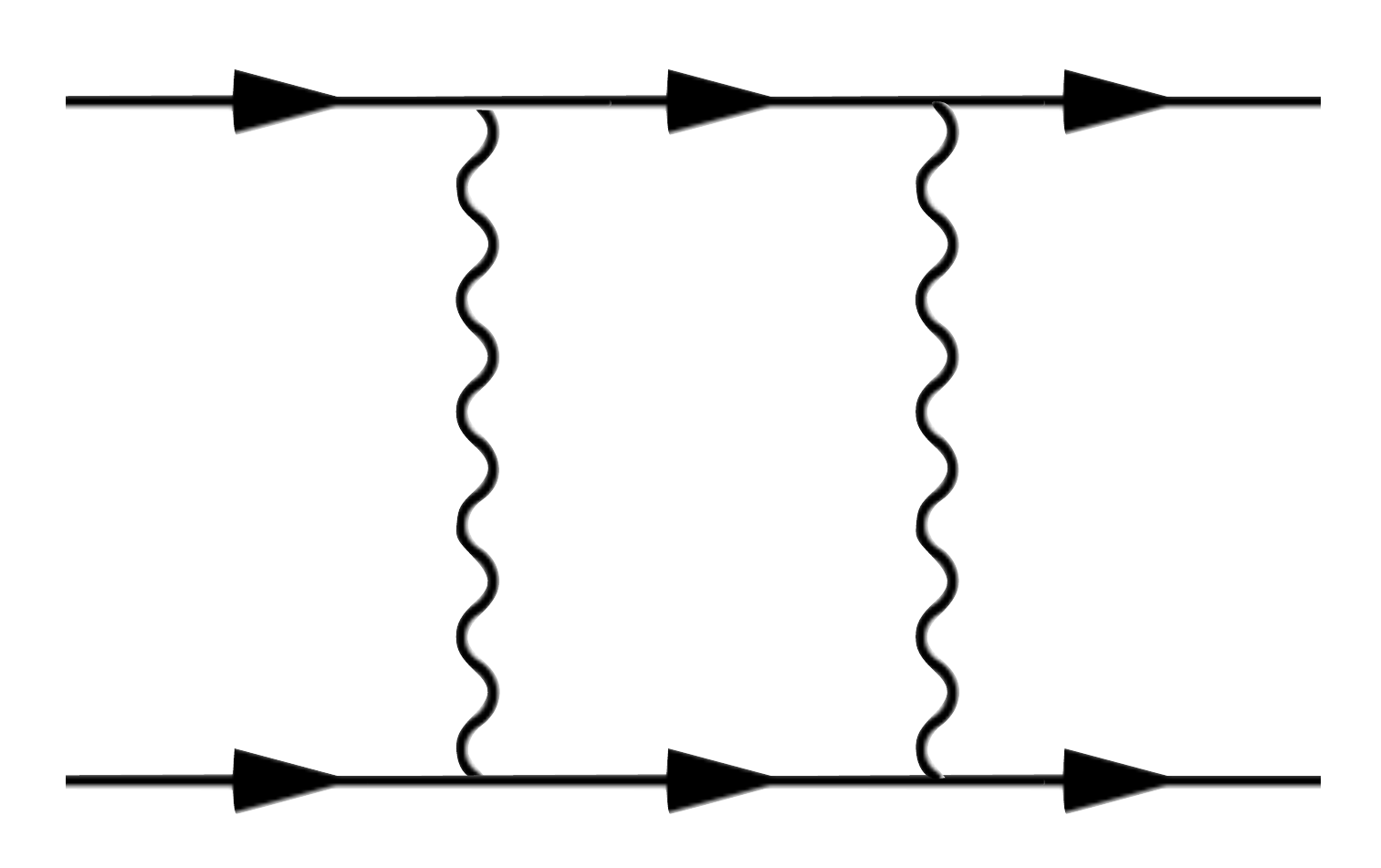}
	\end{center}
	\vspace{-.6cm}
	\caption{\small Sample one-loop diagram in QED.} \label{fig:NRQEDbox}
\end{figure}

Let's now return to our discussion of NRQED, and see how this idea can be implemented in practice. We have already identified two kinematic regions that play a role in NRQED (potential and ultrasoft), but it turns out that there is one more region we should consider. Take for instance the QED loop diagram shown in Fig. \eqref{fig:NRQEDbox}. We will work in the center-of-mass frame, and are interested in the situation where the external fermions are all non-relativistic. Up to an overall constant factor, this loop diagram is equal to the following integral:
\begin{align}
	& \int \frac{d^4 k}{(2\pi)^4} \,  N \, \frac{1}{k_0^2 - \vec{k}^{ 2} + i \epsilon } \, \frac{1}{(p_0 +k_0 - p_0')^2- (\vec p + \vec k - \vec{p}^{\,\prime})^2+ i \epsilon } \nonumber \\ 
	& \qquad \qquad \qquad \qquad \frac{1}{(p_0-k_0)^2 - (\vec p + \vec k)^2 - m^2 + i \epsilon } \, \frac{1}{(p_0+k_0)^2 - (\vec p + \vec k)^2- m^2 + i \epsilon   }, 
\end{align}
with $\pm \vec p$ ($\pm \vec p^{\, \prime}$) the 3-momentum of the incoming (outgoing) particles, and $N$ an analytic function of the momenta that won't play a role in our discussion. It is convenient to rewrite the integrand in terms of the kinetic energy of the fermions, $E = p^0 - m$, and to drop the factors of $i\epsilon$ to simplify the notation. The result is:
\begin{align}
	& \int \frac{d^4 k}{(2\pi)^4} \,  N \, \frac{1}{k_0^2 - \vec{k}^{ 2} + i \epsilon } \, \frac{1}{(E - E' +k^0)^2- (\vec p + \vec k - \vec{p}^{\,\prime})^2+ i \epsilon } \nonumber \\ 
	& \qquad \qquad \qquad \qquad \qquad \qquad \frac{1}{E^2 - 2E k^0 - (\vec p + \vec k)^2 + i \epsilon } \, \frac{1}{E^2 + 2 E k^0 - (\vec p + \vec k)^2+ i \epsilon   }.
\end{align}
According to our previous discussion, the location of any pole in the integrand corresponds to a relevant kinematic region. Given that $E, E' \sim mv^2$ and $p, p' \sim mv$, the second propagator has a pole for $(k^0, \vec k) \sim (mv, mv)$. We will call this the \emph{soft} region.  Notice that when the internal photon lines carry a soft 4-momentum, the internal fermion lines must do so as well by conservation of energy. The need for a soft region arises only when quantum loops are considered. In fact, it doesn't play any role in non-relativistic General Relativity (NRGR)~\cite{Goldberger:2007hy,Porto:2016pyg}, where one is interested in \emph{classical} gravitational bound states. 

It turns out that the soft, potential, and ultrasoft regions are sufficient to systematically implement a velocity expansion in NRQED. To this end, we are going to decompose the fields $\psi$ and $A^\mu$ as follows:
\begin{subequations}\label{psi A decomposition}
\begin{align} 
	\psi &= \psi_{\rm S} + \psi_{\rm P} \\ 
	A^\mu &= A^\mu_{\rm S} + A^\mu_{\rm P} + A^\mu_{\rm U},
\end{align}
\end{subequations}
were the subscripts denote the corresponding regions. Notice that there is no need to introduce an ultrasoft fermion. Interactions between the various fields introduced above must preserve energy and momentum with an accuracy of $\mathcal{O} (mv^2)$. At this stage it is not obvious how to implement this, given the fields in the soft and potential region have energy and/or momentum of $\mathcal{O}(mv)$. Moreover, it is not clear how one should power count interactions between fields in different regions: how is the integration measure $d^4x$ supposed to scale?  To address these problems, we can rephase our soft and potential fields to extract a phase of $\mathcal{O}(mv)$:
\begin{subequations} \label{NRQED rephasing}
\begin{align}
	\psi_{\rm S} (x) &= \sum_p e^{i p \cdot x} \, \psi_{{\rm S}, p} (x) \\
	A_{\rm S}^\mu (x) &= \sum_p e^{i p \cdot x} \, A_{{\rm S}, p}^\mu (x) \\
	\psi_{\rm P} (x) &= \sum_{\vec p} e^{i \vec p \cdot \vec x} \, \psi_{{\rm P}, \vec p} \, (x) \\
	A_{\rm P}^\mu (x) &= \sum_{\vec p} e^{i \vec p \cdot \vec x} \, A_{{\rm P}, \vec p}^\mu \, (x) . 
\end{align}
\end{subequations}
Note that we have extracted a four-momentum from the soft fields, and a three-momentum from the potential fields. The procedure outlined above is not that dissimilar from what we did in Eq. \eqref{Psi tilde def} when we expressed the fermion field $\psi$ as a sum of many fields $\psi_{\vec p_F}$, except that now we had to take into account the existence of multiple regions. Incidentally, the extraction of a soft phase from the soft and potential fields can also be thought of as some sort of multipole expansion~\cite{Labelle:1996en,Grinstein:1997gv}. This makes intuitive sense, since ultrasoft fields can only probe the long wavelength behavior of soft and potential fields. 

\begin{table}[t!]
  \begin{center}
    \caption{Fields in NRQED and their properties.}
    \label{tab: NRQED}
    \vspace{.4cm}
    \begin{tabular}{cccc}
    	\toprule
      	\textbf{Region} & \textbf{Field} & \textbf{Scaling} & \textbf{Propagator} \\
      	\midrule
      Soft & $\begin{array}{c}
      		\psi_{{\rm S},  p} \\
      		A_{{\rm S}, p}^\mu 
      	\end{array}$ \vspace{.2cm} & $\begin{array}{l}
      		v^{3/2} \\
      		v 
      	\end{array}$ & $\begin{array}{c}
      		 i /(p^0 - i \epsilon) \\
      		-i \eta^{\mu\nu} / (p^{\, 2} - i \epsilon)
      	\end{array}$ \\
      	\midrule
      	Potential & $\begin{array}{c}
      		\psi_{{\rm P}, \vec p} \\
      		A_{{\rm P}, \vec p}^\mu 
      	\end{array}$ \vspace{.2cm} & $\begin{array}{l}
      		v^{3/2} \\
      		v^{3/2}
      	\end{array}$ & $\begin{array}{c}
      		 i /(k^0 - \vec{p}^{\, 2}/2m + i \epsilon) \\
      		-i \eta^{\mu\nu} / (\vec{p}^{\, 2}  - i \epsilon)
      	\end{array}$ \\
      	\midrule
      	Ultrasoft & $A_{{\rm U}}^\mu$ & $\!\!\!\! v^2$ & $-i \eta^{\mu\nu}/(k^2-i\epsilon)$ \\
      \bottomrule
    \end{tabular}
  \end{center}
\end{table}

By plugging the expressions \eqref{psi A decomposition} and \eqref{NRQED rephasing} in the the effective action \eqref{NRQED action} we can derive the dominant kinetic terms for the fields $\psi_{{\rm S}, p}, \psi_{{\rm P}, \vec p}, A_{{\rm S}, p}^\mu, A_{{\rm P}, \vec p}^\mu \, ,$ and $A_{{\rm U}}^\mu$, as well as their interactions. Because these fields only contain Fourier modes with residual energies and momenta of $\mathcal{O}(mv^2)$, conservation of four-momentum up to $\mathcal{O}(mv^2)$ is automatically guaranteed. Moreover, the appropriate scaling for the integration measure is now clearly $d^4x \sim 1/v^8$. It is then easy to figure out how the various fields must scale with $v$, and to calculate their propagators. The results are summarized in Table \ref{tab: NRQED}, with soft ({\it i.e.} label) momenta denoted with $p$ and ultrasoft ({\it i.e.} residual) momenta with $k$. Notice in particular that the ultrasoft gauge fields scales like partial derivatives---they are both of $\mathcal{O} (mv^2)$---and therefore can be used to build covariant derivatives that scale homogeneously: $D_\mu = \partial_\mu - i e A_\mu^{\rm U}$ 

\begin{eBox}
{\bf Exercise 4.8:} Derive the results summarized in Table \ref{tab: NRQED}. 
\end{eBox}
A detailed description of all Feynman rules and practical applications of NRQED is beyond the scope of this section, whose main goal was to introduce the method of regions. You can find more details in the references provided in the following section.

\subsection{Additional resources}

The standard reference for the EFT of Fermi liquids are Polchinski's TASI lectures~\cite{Polchinski:1992ed}. For a more exhaustive treatment of fermions at finite density based on the renormalization group, see~\cite{Shankar:1993pf}. A pedagogical early review of Helium 3 superfluidity is~\cite{Leggett:1975te}. The use of NRQED to study bound states was first advocated in~\cite{Caswell:1985ui}. Most of the intricacies of power counting in non-relativistic gauge theories have been sorted out in the context of non-relativistic QCD (NRQCD). For more details on NRQCD, see for instance~\cite{Rothstein:2003mp,Griesshammer:1997wz}. You can also find a calculation of the Lamb shift based on NRQED in~\cite{Rothstein:2003mp}.

\newpage

\section*{Acknowledgments}

I would like to thank the organizers of the 2nd Joint ICTP-Trieste/ICTP-SAIFR school on Particle Physics (Enrico Bertuzzo, Joan Elias Mir\'o, Rogerio Rosenfeld, and Giovanni Villadoro) for inviting me to give these lectures, and all the participants and other lecturers for the stimulating discussions. In preparing these notes, I have benefited greatly from reading the many excellent reviews on the subject, most of which I have cited where appropriate. I would also like to thank Angelo Esposito, Walter Goldberger, Garrett Goon, Alberto Nicolis, Rachel Rosen, Adam Solomon, and especially Ira Rothstein for helpful discussions and comments on earlier drafts of these notes. This work was supported in part by the National Science Foundation under Grant No. PHY-1915611.

\appendix

\bibliography{$HOME/Dropbox/Bibtex/library.bib}{}
\bibliographystyle{hunsrt}

\end{fmffile}

\end{document}